%% file: paper.tex
\documentclass[aps,prd,preprint,superscriptaddress]{revtex4-1}
\usepackage[english]{babel}
\usepackage[utf8]{inputenc}
\usepackage[T1]{fontenc}
\usepackage{amsmath,amssymb,braket,slashed,mathrsfs}
\usepackage{pifont}
\usepackage{hyperref}
\usepackage{graphicx}
\graphicspath{{./figures/}}
\usepackage{tikz}
\usetikzlibrary{shapes,arrows,positioning}
\tikzset{decision/.style={diamond, draw, fill=blue!20, text width=4.5em, text badly centered, inner sep=0pt}}
\tikzset{block/.style={rectangle, draw, fill=blue!20, text width=10em, text centered, rounded corners, minimum width=3.5cm}}
\tikzset{block1/.style={rectangle, draw, fill=blue!20, text width=18.5em, text centered, rounded corners, minimum width=3.5cm}}
\tikzset{line/.style={draw, -latex, thick}}
\usepackage{color,hyperref}
\usepackage{cleveref}
\newcommand{\ba}{\begin{eqnarray}}
\newcommand{\ea}{\end{eqnarray}}
\newcommand{\bi}{\begin{itemize}}
\newcommand{\ei}{\end{itemize}}

\newcommand{\nn}{\nonumber}
\newcommand{\MS}{\overline{\mathrm{MS}}}
\newcommand{\MSB}{\overline{\textbf{MS}}}
\newcommand{\lat}{\mathrm{lat}}

\newcommand{\muRI}{\mu_0}
\newcommand{\RI}{\mathrm{RI}}
\newcommand{\RIB}{\textbf{RI}}
\newcommand{\muMS}{\mu_{\overline{\mathrm{MS}}}}

\usepackage{multirow,array}

\definecolor{linkcolor}{rgb}{.17578125,.1875,.5703125}
\hypersetup{
    pdfstartview={FitH},    
    pdftitle={Rare kaon decays: a lattice study},    
    pdfauthor={RBC and UKQCD collaborations},           
    pdfsubject={Physics article},                
    pdfcreator={Xu Feng},          
    pdfkeywords={particle} {physics} {quantum} {chromodynamics} {QCD}  {hadrons} {kaons} {rare decays} {flavor} {FCNC} {lattice}, 
    colorlinks=true,        
    linkcolor=linkcolor,    
    citecolor=linkcolor,    
    filecolor=black,        
    urlcolor=linkcolor
}

\newcommand{\cu}{Physics Department, Columbia University, New York, NY 10027, USA}
\newcommand{\soton}{Department of Physics and Astronomy, University of Southampton, Southampton SO17 1BJ, UK}
\newcommand{\pkuphy}{School of Physics and State Key Laboratory of Nuclear
Physics and Technology, Peking University, Beijing 100871, China}
\newcommand{\innovation}{Collaborative Innovation Center of Quantum Matter, Beijing 100871, China}
\newcommand{\chep}{Center for High Energy Physics, Peking University, Beijing 100871, China}
\newcommand{\uoe}{School of Physics and Astronomy, University of Edinburgh, Edinburgh EH9 3JZ, UK}

\begin{document}
\title{Exploratory lattice QCD study of the rare kaon decay \boldmath{$K^+\to\pi^+\nu\bar{\nu}$}}

\author{Ziyuan~Bai}\affiliation{\cu}
\author{Norman~H.~Christ}\affiliation{\cu}
\author{Xu~Feng}\affiliation{\pkuphy,\innovation,\chep}
\author{Andrew~Lawson}\affiliation{\soton}
\author{Antonin~Portelli}\affiliation{\uoe,\soton}
\author{Christopher~T.~Sachrajda}\affiliation{\soton}
\collaboration{RBC and UKQCD collaborations}
\pacs{PACS}

\date{\today}

\begin{abstract}
In Ref\,\cite{Bai:2017fkh} we have presented the results of an exploratory lattice QCD computation of the 
long-distance contribution to the $K^+\to\pi^+\nu\bar{\nu}$ decay amplitude. 
In the present paper we describe the details of this calculation, which includes the implementation of a number of novel techniques.
The $K^+\to\pi^+\nu\bar{\nu}$ decay amplitude is dominated by short-distance contributions 
which can be computed in perturbation theory with the 
only required non-perturbative input being the relatively well-known form factors of semileptonic kaon decays. 
The long-distance contributions, which are the target of this work, are expected to be of $O(5\%)$ 
in the branching ratio. Our study demonstrates the feasibility of lattice QCD computations of the
$K^+\to\pi^+\nu\bar{\nu}$ decay amplitude, and in particular of the long-distance component.
Though this calculation is performed on a small lattice ($16^3\times 32$) and at unphysical pion, kaon and charm quark masses,
$m_\pi=420$\,MeV, $m_K=563$\,MeV and $m_c^{\MS}(\mbox{2 GeV})=863$\,MeV,
the techniques presented in this work can readily be applied to a future realistic
calculation.

\end{abstract}

\maketitle

\tableofcontents

\section{Introduction}
\input{Introduction}

\section{Numerical evaluation of hadronic matrix elements}
\label{sec:matrix_element}
\input{Matrix_element}

\input{WW}
\input{Z_exchange}

\section{Removal of the short-distance divergence using nonperturbative
renormalization}
\label{sec:SD_NPR}
\input{NPR}

\section{Perturbative elements in the determination of the decay amplitude}
\label{sec:SD_PT}
\input{PT_results}

\section{Lattice results and a discussion of systematic uncertainties}
\label{sec:results}
\input{Summary_results}

\section{Conclusions and future prospects}
\label{sec:conclusion}
\input{Conclusion}

\appendix
\section{Free lepton propagator using the overlap formalism}
\input{Appendix_lepton_prop}

\label{sec:applepton_prop}
\section{Evaluation of $\RIB\boldsymbol{\to}\MSB$ conversion for the bilocal
operator: \boldmath$Y_{AB}(\mu,\mu_0)$}\label{sec:appYAB}
\input{Appendix_YAB}

\section{Finite volume effects in the \boldmath{$W$-$W$} diagrams}\label{sec:appFVWW}
\input{Appendix_FV}

\bibliography{paper}
\end{document}

%% file: Introduction.tex
$K\to\pi\nu\bar{\nu}$ decays provide an excellent probe for searching for new physics 
(as recalled in Sec.\,\ref{subsec:probing} below). The decays are dominated by short-distance contributions (from top-quark loops with also a significant contribution 
from the charm quark in $K^+\to\pi^+\nu\bar\nu$ decays) which can be calculated to a good precision using perturbation theory with the 
only required non-perturbative input being the relatively well-known form factors of semileptonic kaon decays. The target of the current study is the evaluation of the long-distance (LD) contributions to the 
$K^+\to\pi^+\nu\bar\nu$ decay amplitude and 
phenomenological estimates suggest that they are of the order of about 5\%\,\cite{Isidori:2005xm}. 

The techniques required to compute the long-distance contributions to $K^+\to\pi^+\nu\bar\nu$ decay amplitudes were developed in Ref.\,\cite{Christ:2016eae}. They have subsequently been applied to an exploratory computation on a $16^3\times 32$ lattice at unphysical pion, kaon and charm quark masses
($m_\pi=420$\,MeV, $m_K=563$\,MeV and $m_c^{\MS}(\mbox{2 GeV})=863$\,MeV) and the results were reported in 
the letter\,\cite{Bai:2017fkh}. The purpose of this paper is to present the details of this computation, demonstrating how the various novel ideas from Ref.\,\cite{Christ:2016eae} can be implemented in an actual calculation. Our study 
demonstrates the feasibility of lattice QCD computations of the
$K^+\to\pi^+\nu\bar{\nu}$ decay amplitude, and in particular its long-distance component so that 
these techniques can readily be applied to a future realistic calculation.

As a strangeness ($S$) changing second-order weak interaction process, within the Standard Model
the calculation of the $K^+\to\pi^+\nu\bar{\nu}$ decay amplitude involves diagrams with the exchange of two $W$ 
bosons ($W$-$W$ diagrams), or those with the exchange of one $W$ and one $Z$ boson ($Z$-exchange diagrams) 
or those with a loop containing a $W$-$W$-$Z$ vertex. The long-distance contributions are given by the $W$-$W$ 
and $Z$-exchange diagrams. Their evaluation requires the computation of the matrix elements of bilocal operators 
composed of two local operators of the effective Hamiltonian (in which the $W$s and $Z$s are contracted to a point) 
and we include all the connected, closed quark-loop and disconnected
contractions in the correlation functions. 
The three main difficulties which had to be overcome, and which will be described in detail in the following sections, are:
\begin{enumerate}
\item[i)] the removal of the
unphysical terms which appear in second-order Euclidean correlation functions. When there are
intermediate states propagating between the two local operators which are lighter than the mass of the 
kaon, $m_K$ (we take the kaon 
to be at rest), then these terms grow exponentially with the range of the integration over the temporal separation of 
the two operators (see Sec.\,\ref{subsubsec:exponential});
\item[ii)] the subtraction of the additional ultraviolet divergences which arise from the integration
region where the two local operators comprising the bilocal operator approach
each other (see Secs.\,\ref{subsec:LD},\,\ref{subsec:method} and \ref{sec:SD_NPR}) and 
\item[iii)] the finite-volume corrections associated with
on-shell intermediate states with energies smaller than $m_K$ (see Sec.\,\ref{subsubsec:FV}).
\end{enumerate}

The plan for the remainder of this paper is as follows. In the following section we present an overview of 
the importance of $K\to\pi\nu\bar{\nu}$ decays as a probe for possible new physics, explain what we mean by long-distance contributions and give an outline of how lattice computations can be used to compute their contribution to the decay amplitude.
The following three sections contain the details of the three main elements of the computation of the long-distance contributions to the amplitude for the rare-kaon decay $K^+\to\pi^+\nu\bar\nu$. Sec.\,\ref{sec:matrix_element} contains a description of the computation of the matrix element of bare lattice bilocal operators, i.e. of the product of the two local weak operators in the effective Hamiltonian. As the two operators approach each other, new ultraviolet divergences appear and we discuss the subtraction of these divergences in Sec.\,\ref{sec:SD_NPR}. In the next section,
Sec.\,\ref{sec:SD_PT}, we discuss two perturbative aspects of the calculation. One of these is the calculation of the matching factor relating the
matrix elements computed non-perturbatively to those in the (purely perturbative) $\MS$ scheme. In this section we also 
follow the standard procedure of integrating out the charm quark so that the amplitude can be obtained using perturbation theory and the form-factors from $K_{\ell 3}$ decays. We compare this result with the non-perturbative lattice determination of the amplitude in Sec.\,\ref{sec:results} where we combine the elements from the earlier sections to obtain our final results. 
In Sec.\ref{sec:conclusion} we present a brief summary and discuss prospects for our future calculations at physical quark masses. There are three
appendices in which we discuss the free lepton propagator in the overlap formalism\,(Appendix\,\ref{sec:applepton_prop}); the details of the evaluation of the matching constant for bilocal operators in the RI-SMOM and $\MS$ renormalization schemes\,(Appendix\,\ref{sec:appYAB}) and finally a discussion of the finite-volume effects for the $W$-$W$ class of diagrams\, (Appendix\,\ref{sec:appFVWW}).

\section{Brief overview of \boldmath{$K^+\to\pi^+\nu\bar\nu$ decays}}\label{sec:overview}
We begin this section with a brief overview of the importance of $K\to\pi\nu\bar{\nu}$ decays as 
a probe for possible new physics and summarize the current status of experimental measurements of their decay widths. We then explain what we mean by the long-distance contributions to the $K^+\to\pi^+\nu\bar\nu$ decay amplitude in Sec.\,\ref{subsec:LD} and 
quote phenomenological estimates that they are of the order of a few percent\,\cite{Isidori:2005xm}. 
In Sec.\,\ref{subsec:method} we outline the procedure for calculating the long-distance contributions non-perturbatively 
in lattice simulations, focussing in particular on the renormalization of bilocal operators. More details are then given in the following sections.

\subsection{Probing new physics with the rare kaon decays \boldmath{$K\to\pi\nu\bar{\nu}$}}\label{subsec:probing}
As flavor-changing-neutral-current (FCNC) processes, the leading contributions to $K\to\pi\nu\bar{\nu}$ decay amplitudes are genuine one-loop electroweak effects, 
usually described by the following $O(G_F^2)$ effective Hamiltonian~\cite{Buchalla:1993wq,Buchalla:1998ba}
\begin{equation}
    \label{eq:eff_Hamiltonian}
{\mathcal
H}_{\mathrm{eff},0}=\frac{G_F}{\sqrt{2}}\frac{\alpha}{2\pi\sin^2\theta_W}\sum_{\ell=e,\mu,\tau}\left[\lambda_tX_t(x_t)+\lambda_cX_c^\ell(x_c)\right]\,[(\bar{s}d)_{V-A}(\bar{\nu_\ell}\nu_\ell)_{V-A}],
\end{equation}
where $X_t(x_t)$ and $X_c^\ell(x_c)$ indicate the top and charm quark
contributions respectively and the label $\ell$ indicates the leptonic flavor quantum number.
The loop functions $X_q(x_q)$ behave as $X_q(x_q)\propto x_q\equiv\frac{m_q^2}{M_W^2}$~\cite{Inami:1980fz} leading to a quadratic Glashow-Iliopoulos-Maiani (GIM) mechanism.
Thus the dominant contribution to the $K\to\pi\nu\bar{\nu}$ amplitude comes from the internal top quark loop.
From Eq.\,(\ref{eq:eff_Hamiltonian}) we see that compared to the tree-level
semi-leptonic decay $K\to\pi\ell\nu_\ell$, the rare kaon decay is suppressed by a factor of 
${\mathcal N}\simeq\frac{\alpha}{2\pi\sin^2\theta_W}\frac{\lambda_t}{\lambda}X_t(x_t)$.
The Cabibbo-Kobayashi-Maskawa (CKM) factor $\lambda_q$ is defined as $\lambda_q=V_{qs}^*V_{qd}$,
$\lambda=|V_{us}|$ and numerically $\frac{\lambda_t}{\lambda}=O(\lambda^4)$. $\alpha$
is the electromagnetic fine-structure constant and $\theta_W$ is the Weinberg
angle.
 The top-quark loop function $X_t(x_t)$ is known up to NLO QCD corrections~\cite{Buchalla:1998ba,Misiak:1999yg}
and two-loop EW contributions~\cite{Brod:2010hi}.
The estimate of $X_t(x_t)=1.481(9)$~\cite{Buras:2015qea} suggests a suppression of ${\mathcal N}\simeq 2\times 10^{-5}$
in the Standard Model (SM). Thus this decay channel can be used to probe the new physics at the scales of
${\mathcal N}^{-\frac{1}{2}}M_W=O(10\mbox{ TeV})$ or higher.

The theoretical cleanliness described above is an important reason making $K\to\pi\nu\bar{\nu}$
decays among the most interesting processes in the phenomenology of rare decays. 
The loop functions $X_t(x_t)$ and $X_c^\ell(x_c)$ can be calculated using QCD and electroweak perturbation 
theory~\cite{Buchalla:1998ba,Misiak:1999yg,Brod:2010hi,Buras:2005gr,Buras:2006gb,Brod:2008ss}.
The non-perturbative hadronic matrix element of the local four-fermion
operator in Eq.\,(\ref{eq:eff_Hamiltonian}) can be determined accurately from the
experimental measurement of the semileptonic decay $K\to\pi \ell\nu_\ell$ using an isospin rotation~\cite{Mescia:2007kn}.
As a result, the SM predictions for the branching ratios of $K\to\pi\nu\bar{\nu}$ decays,~\cite{Buras:2015qea}
\begin{eqnarray}
\mathrm{Br}(K^+\to\pi^+\nu\bar{\nu})_{\mathrm{SM}}&=&9.11(72)\times10^{-11},
\nonumber\\
\mathrm{Br}(K_L\to\pi^0\nu\bar{\nu})_{\mathrm{SM}}&=&3.00(30)\times10^{-11},
\end{eqnarray}
can be determined to a precision of about 10\%. This is considerably better than 
the precision of the previous experimental
measurements~\cite{Adler:1997am,Adler:2000by,Adler:2001xv,Adler:2002hy,Anisimovsky:2004hr,Artamonov:2008qb,Ahn:2009gb}
\begin{eqnarray}
\mathrm{Br}(K^+\to\pi^+\nu\bar{\nu})_{\mathrm{exp}}&=&1.73^{+1.15}_{-1.05}\times10^{-10},
\nn\\
\mathrm{Br}(K_L\to\pi^0\nu\bar{\nu})_{\mathrm{exp}}&\le&2.6\times10^{-8},
\end{eqnarray}
motivating the new generation of experiments designed to search for these rare decay events. 
The NA62 experiment at CERN aims to obtain $O(100)$ events in 2-3 years
and will thus test the SM at a 10\% precision~\cite{fortheNA62:2013jsa}. 
The search for $K_L\to\pi^0\nu\bar{\nu}$ decays is more challenging, since all the particles
in the initial and final state are neutral. The KOTO experiment at
J-PARC is designed to search for $K_L$ decays~\cite{Yamanaka:2012yma}.
It has observed one candidate event while expecting 0.34(16) background events
and set an upper limit of
$5.1\times10^{-8}$ for the branching ratio at 90\% confidence
level~\cite{Ahn:2016kja}.

\subsection{Long-distance contributions to \boldmath{$K\to\pi\nu\bar{\nu}$} decays}\label{subsec:LD}

We have seen that the dominant contribution to $K\to\pi\nu\bar{\nu}$ decay amplitudes comes from the top quark loop.
As a CP-violating decay, whose amplitude is proportional to the imaginary parts of the $\lambda_q$, the $K_L\to\pi^0\nu\bar{\nu}$ process is completely short-distance
(SD) dominated and thus does not require a lattice QCD calculation of 
long-distance effects.
On the other hand, for the CP-conserving $K^+\to\pi^+\nu\bar{\nu}$
decay, there is an enhancement of the charm-quark contribution, since the corresponding CKM factor, $\lambda_c$,
is much larger than that for the top-quark loop, $\lambda_c\gg\lambda_t$. This enhancement
makes the charm quark contribution important; neglecting it would reduce the 
theoretical estimate for the branching ratio by a factor of about 2. At leading order of QCD perturbation theory, i.e. $O(\alpha_S^0)$, Inami and Lim's
calculation~\cite{Inami:1980fz} suggested that
the charm-quark contribution is dominated by SD physics,
which receives contributions from energy scales ranging from
the mass of the $W$-boson, $\mu=O(M_W)$, to that of the charm quark, 
$\mu=O(m_c)$, leading to an enhancement factor of $\ln(M_W^2/m_c^2) \approx
8.4$. 
However, when higher-order QCD corrections are included,
this enhancement is
significantly reduced~\cite{Buchalla:1993wq}. As a consequence, the precise
determination of the long-distance (LD) contribution becomes more important.

We now clarify what we mean by the LD contributions by sketching the general procedure used to perform the calculation. We start by integrating out the $W$ and $Z$ bosons in order to explore the bilocal structure of the charm-quark contribution
to the $K^+\to\pi^+\nu\bar{\nu}$ decay amplitude.
The transition amplitude takes the form:
\begin{equation}
\label{eq:amplitude}
\langle\pi^+\nu\bar{\nu}|\{C_A^{\MS}Q_A^{\MS}\,C_B^{\MS}Q_B^{\MS}\}_\mu^{\MS}|K^+\rangle +C_0^{\MS}(\mu)\langle\pi^+\nu\bar{\nu}|Q_0^{\MS}(\mu)|
K^+\rangle,
\end{equation}
where we have used the notation $\{Q_A^{S}Q_B^{S}\}^{S'}=\int d^4x\,T\{Q_A^S(x)Q_B^S(0)\}^{S'}$. 
Here $Q_{A,B}$ are local operators appearing in the first-order effective weak Hamiltonian from $W$ and $Z$ exchange, the superscript 
$S$ indicates the renormalization scheme used to define them and $C_{A,B}$ are the corresponding Wilson coefficient functions. The label $S'$
specifies the scheme used to define the bilocal operator and to remove the additional ultraviolet divergence present when $x\to0$. A sum over the 
relevant operators $Q_{A,B}$ is implied.
In Eq.~(\ref{eq:amplitude}) both $S$ and $S'$ denote the
$\MS$ scheme, but in order to obtain the matrix elements in the $\MS$ scheme from a lattice simulation we need to introduce intermediate renormalization schemes as discussed in the following subsection. At the scale $\mu$ (at this stage $m_c<\mu<M_W$), the transition amplitude is separated into a 
bilocal component $\{C_A^{\MS}Q_A^{\MS}\,C_B^{\MS}Q_B^{\MS}\}_\mu^{\MS}$ and the local term $C_0^{\MS}(\mu)Q_0^{\MS}(\mu)$.
The local operator $Q_0^{\MS}=(\bar{s}d)_{V-A}(\bar{\nu}\nu)_{V-A}$ and the second term on the right-hand side of Eq.\,(\ref{eq:amplitude}) is required to fully match the SM, and in particular the SD contributions, to the effective theory.
The coefficients $C_A(\mu)$, $C_B(\mu)$ and $C_0(\mu)$ can be
determined using NNLO QCD perturbation theory~\cite{Buras:2006gb}. 

The next step in the conventional approach is to 
integrate out the charm quark field in the bilocal term; this is schematically represented by 
\begin{equation}
\{C_A^{\MS}Q_A^{\MS}\,C_B^{\MS}Q_B^{\MS}\}_\mu^{\MS}\quad \to\quad
C_A^{\MS}(\mu)C_B^{\MS}(\mu)r_{AB}^{\MS}(\mu)Q_0^{\MS}(\mu),
\end{equation}
where the parameter $r_{AB}^{\MS}(\mu)$ can be calculated using QCD perturbation theory
and the hadronic matrix element of $Q_0^{\MS}(\mu)$ can be determined from the
experimental measurement of $K_{\ell3}$ decays.
To estimate the remaining LD contributions, the authors of Ref.\,\cite{Isidori:2005xm}
have taken into account and estimated the matrix elements of local FCNC operators of dimension eight, such as
$(\bar{s}\Gamma\partial_\mu
d)\times(\bar{\nu}\Gamma\partial^\mu\nu)$, where $\Gamma$ represents a Dirac matrix, and used chiral perturbation theory. They find that this contribution is 
$\delta P_{c}=0.04\pm0.02$ which 
enhances the branching ratio $\textmd{Br}(K^+\to\pi^+\nu\bar{\nu})_{\textmd{SM}}$ by 6\%. However, at the charm quark mass scale $\mu=O(\mbox{1 GeV})$,
it is doubtful whether the operator product expansion converges very well and one can also 
have reservations about the precision of perturbation theory.
Integrating out the charm quark may therefore constitute a source of uncontrolled
theoretical uncertainty. We therefore,
proposed in Ref.~\cite{Christ:2016eae} to keep the charm quark as a dynamical degree of freedom and to calculate the bilocal matrix element 
$\langle\pi^+\nu\bar{\nu}|\{C_A^{\MS}Q_A^{\MS}\,C_B^{\MS}Q_B^{\MS}\}_\mu^{\MS}|K^+\rangle$ directly using
lattice QCD at a scale $\mu>m_c$ where perturbation theory can be used more reliably. In this way we calculate the transition amplitude in Eq.\,(\ref{eq:amplitude}) fully and directly. In principle therefore, we do not need to talk about the separation of long- and short- distance contributions, but to be definite we simply call the long-distance contributions to be the bilocal term $\langle\pi^+\nu\bar{\nu}|\{C_A^{\MS}Q_A^{\MS}\,C_B^{\MS}Q_B^{\MS}\}_\mu^{\MS}|K^+\rangle$ in Eq.\,(\ref{eq:amplitude}). This matrix element of the bilocal operator is of course scale dependent; here we
simply require that $\mu>m_c$ and is sufficiently large for perturbation theory to be reliable.

An interesting question is to what extent is $P_c-P_c^{\textrm{PT}}$, the difference between the full lattice result of the charm-quark contribution to the amplitude $P_c$ and that obtained using perturbation theory $P_c^{\textrm{PT}}$ combined with the matrix element of $Q_0^{\MS}$ from $K_{\ell 3}$ decays, estimated reliably. Lattice computations will be able to answer this question. We have seen above that a phenomenological study has estimated a correction of $\delta P_{c}=0.04\pm0.02$~\cite{Isidori:2005xm}.   

Using the results from NNLO QCD perturbation theory~\cite{Buras:2006gb}, we find that at a scale of $\mu=2.5$ GeV,
the bilocal contribution $C_A^{\MS}(\mu)C_B^{\MS}(\mu)r_{AB}^{\MS}(\mu)$ is of similar size to the local contribution
$C_0^{\MS}(\mu)$. Thus we would expect that the lattice calculation of the bilocal operator at such scales would account for approximately half of the full charm quark contribution.

The operators in Eq.\,(\ref{eq:amplitude}) are defined in the $\MS$ scheme. Since this scheme is purely perturbative, we cannot compute matrix elements of operators defined in the $\MS$ scheme directly using lattice QCD. In the following subsection we explain the procedure used to overcome this.

\subsection{Introduction to the lattice methodology}
\label{subsec:method}

There has been a series of lattice QCD studies of rare kaon 
decays~\cite{Isidori:2005tv,Sachrajda:2013vqa,Sachrajda:2013fxa,Feng:2015kfa,Christ:2015aha,Christ:2016eae,Christ:2016psm,Christ:2016awg,Christ:2016lro,Christ:2016mmq,Bai:2017fkh}.
The general lattice QCD method to calculate second-order electroweak amplitudes
has been developed in Refs.~\cite{Christ:2010gi,Christ:2012np,Christ:2015pwa}. 
It has been successfully applied to
the lattice calculation of the $K_L$-$K_S$ mass difference~\cite{Christ:2012se,Bai:2014cva}
and is currently being applied to the evaluation of the LD contribution 
to the indirect CP-violating parameter $\epsilon_K$\,\cite{Bai:2016gzv}.
The possibility of calculating rare kaon
decay amplitudes using lattice QCD was first proposed in 
Ref.~\cite{Isidori:2005tv}.  A more detailed method to calculate the
$K\to\pi\ell^+\ell^-$ decay amplitude was later developed in Ref.~\cite{Christ:2015aha} 
and applied to a first exploratory lattice QCD
calculation in Ref.~\cite{Christ:2016mmq}. 
These same techniques were also applied to the calculation of the
LD contribution to the $K^+\to\pi^+\nu\bar\nu$ decay amplitude
in Ref.\,\cite{Christ:2016eae}, in which a method was presented to combine the LD
contribution computed using lattice QCD with the SD components
determined using perturbation theory, including a consistent
treatment of the logarithmic singularities present in the LD and SD
contributions.

The discussion below follows Ref.~\cite{Christ:2016eae}. 
Since the $\MS$ scheme is purely perturbative, we cannot compute matrix elements of operators defined in the $\MS$ scheme directly using lattice QCD.  We therefore employ an intermediate RI/SMOM scheme and write
the $\MS$ bilocal operator in (\ref{eq:amplitude}) as
\begin{equation}
\label{eq:bilocal_MS_RI}
\{Q_A^{\MS}\,Q_B^{\MS}\}_\mu^{\MS}=Z_{Q_A}^{\RI\to\MS}(\mu/\mu_0)Z_{Q_B}^{\RI\to\MS}(\mu/\mu_0)\{Q_A^{\RI}\,Q_B^{\RI}\}_{\mu_0}^{\RI}
+Y_{AB}(\mu,\mu_0)Q_0^{\RI}(\mu_0).
\end{equation}
Given an operator $Q$, $Z_Q^{\RI\to\MS}$ is a conversion factor from the RI/SMOM to the $\MS$ scheme: 
$Q^{\MS}(\mu)=Z_Q^{\RI\to\MS}(\mu/\mu_0)Q^{\RI}(\mu_0)$ (more generally, when there is mixing of operators, as in the present case, $Z$ is a matrix). For compactness of notation we denote operators renormalised in the RI/SMOM scheme with the superfix {\footnotesize RI} and the precise choice of momenta used to define this scheme will be presented in Sec.\,\ref{sec:SD_NPR}.
The local term $Y_{AB}(\mu,\mu_0)Q_0^{\RI}(\mu_0)$ accounts for the difference between the bilocal operators in the $\MS$
and RI/SMOM scheme.
The bilocal operator $\{Q_A^{\RI}\,Q_B^{\RI}\}_{\mu_0}^{\RI}$ is defined as
\begin{equation}
\{Q_A^{\RI}\,Q_B^{\RI}\}_{\mu_0}^{\RI}\equiv Z_{Q_A}^{\lat\to\RI}(a\mu_0)Z_{Q_B}^{\lat\to\RI}(a\mu_0)\{Q_A^{\lat}Q_B^{\lat}\}_a^{\lat}-X_{AB}(\mu_0,a)Q_0^{\RI}(\mu_0).
\end{equation}
Here $Q_A^{\lat}$ and $Q_B^{\lat}$ are bare lattice operators and $a$ is the lattice spacing. A counterterm $X_{AB}(\mu_0,a)Q_0^{\RI}(\mu_0)$ is introduced 
to remove the SD singularity in the product $Q_A^{\lat}(x)Q_B^{\lat}(0)$
as $x\to0$. After including the counterterm the bilocal operator $\{Q_A^{\RI}\,Q_B^{\RI}\}_{\mu_0}^{\RI}$ 
is independent of the ultraviolet cut-off $1/a$.
The explicit renormalization conditions used to determine the coefficient $X_{AB}(\mu_0,a)$ and $Y_{AB}(\mu,\mu_0)$ are given in Ref.~\cite{Christ:2016eae}.

%% file: Matrix_element.tex
In this section we describe the details of the computation of the bilocal operators in lattice simulations. 
We start by presenting the parameters and details of our exploratory simulation in Sec.\,\ref{subsec:simulationdetails}. 
We then, in Sec.\,\ref{subsec:kinematics}, discuss the 
kinematics of the $K^+\to\pi^+\nu\bar \nu$ decays and explain our choice of the momenta of the external particles. 
The bilocal operators relevant for these rare decays are explicitly introduced in Sec.\,\ref{subsec:bilocal}. 
The evaluation of the amplitude also requires the determination of a number of matrix elements of local operators; these 
are identified in Sec.\,\ref{subsec:local_matrix} together with a detailed discussion of their evaluation. The evaluation of the matrix
elements of the bilocal operators for the $W$-$W$ and $Z$-exchange diagrams (introduced in Sec.\,\ref{subsec:bilocal} below) is presented in Secs.\,\ref{subsec:bilocalWW} and \ref{subsec:bilocalZ} respectively.

\subsection{Details of the simulation}\label{subsec:simulationdetails}
   In this work we use configurations generated by the RBC-UKQCD collaborations with $2+1$ flavors of domain wall fermions 
   and the Iwasaki gauge action. Because of the importance of the GIM cancellation in this decay, we use four flavors 
   of valence quarks including an active charm quark.  However, we neglect the contribution of the charm quark to the 
   fermion determinant.
   The results presented here are from an ensemble on $16^3\times 32\times 16$ lattices with an inverse lattice spacing of
   $a^{-1}=1.729(28)$\,GeV and a box size of $L=16\, a=1.83$\,fm~\cite{Blum:2011pu}.
   The residual mass is determined to be $m_{\mathrm{res}}\,a=0.00308(4)$ and
   the extent of the fifth-dimension is $L_s=16$.
   The pion and kaon masses are $m_\pi=421(1)(7)$\,MeV and $m_K=563(1)(9)$\,MeV and the corresponding input bare light
   and strange quark masses are $am_l=0.010$ and $am_s=0.032$.
   The valence charm quark mass is  $a m_c=0.330$, which corresponds to the $\MS$
   mass $m_c^{\MS}(\mbox{2 GeV})=863(24)$\,MeV with the mass renormalization factor
   $Z_m^{\MS}(\mbox{2 GeV})=1.498(34)$~\cite{Aoki:2010dy}, where $m_c^{\MS}(\mbox{2 GeV})=Z_m^{\MS}(\mbox{2 GeV})\,
   (m_c+m_{\textrm{res}})$.
   To achieve a high statistical precision, we use 800 configurations,
   each separated by 10 trajectories. For simplicity, all the results
   presented below are given in lattice units
   unless otherwise specified.

\subsection{The kinematics}\label{subsec:kinematics}
   \begin{figure}
   \centering
   \includegraphics[width=.8\textwidth]{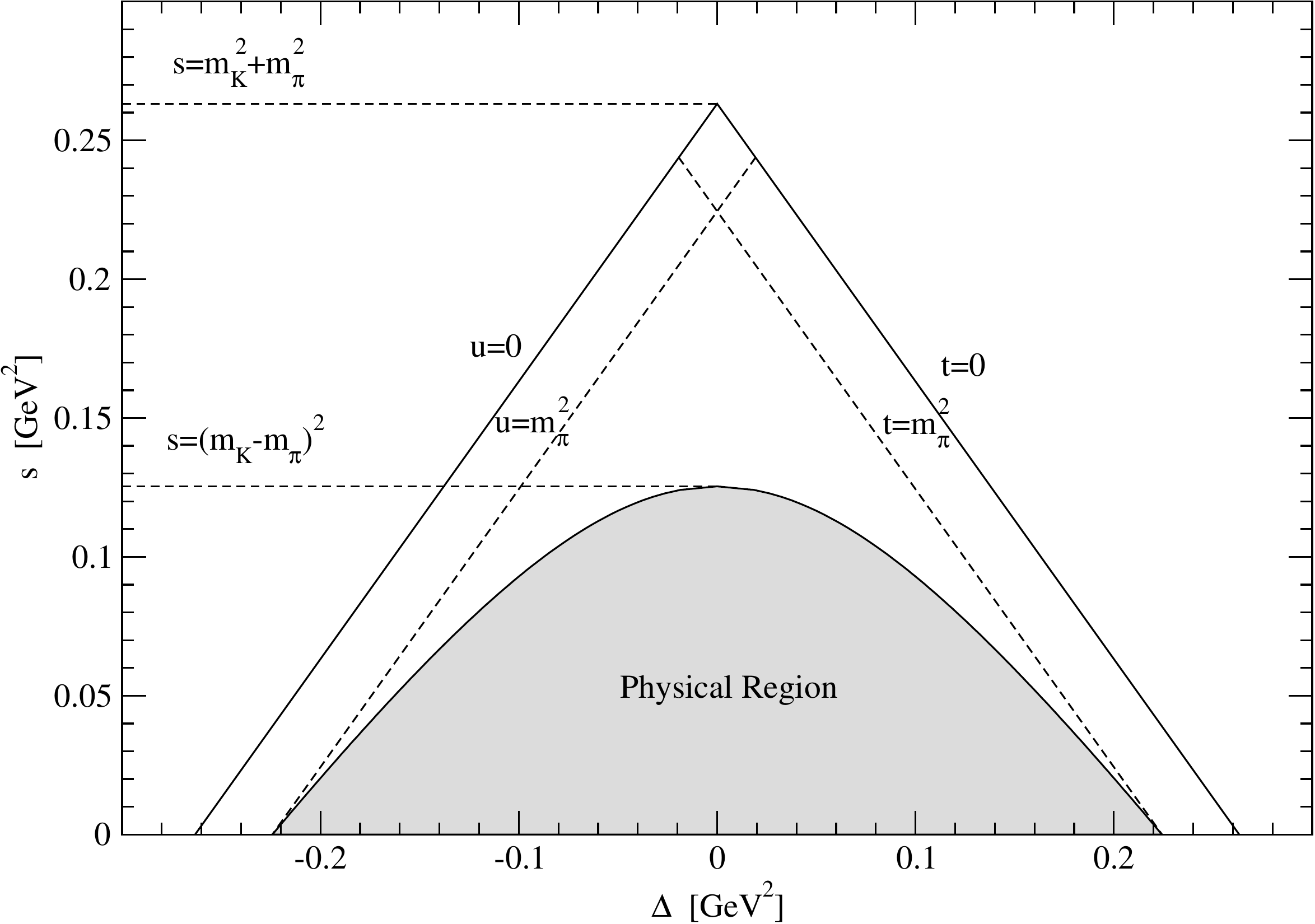}
   \caption{Dalitz plot for $K\to\pi\nu\bar{\nu}$.}
   \label{fig:Dalitz}
   \end{figure}

   Given the momenta $p_K$, $p_\pi$, $p_\nu$ and $p_{\bar{\nu}}$, one can define three Lorentz
   invariants
   \begin{eqnarray}
   \label{eq:Lorentz_invariant}
   s=-(p_K-p_\pi)^2,\quad t=-(p_K-p_\nu)^2,\quad u=-(p_K-p_{\bar{\nu}})^2,
   \end{eqnarray}
   where two of them are independent: $s+t+u=m_K^2+m_\pi^2$. Here we use a Euclidean metric with the signature (++++) 
   so that an on-shell momentum is written as
   $p_\pi=(iE_\pi,{\bf p}_\pi)$ for a pion, and a minus sign appears in the definition
   for $s$, $t$ and $u$. Defining $\Delta\equiv u-t$, the physical region for $(\Delta,s)$ is denoted by the bounds
   \begin{eqnarray}
   s\ge0,\quad \Delta^2\le(m_K^2+m_\pi^2-s)^2-4m_K^2m_\pi^2
   \end{eqnarray}
   and is illustrated in Fig.~\ref{fig:Dalitz}.

   In our lattice calculation we take the kaon to be at rest so that $p_K=(im_K,{\bf 0})$.
   The pion's three-momentum is then given by
   \begin{eqnarray}
   \label{eq:mom_pi}
   |{\bf p}_\pi|=\frac{\sqrt{s^2-2(m_K^2+m_\pi^2)s+(m_K^2-m_\pi^2)^2}}{2m_K}\,.
   \end{eqnarray}
   Without loss of generality, we choose the direction of the pion's momentum to be
   ${\bf p}_\pi=\frac{|{\bf p}_\pi|}{\sqrt{3}}({\bf e}_x+{\bf e}_y+{\bf e}_z)$, where ${\bf e}_i$ is the unit vector in the $i$-direction.
   We decompose the spatial momenta of the neutrino and anti-neutrino into components parallel and perpendicular to ${\bf p}_\pi$ writing 
   \begin{eqnarray}
   \label{eq:mom_neutrino}
   {\bf p}_\nu={\bf p}_\parallel+{\bf p}_\perp,\quad
   {\bf p}_{\bar{\nu}}=-{\bf p}_\pi-{\bf p}_\parallel-{\bf p}_\perp\,,
   \end{eqnarray}
   where ${\bf p}_{\parallel(\perp)}$ is parallel (perpendicular) to ${\bf p}_\pi$.
   The values of ${\bf p}_\parallel$ and ${\bf p}_\perp$ are given by    
   \begin{eqnarray}
    \label{eq:mom_para_perp}
   {\bf p}_\parallel&=&-\frac{1}{2}\left\{\pm\frac{(m_K-E_\pi)\Delta}{2m_K|{\bf
   p}_\pi|^2}+1\right\}\, {\bf p}_\pi,
   \nn\\
   {\bf
   p}_{\perp}&=&\frac{1}{2}\left\{s+\left(\frac{\Delta}{2m_K}\right)^2-\left(\frac{(m_K-E_\pi)\Delta}{2m_K|{\bf
   p}_\pi|}\right)^2\right\}^{\frac{1}{2}}\, {\bf e}_{\perp},
   \end{eqnarray}
   where ${\bf e}_{\perp}$ is any unit vector perpendicular to ${\bf p}_\pi$.
   We use twisted boundary conditions 
   to implement the momenta given by Eqs.\,(\ref{eq:mom_pi})\,-\,(\ref{eq:mom_para_perp}).

   Using the Dirac equation for the massless neutrinos,
   one can show that the magnitude of the decay amplitude vanishes at the edge of
    the physically-allowed region, where the momenta satisfy the condition
   $\Delta^2=(m_K^2+m_\pi^2-s)^2-4m_K^2m_\pi^2$.
   We are therefore more interested in momenta
   that are well inside the region and a natural choice is $(\Delta,s)=(0,0)$, which corresponds to
   the case in which the $\nu$ and $\bar{\nu}$ carry
   the same spatial momentum and the pion moves in the opposite direction with
   twice the momentum of each of the $\nu$ and $\bar{\nu}$.
   Since we perform the calculation at $m_\pi=420$\,MeV,
   the allowed momenta for the final-state particles are constrained to lie in a small region.
   Given this small momentum range we expect that it will be
   be difficult to extract reliably the momentum dependence. 
   For this reason, in this exploratory study we devote our computational resources to evaluating the amplitude at the 
   single kinematical point with $(\Delta,s)=(0,0)$.
   The situation is expected to change once we perform the calculation at physical quark masses.
   In that case we will need to compute the $K^+\to\pi^+\nu\bar{\nu}$ amplitude at several values of
   $(\Delta,s)$ to gain a better understanding of the momentum dependence.
   Another consequence of the heavy pion mass is that the momenta of the 
   pion and the neutrinos are very small. For $(\Delta,s)=(0,0)$ these are
   \begin{equation}
   \label{eq:mom_setup}
   {\bf p}_{\nu}={\bf p}_{\bar{\nu}}=(0.0207,0.0207,0.0207),\quad {\bf p}_\pi=(-0.0414,-0.0414,-0.0414)\,.
   \end{equation}
   Here $|{\bf p}_\pi|=0.0717$ is only about 18\% of the lowest lattice momentum with periodic boundary conditions, 
   $2\pi/L=0.3927$.

\subsection{The bilocal operators}\label{subsec:bilocal}
   \begin{figure}
   \centering
        \shortstack{
	\shortstack{\includegraphics[width=.3\textwidth]{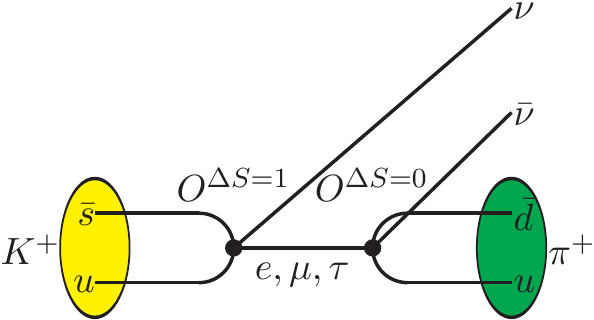}\\Type 1}
	\hspace{0.3cm}
        \shortstack{\includegraphics[width=.3\textwidth]{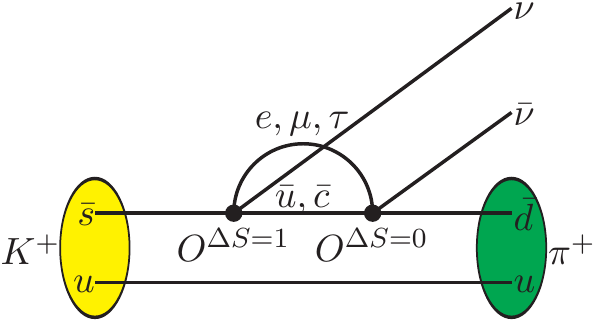}\\Type 2}
	\\$W$-$W$ diagrams}\\
	\ \\
	\shortstack{
    \shortstack{\includegraphics[width=.3\textwidth]{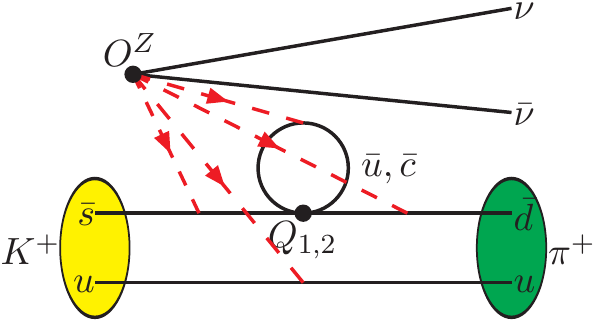}\\with
    self-loop}
    \hspace{0.3cm}
    \shortstack{\includegraphics[width=.3\textwidth]{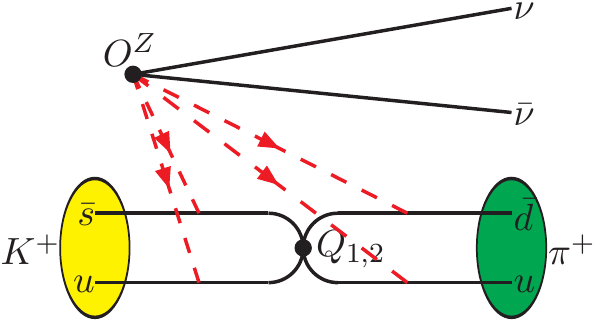}\\without
    self-loop}
    \\connected $Z$-exchange diagrams}\\
    \ \\
    \shortstack{
    \shortstack{\includegraphics[width=.3\textwidth]{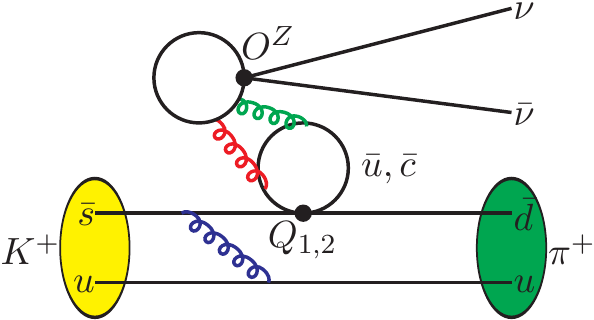}\\with
    self-loop}
    \hspace{0.3cm}
    \shortstack{\includegraphics[width=.3\textwidth]{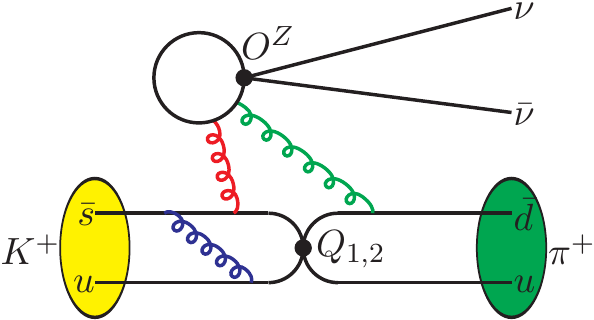}\\without
    self-loop}
    \\disconnected $Z$-exchange diagrams}
   \caption{From top to bottom: quark and lepton contractions for $W$-$W$, connected and disconnected $Z$-exchange diagrams.}
   \label{fig:contraction}
   \end{figure}

   There are two classes of diagrams which contribute to $K^+\to\pi^+\nu\bar{\nu}$ decays, we call these the $W$-$W$ and
   $Z$-exchange diagrams. In the $W$-$W$ diagrams the second-order weak transition proceeds through the exchange of two
   $W$-bosons , while for the $Z$-exchange diagrams the decay occurs 
   through the exchange of one $W$-boson and one $Z$-boson; both classes of diagrams are illustrated in
   Fig.~\ref{fig:contraction}. 
   The bilocal contribution to the decay amplitude is a combination of these two types of diagrams so that it can be written in terms of
   the matrix element $\langle\pi^+\nu\bar{\nu}|{\mathcal B}(0)|K^+\rangle$,
   where the bilocal operator ${\mathcal B}(y)$ receives contributions from
   both ${\mathcal B}_{WW}(y)$ and ${\mathcal B}_{Z}(y)$
    \begin{equation}\label{eq:bilocal}
    {\mathcal B}(y)=\frac{G_F}{\sqrt{2}}\frac{\alpha}{2\pi\sin^2\theta_W}\frac{\pi^2}{M_W^2}\lambda_c\,
    \Big({\mathcal B}_{WW}(y)+{\mathcal B}_{Z}(y)\Big).
    \end{equation}
    Here 
    \begin{equation}
    {\mathcal B}_{WW}(y)=\sum_{\ell=e,\mu,\tau}{\mathcal B}^{(\ell)}_{WW}(y),\quad {\mathcal B}_{Z}(y)=\sum_{\ell=e,\mu,\tau}{\mathcal B}^{(\ell)}_Z(y)\end{equation} 
    and ${\mathcal B}^{(\ell)}_{WW}(y)$ and ${\mathcal B}^{(\ell)}_{Z}(y)$ are defined as 
    \begin{equation}\label{eq:B_WW}
    {\mathcal B}_{WW}^{(\ell)}(y)=\int d^{\hspace{1.5pt}4}\hspace{-1pt}x\,T[O_{u\ell}^{\Delta S=1}(x)O_{u\ell}^{\Delta S=0}(y)]
    -\{u\to c\}
    \end{equation}
    and
    \begin{equation}
     \label{eq:B_Z}
    {\mathcal B}_{Z}^{(\ell)}(y)=\int d^{\hspace{1.5pt}4}\hspace{-1pt}x\,T[O_u^W(x)O_\ell^Z(y)]-\{u\to c\}.
    \end{equation}
    Here, as in Ref.\,\cite{Bai:2017fkh,Christ:2016eae}, we find it convenient to use the letter $O$ to represent an operator which incorporates a Wilson coefficient and the letter $Q$ for an operator which does not include such a coefficient. In Eq.\,(\ref{eq:B_WW}) 
     $O_{q\ell}^{\Delta S=1}$ and $O_{q\ell}^{\Delta
    S=0}$ are the appropriate products,  $C_A^{\MS}Q_A^{\MS}$ and
    $C_B^{\MS}Q_B^{\MS}$, for the $W$-$W$ diagrams. We can write them in terms of bare lattice operators as
   \begin{equation}
   O_{q\ell}^{\Delta S=1}=Z_V\,(\bar{s}q)_{V-A}(\bar{\nu}\ell)_{V-A},\quad
   O_{q\ell}^{\Delta S=0}=Z_V\,(\bar{q}d)_{V-A}(\bar{\ell}\nu)_{V-A},\label{eq:OWW}
   \end{equation}
    where $Z_V=Z_A$ is the renormalization factor relating the local lattice vector or axial-vector current (which we use) 
    to the conserved or partially conserved ones and is effectively the corresponding Wilson coefficient. 
    By taking the ratio of two-point functions computed with the local and conserved axial currents we obtain $Z_A = 0.7163$,
    which is consistent with the result quoted in Ref.\,\cite{Aoki:2007xm}.
    
    The two effective operators for the $Z$-exchange diagrams are given by
    \begin{equation}
    \label{eq:operator_Z}
    O_q^W=C_1\,Q_{1,q}+C_2\, Q_{2,q},\quad O^Z_\ell=Z_V\, J_\mu^Z\, \big[\bar{\nu}_\ell\gamma^\mu(1-\gamma_5) \nu_\ell\big]
    \end{equation}
    with $Q_{1,q}$ and $Q_{2,q}$ the conventional current-current operators and
    $J_\mu^Z$ the quark current which couples to the $Z^0$. Their definition is given in Eq.\,(15) of Ref.~\cite{Christ:2016eae}, where a discussion of the 
    corresponding operator renormalization
    from the lattice to the $\MS$ scheme is also presented. 
 
\subsection{Matrix elements of local operators}
  \label{subsec:local_matrix}
  In addition to the evaluation of the matrix elements of the bilocal operators discussed in Sec.\,\ref{subsec:bilocal}, which is the 
  main task of this work, there are three types of matrix elements of local operators which must be computed in order to determine the 
  $K^+\to\pi^+\nu\bar{\nu}$ decay amplitude.
  \begin{itemize}
      \vspace{-0.1in}\item Matrix elements for the SD contributions, $\langle\pi^+\nu\bar{\nu}|Q_0|K^+\rangle$, for which the hadronic effects are obtained from matrix elements of the form
      $\langle\pi^+ (\vec{p}_f)|\bar{s}\gamma_\mu d|K^+(\vec{p}_i)\rangle$. The labels $i$ and $f$ indicate the initial and final states, and are
      used to distinguish these states from the intermediate states discussed below.
      \vspace{-0.1in}\item Matrix elements for low-lying intermediate states. This type
      of matrix element corresponds to unphysical contributions which grow exponentially in $T$, the time interval over which 
      the separation of the two local operators $Q_A^S$ and $Q_B^S$ are integrated (see the discussion around Eq.\,(\ref{eq:amplitude})). 
      Such terms arise when there are intermediate states whose energies are smaller than the kaon mass
      \,\cite{Christ:2016eae}.
  For the $W$-$W$ diagrams, see Fig.\,\ref{fig:contraction}, we study the effects from the lowest two intermediate states: $|\bar{\ell}\nu\rangle$ and $|\pi^0\bar{\ell}\nu\rangle$.
  The unphysical contribution from the multi-hadron state
  $|\pi\pi\bar{\ell}\nu\rangle$ can be neglected due to phase space suppression. For the $Z$-exchange diagrams we examine and subtract
  the exponentially growing effects from $|\pi^+\rangle$ and $|(\pi^+\pi^0)_{I=2}\rangle$
  states, where $I$ is the total isospin of the two-pion state. Note that because of charge and angular momentum conservation only the $I=2$ $\pi\pi$ 
  state can contribute to
  $Z$-exchange diagrams.
  \item Matrix elements of the local scalar density $\bar{s}d$, $\langle\pi^+ |\bar{s}d|K^+\rangle$. Since the 
  scalar density operator does not contribute to the on-shell matrix
  element, we can shift the effective Hamiltonian by $H_W\to H_W'=H_W-c_s\,\bar{s}d$ without changing the amplitude\,\cite{Christ:2012se}.
  By choosing an appropriate value for $c_s$ we remove the unphysical
  contribution from the $|\pi^+\rangle$ intermediate state in the $Z$-exchange
  diagrams. We will discuss this in more detail in the following sections. In other applications, one also frequently subtracts a term proportional 
  to the pseudoscalar density from the effective Hamiltonian to remove a low-lying state from the correlation 
  function. However, in this case there is no contribution from the vacuum state and the operator $\bar{s}\gamma^5 d$ 
  cannot mediate transitions to $I=2$ two-pion states (by isospin conservation). We therefore do not make the subtraction 
  $H_W\to H_W'=H_W-c_p\,\bar{s}\gamma^5 d$  here.
   
%
%
  \end{itemize}
  The three types of hadronic matrix elements are summarized in
  Table~\ref{tab:local_matrix_element} and will be used below for the 
  analysis of the second-order weak transition amplitude. We now proceed to a discussion of the evaluation of the 
  matrix elements of these local operators.
\begin{table}
  \centering
  \begin{tabular}{c|c|c}
    \hline
    \multicolumn{3}{l}{Matrix element for the SD contribution}\\
    \hline
    $Q_0$ & $\langle\pi^+\nu\bar{\nu}|Q_0|K^+\rangle$ & $\langle\pi^+(\vec{p}_f)|\bar{s}\gamma_\mu d|K^+(\vec{p}_i)\rangle$ \\
    \hline
    \hline
    \multicolumn{3}{l}{Matrix element relevant for low-lying intermediate states}\\
    \hline
    \multirow{2}{*}{$W$-$W$} & $\langle\pi^+\nu\bar{\nu}|O_u^{\Delta S=0}|\bar{\ell}\nu\rangle\langle\bar{\ell}\nu|O_u^{\Delta S=1}|K^+\rangle$
            & $\langle\pi^+(\vec{p}_f)|\bar{u}\gamma_\mu \gamma_5 d|0\rangle$, $\langle0|\bar{s}\gamma_\mu \gamma_5u|K^+(\vec{p}_i)\rangle$\\
    	    & $\langle\pi^+\nu\bar{\nu}|O_u^{\Delta S=0}|\pi^0\bar{\ell}\nu\rangle\langle\pi^0\bar{\ell}\nu|O_u^{\Delta S=1}|K^+\rangle$
    	    & $\langle\pi^+(\vec{p}_f)|\bar{u}\gamma_\mu d|\pi^0\rangle$, $\langle\pi^0|\bar{s}\gamma_\mu u|K^+(\vec{p}_i)\rangle$ \\
    \hline
    \multirow{6}{*}{$Z$-exchange} & \multirow{2}{*}{$\langle\pi^+\nu\bar{\nu}|O_\ell^Z|\pi^+\rangle\langle\pi^+|O_q^W|K^+\rangle$} & 
    $\langle\pi^+(\vec{p}_f)|\bar{u}\gamma_\mu u|\pi^+\rangle$, $\langle\pi^+(\vec{p}_f)|\bar{d}\gamma_\mu d|\pi^+\rangle$ \\
    & & $\langle\pi^+|Q_{1,q}|K^+(\vec{p}_i)\rangle$, $\langle\pi^+|Q_{2,q}|K^+(\vec{p}_i)\rangle$	\\
    & \multirow{4}{*}{$\langle\pi^+\nu\bar{\nu}|O_\ell^Z|(\pi^+\pi^0)_{I=2}\rangle\langle(\pi^+\pi^0)_{I=2}|O_q^W|K^+\rangle$} &
    $\langle\pi^+(\vec{p}_f)|\bar{u}\gamma_\mu\gamma_5 u|(\pi^+\pi^0)_{I=2}\rangle$ 
    \\ &&$\langle\pi^+(\vec{p}_f)|\bar{d}\gamma_\mu\gamma_5 d|(\pi^+\pi^0)_{I=2}\rangle$ 
    \\
    & & $\langle(\pi^+\pi^0)_{I=2}|Q_{1,u}|K^+(\vec{p}_i)\rangle$ 
    \\ && $\langle(\pi^+\pi^0)_{I=2}|Q_{2,u}|K^+(\vec{p}_i)\rangle$\\
    \hline
    \hline
    \multicolumn{3}{l}{Matrix element for the shift in the Hamiltonian}\\
    \hline
    $Z$-exchange & $H_W\to H_W-c_s\,\bar{s}d$ & $\langle\pi^+|\bar{s}d|K^+(\vec{p}_i)\rangle$\\
    \hline
    \end{tabular}
  \caption{Hadronic matrix elements of local operators required for the calculation of the $K^+\to\pi^+\nu\bar{\nu}$ amplitude (third column). 
  $\vec{p}_{i}$ and $\vec{p}_{f}$ are the momenta of the initial state kaon and final state pion, whereas the momenta of the intermediate 
   states are not shown explicitly. The second column includes the neutrinos and for the $W$-$W$ and
$Z$-exchange diagrams displays the corresponding contributions to the bilocal matrix elements.}
  \label{tab:local_matrix_element}
\end{table}

\subsubsection{Correlators and propagators}
  In Table~\ref{tab:local_matrix_element}, except for the matrix elements 
  $\langle\pi^+(\vec{p}_f)|\bar{u}\gamma_\mu\gamma_5 d|0\rangle$ and 
  $\langle0|\bar{s}\gamma_\mu \gamma_5 u|K^+(\vec{p}_i)\rangle$ which are proportional to the 
  the leptonic decay constants and can be determined from 2-point correlation
  functions, the remaining matrix elements of local operators can be
  extracted from 3-point correlation functions of the general form 
  $\langle \phi_A(t_A)\, O(t_O)\, \phi_B^\dagger(t_B)\rangle$, where $\phi_A$ and $\phi^\dagger_B$ are 
  interpolating operators which can annihilate hadron A or create hadron B. We define the quantity  
  \begin{equation}\label{eq:MAOB}
  \mathcal{M}_{AOB}(t_A,t_O,t_B)=\frac{2E_A\,2E_B}{N_A\,N_B^\dagger}\langle \phi_A(t_A)
  O(t_O) \phi_B^\dagger(t_B)\rangle e^{E_A(t_A-t_O)} e^{E_B(t_O-t_B)},
  \end{equation}
  where we do not exhibit the dependence of the operators on  the spatial coordinates.
  Here $A$ and $B$ indicate initial-, intermediate- or final-state particles, i.e. $K^+$, $\pi^{+,0}$ and $(\pi^+\pi^0)_{I=2}$.
   We use Coulomb gauge-fixed wall sources for the $\phi_A$ and $\phi_B$ interpolating
   operators. Such wall-source operators have a good overlap with the $\pi$, $K$ and $(\pi^+\pi^0)_{I=2}$ ground states.
  The coefficients $N_A$ and $N_B$ can be extracted from the corresponding 2-point correlation functions using the same
  wall-source operators. 
  $E_A$ and $E_B$ are the ground-state energies which can also be determined
  from 2-point functions. The matrix element $\langle A|O|B\rangle=\mathcal{M}_{AOB}(t_A,t_O,t_B)$ 
  can then be determined from the three-point correlation functions using Eq.\,(\ref{eq:MAOB}) at large 
  $t_A-t_O\gg0$ and $t_O-t_B\gg0$.
  
  In Eq.\,(\ref{eq:MAOB}) the operator $O$ can be a vector or axial-vector current, the current-current 
  operators $Q_{1q}$ and  $Q_{2q}$ or the scalar density $\bar{s}d$. The interpolating operators 
  $\phi_{A,B}$ are constructed using twisted boundary conditions to ensure that the corresponding states 
  have the required momenta.
  Translation invariance then implies that the correlation functions in Eq.(\,\ref{eq:MAOB}) do not depend on 
  the   spatial position $\vec{x}$  of the operator $O(t_O,\vec{x}\,)$. In order to obtain a better precision 
  we treat $\vec{x}$ 
   as the sink of the quark propagators and sum over $\vec{x}$ with the appropriate phase factor to account for the momentum transfer between states $A$ and $B$. 
  The resulting volume factor in the 3-point function cancels with that from the 2-point functions used to determine $N_A$ and $N_B$.
  
  The operators $Q_{1,q}$ and $Q_{2,q}$
 can induce closed quark loops in the contractions.
   We therefore need to calculate
   the light and charm quark propagators $D^{-1}_{u,c}(x,x)$ for all possible $x$ and 
   using random-source propagators is a natural way to evaluate these
   quark loops~\cite{Christ:2016mmq}.
   For a similar cost, one can either put one random wall source at each of the $T$ time slices or use
   $N_r=T$ random volume sources with no dilution in the time slices.
   Although the cost of these two choices is almost the same,
   the latter one reduces the error by a factor of $1.5$ compared to the former.
   We thus use $N_r=T=32$ random volume source propagators to calculate
   the light and charm quark propagator $D^{-1}_{u,c}(x,x)$ for all possible
   $x$.
   We also make use of the time translation invariance and average the correlator over all $T$ time translations
   \begin{equation}
    \label{eq:time_average}
   \bar{\mathcal M}_{AOB}(t_2,t_1)=\frac{1}{T}\sum_{t=0}^{T-1}{\mathcal M}_{AOB}(t_2+t,t_1+t,t).
   \end{equation}
   By doing this, our results show
   that the statistical error can be efficiently reduced by nearly a factor of $\sqrt{T}$. 
   The time translation average requires
   the wall-source propagators to be generated on all time slices. This can be
   achieved in an efficient way by calculating the low-lying eigenvectors of the Dirac 
   operator using the Lanczos method
   and then using low-mode deflation
   to accelerate the light-quark inversions. Working on the $16^3\times32$
   lattice, we find that by using 100 eigenvectors in low-mode deflation the light-quark
   conjugate gradient (CG) time is reduced to 16\% of that required for the CG inversions without low-mode deflation. 

\subsubsection{Exploiting isospin symmetry to simplify the derivation of the contractions}
Since this computation is performed in the isospin-symmetric limit, we can exploit this symmetry to derive the necessary contractions more readily. For example, we have the following relations between the matrix elements: 
   \begin{eqnarray}\label{eq:isospin_rotation}
   \langle\pi^0|\bar{s}\gamma_\mu u|K^+\rangle&=&\frac{1}{\sqrt{2}}\langle\pi^+|\bar{s}\gamma_\mu d|K^+\rangle
   \nn\\
   \langle\pi^+|\bar{u}\gamma_\mu d|\pi^0\rangle&=&\sqrt{2}\langle\pi^+|\bar{d}\gamma_\mu d|\pi^+\rangle\label{eq:isospin}
   \\
   \langle\pi^+|\bar{u}\gamma_\mu\gamma_5 u-\bar{d}\gamma_\mu\gamma_5d|(\pi^+\pi^0)_{I=2}\rangle
   &=&\langle\pi^+|\bar{d}\gamma_\mu\gamma_5u|(\pi^+\pi^+)_{I=2}\rangle.\nn
   \end{eqnarray}
   The matrix elements on the right-hand side have simpler contractions since they do not involve 
   the neutral pion, the $\pi^0$. More precisely, although the final set of contractions is of course the same, 
   by using the relations in Eqs.\,(\ref{eq:isospin}) there are fewer cancelations of diagrams in intermediate 
   steps of the calculation.
   
   We now express some of the matrix elements in Table~\ref{tab:local_matrix_element} in terms of 
   invariant form factors:
   \begin{eqnarray}
   \label{eq:local_form_factor}
   Z_V\langle\pi^+(p_\pi)|\bar{s}\gamma_\mu
   d|K^+(p_K)\rangle&=&i\left\{\,(p_K+p_\pi)_\mu f_+(s)+(p_K-p_\pi)_\mu f_-(s)\,\right\}\\
   (m_s-m_d)\langle\pi^+(p_\pi)|\bar{s}d|K^+(p_K)\rangle&=&(m_K^2-m_\pi^2)f_0(s)\label{eq:kl32}\\ 
   Z_V\langle\pi^+(p_2)|\bar{d}\gamma_\mu
   d|\pi^+(p_1)\rangle&=&i\,F_\pi(s)(p_1+p_2)_\mu\,,\label{eq:pionff}
   \end{eqnarray}
   where $s=-(p_K-p_\pi)^2$ for the $K_{\ell3}$ form factors $f_{+,-,0}(s)$ and
   $s=-(p_1-p_2)^2$ for the pion form factor $F_\pi(s)$. In Eqs.\,(\ref{eq:local_form_factor}) and (\ref{eq:pionff}),
    $Z_V$ is the renormalization constant relating the local vector current to the conserved one. 
    The momentum $p_i$ is a Euclidean four-momentum defined as
   $p_i=(iE_i,{\bf p}_i)$ with $E_i$ and ${\bf p}_i$ the energy and spatial momentum of the corresponding
    on-shell particle.
   The scalar form factor is a linear combination of $f_+(s)$ and $f_-(s)$:
   \begin{equation}
   f_0(s)=f_+(s)+\frac{s}{m_K^2-m_\pi^2}f_-(s)\,,
   \end{equation}
   which follows from Eqs.\,(\ref{eq:local_form_factor}) and (\ref{eq:kl32}) and a chiral Ward identity.

   The current-current operators $Q_{i,u}$ in Eq.\,(\ref{eq:operator_Z}) are linear combinations 
   of $\Delta I=3/2$ and $\Delta I=1/2$ operators. Only the $\Delta I=3/2$
   component contributes to the $K^+\to(\pi^+\pi^0)_{I=2}$ transition. For the $K\to(\pi\pi)_{I=2}$ transition we have
   \begin{equation}
   \langle(\pi^+\pi^0)_{I=2}|Q_{i,u}|K^+\rangle=\frac{1}{\sqrt{3}}\langle(\pi^+\pi^0)_{I=2}|Q^{\Delta
   I=\frac{3}{2},\Delta I_z=\frac{1}{2}}|K^+\rangle,\quad i=1,2
   \end{equation}
   where the operator with isospin $\Delta I=3/2$, $\Delta I_z=1/2$ is given by
   \begin{equation}
   Q^{\Delta I=\frac{3}{2},\Delta
   I_z=\frac{1}{2}}=\frac{1}{\sqrt{3}}\left(-(\bar{s}d)_{V-A}(\bar{d}d)_{V-A}+(\bar{s}d)_{V-A}(\bar{u}u)_{V-A}+(\bar{s}u)_{V-A}(\bar{u}d)_{V-A}\right).
   \end{equation}
   One can now use the Wigner-Eckhart theorem for isospin symmetry and write the matrix element for the $K\to(\pi^+\pi^0)_{I=2}$ decay in terms of that into the maximally extended state $|\pi^+\pi^+\rangle$:
   \begin{equation}\label{eq:WE}
   \langle(\pi^+\pi^0)_{I=2}|Q_{i,u}|K^+\rangle=\frac{1}{2}\langle(\pi^+\pi^+)_{I=2}|(\bar{s}d)_{V-A}(\bar{u}d)_{V-A}|K^+\rangle,
   \end{equation}
   where $(\bar{s}d)_{V-A}(\bar{u}d)_{V-A}$ is a $\Delta I=3/2$, $\Delta I_z=3/2$ operator. The determination of 
   the necessary contractions is simpler using the matrix element for the $K^+\to(\pi^+\pi^+)_{I=2}$ decay than for
   the $K^+\to(\pi^+\pi^0)_{I=2}$ transition.
   (Note that Eq.\,(\ref{eq:WE}) was used throughout the RBC-UKQCD collaborations' computations
   of the $\Delta I=\frac32,\,K\to\pi\pi$ amplitude $A_2$\,\cite{Blum:2011ng,Blum:2012uk,Blum:2015ywa}.  
   The motivation in Refs.\,\cite{Blum:2011ng,Blum:2012uk,Blum:2015ywa} was different however; there it 
   was to use antiperiodic boundary conditions on the $u$ quark to match the $I=2$, $\pi\pi$ ground-state 
   energy to the mass of the kaon, $m_K$.)

\subsubsection{Around-the-world effects}
  To extract the matrix elements one needs to determine the coefficients $N_A$ and $N_B$ for 
  $A,B=K^+,\pi^{+,0},(\pi^+\pi^0)_{I=2}$. For the case when $A=B=(\pi^+\pi^0)_{I=2}$ one has to consider the 
  subtlety of round-the-world effects. The corresponding two-point function is given by 
    \begin{equation}\label{eq:Cpipi}
      C_{\pi\pi}(t)=\langle\phi_{\pi\pi}(t)\phi_{\pi\pi}^\dagger(0)\rangle\xrightarrow[]{T\gg t\gg0}\frac{N_{\pi\pi}^2}{2E_{\pi\pi}}\left(e^{-E_{\pi\pi}t}+e^{-E_{\pi\pi}(T-t)}\right)+N_0(T).
  \end{equation}
  Here an unwanted term, $N_0(T)$ (proportional to $e^{-E_\pi T}$ where $E_\pi$ is 
  the energy of a single pion), is induced by the around-the-world effects in which each of 
  $\phi_{\pi\pi}$  interpolating operators in Eq.\,(\ref{eq:Cpipi}) creates one pion and annihilates another.
  We can remove this term by performing the subtraction through
  \begin{equation}
  C_{\pi\pi}(t)-C_{\pi\pi}(t+1)=\frac{N_{\pi\pi}^2}{2E_{\pi\pi}}\left(-4e^{-\frac{E_{\pi\pi}}{2}T}\right)\sinh(E_{\pi\pi}t')\sinh\frac{E_{\pi\pi}}{2}.
  \end{equation}
  where $t'=t+1/2-T/2$.
  For the single-pion 2-point function, $C_\pi(t)$, where the pion has energy $E_\pi$, we have
  \begin{equation}
  C_\pi^2(t)-C_\pi^2(t+1)=\frac{N_\pi^4}{(2E_\pi)^2}\left(-4e^{-E_\pi T}\right)\sinh(2E_{\pi}t')\sinh E_\pi\,.
  \end{equation}
  By constructing the ratio $R(t+\frac{1}{2})\equiv\frac{C_{\pi\pi}(t)-C_{\pi\pi}(t+1)}{C_\pi^2(t)-C_\pi^2(t+1)}$, we can 
  determine $N_{\pi\pi}$ and $\delta E\equiv E_{\pi\pi}-2E_\pi$ from\,\cite{Feng:2009ij}
  \begin{eqnarray}
      R(t+1/2)&=&A_R\left(\cosh(\delta E\,t')+\sinh(\delta
  E\,t')\coth(2E_\pi\,t')\right),\quad \textrm{where}
  \nn\\
   A_R&=&\frac{N_{\pi\pi}^2}{2E_{\pi\pi}}\frac{(2E_\pi)^2}{N_\pi^4}e^{-\frac{\delta E}{2}T}\frac{\sinh\frac{E_{\pi\pi}}{2}}{\sinh E_\pi}.
  \end{eqnarray}
  At threshold (i.e. with $E_\pi=m_\pi$)  we obtain $\delta E=0.01803(32)$ from which, using
  L\"uscher's finite-size formula\,\cite{Luscher:1986pf}, we find $m_\pi a_{\pi\pi}=-0.2816(43)$,
  where $a_\pi$ is the $\pi$-$\pi$ scattering length. This result is close to
  the estimate $m_\pi a_{\pi\pi}^{\mathrm{LO}}=-\frac{m_\pi^2}{8\pi f_\pi^2}=-0.2978(23)$ from 
  leading-order chiral perturbation 
  theory (ChPT)\,\cite{Weinberg:1966kf}. Here we have used the values $am_\pi=0.24360(47)$ and $af_\pi=0.08904(19)$ 
  from our simulation.
  The difference between the values deduced from $\delta E$ and LO ChPT is expected to be due to higher-order 
  terms in ChPT, as well as to possible systematic effects.

\subsubsection{Lattice results}
\begin{table}[t]
\centering
\begin{tabular}{c c | c c}
\hline
\multicolumn{4}{c}{Matrix elements for the SD contribution}\\
\hline
$\langle\pi^+(p)|\bar{s}\gamma_id|K^+(0)\rangle$ & $-i~0.06014(77)$ &
$f_+(s)$ & $0.993(3)$ \\
$\langle\pi^+(p)|\bar{s}\gamma_td|K^+(0)\rangle$ & $-0.7970(14)$ & $f_-(s)$
& $-0.048(12)$ \\
   & & $f_0(s)$ & $0.993(3)$ \\
 $\langle\pi^+(0)|\bar{s}\gamma_t
d|K^+(0)\rangle$ & $-0.7992(15)$ & $f_0(s_{\mathrm{max}})$ & $1.006(3)$ \\
\hline
\hline
\multicolumn{4}{c}{Matrix elements relevant for the contributions of low-lying intermediate states}\\
\hline
$W$-$W$\\
\hline
$\langle\pi^0(0)|\bar{s}\gamma_t
u|K^+(0)\rangle$ & $-0.7992(15)$ & $f_0(s_{\mathrm{max}})$ & $1.006(3)$ \\
 $\langle\pi^+(p)|\bar{u}\gamma_i d|\pi^0(0)\rangle$ & $-i~0.05612(62)$ 
& $F_\pi(s)$& $0.971(11)$ \\
$\langle\pi^+(p)|\bar{u}\gamma_t d|\pi^0(0)\rangle$ & $-0.6830(15)$ &
 $F_\pi(s)$& $0.986(2)$ \\
\hline
$Z$-exchange\\
\hline
$\langle\pi^+(0)|Q_{1,q}|K^+(0)\rangle$ & $1.697(87)\times10^{-4}$ & $c_s^{(1)}$ & $0.795(41)\times10^{-4}$ \\
$\langle\pi^+(0)|Q_{2,q}|K^+(0)\rangle$ &                                        
$3.828(98)\times10^{-4}$ & $c_s^{(2)}$ & $1.794(46)\times10^{-4}$  \\
 $\langle(\pi^+\pi^0)_{I=2}(0)|Q_{i,q}|K^+(0)\rangle$ & $-i~4.165(18)\times10^{-4}$ & & \\
$\langle\pi^+(0)|\bar{u}\gamma_t\gamma_5u-\bar{d}\gamma_t\gamma_5d|(\pi^+\pi^0)_{I=2}(0)\rangle$ & $i~2.4930(84)$ & & \\
\hline
\hline
\multicolumn{4}{c}{Matrix element for the subtraction in the effective Hamiltonian} \\
\hline
$\langle\pi^+(0)|\bar{s}d|K^+(0)\rangle$ & $2.1335(58)$ & $f_0(s_{\mathrm{max}})$ & $1.007(2)$ \\
\hline
\end{tabular}
\caption{Lattice results for the local matrix elements. 
The state $|\pi^+(p)\rangle$ denotes a $\pi^+$ with momentum $p=|{\bf p}_{\pi}|$ where ${\bf
p}_{\pi}$ given in Eq.~(\ref{eq:mom_setup}). For the matrix element $\langle\pi^+(p)|\bar{s}\gamma_\mu d|K^+(0)\rangle$, 
$s=(m_K-E_\pi)^2-p^2$ whereas for $\langle\pi^+(p)|\bar{u}\gamma_\mu d|\pi^0(0)\rangle$,
 $s=(E_\pi-m_\pi)^2-p^2$. Similarly, when the $\pi^{+,0}$ in the intermediate state is at rest, $s=s_{\textrm{max}}=(m_K-m_\pi)^2$.
 The matrix elements $\langle(\pi^+\pi^0)_{I=2}(0)|Q_{1,q}|K^+(0)\rangle$ and $\langle(\pi^+\pi^0)_{I=2}(0)|Q_{2,q}|K^+(0)\rangle$
 are equal.}
\label{tab:lattice_local_matrix}
\end{table}

  Consider the time-dependent amplitude $\bar{\mathcal M}_{AOB}(t_2,t_1,0)$
  defined in Eq.~(\ref{eq:time_average}). We require $t_2-t_1$ and $t_1-0$
  to be sufficiently large to suppress the contamination from excited states and
  $t_2\ll T$ to suppress around-the-world effects. 
  In practice we define 
  $\mathcal M_{AOB}^{\mathrm{mid}}(t)\equiv\bar{\mathcal
  M}_{AOB}(t,\frac{t}{2},0)$ (or if $t$ is 
  odd, then $\mathcal M_{AOB}^{\mathrm{mid}}(t)\equiv\frac{1}{2}[\bar{\mathcal
  M}_{AOB}(t,\frac{t-1}{2},0)+\bar{\mathcal M}_{AOB}(t,\frac{t+1}{2},0)]$)
  and choose appropriate values for $t$ to control both the excited-state and
  around-the-world effects.
    By studying the $t$ dependence 
  of $\mathcal M_{AOB}^{\mathrm{mid}}(t)$ we determine
  the local matrix element $\langle A|O|B\rangle$ and present the corresponding
  results in Table~\ref{tab:lattice_local_matrix}.
  In the table we present the values of the $K\to\pi$, $\pi\to\pi$, $K\to(\pi^+\pi^0)_{I=2}$ and $(\pi^+\pi^0)_{I=2}\to\pi$ 
  matrix elements required for the analysis, and in particular for the subtraction of the exponentially growing 
  contributions from low-lying states. Although in 
  this simulation $m_K<2m_\pi$, so that there are no exponentially growing contributions from two-pion 
  intermediate states, we include below an explicit discussion of the $|(\pi^+\pi^0)_{I=2}\rangle$ state and the evaluation of the 
  corresponding matrix elements in preparation for simulations with physical quark masses for which $m_K>2m_\pi$.
  In the final two columns of Table~\ref{tab:lattice_local_matrix} we present the $K_{\ell3}$ form factors
  $f_+(s)$, $f_-(s)$ and $f_0(s)$,
  the pion form factors $F_\pi(s)$, and the coefficient $c_s^{(i)}$ from the ratio
  $c_s^{(i)}=\frac{\langle\pi|Q_{i,q}|K\rangle}{\langle\pi|\bar{s}d|K\rangle}$.
  We determine $f_0(s_{\mathrm{max}})$ with $s_{\mathrm{max}}=(m_K-m_\pi)^2$
  from both $\langle\pi^+(0)|\bar{s}\gamma_td|K^+(0)$ and $\langle\pi^+(0)|\bar{s}d|K^+(0)$ and obtain consistent results.
  The matrix element $\langle\pi^+(p)|\bar{u}\gamma_\mu d|\pi^0(0)\rangle$
  yields consistent results for $F_\pi(s)$ from the spatial and temporal polarization
  directions, although the former one is much
  noisier. 
  
  For the $\pi^+\pi^0$ contribution to the $Z$-exchange diagrams, we determine the matrix element
   $\langle\pi^+(0)|(\bar{u}\gamma_t\gamma_5u-\bar{d}\gamma_t\gamma_5d)|(\pi^+\pi^0)_{I=2}(0)\rangle=i~2.4930(84)$
  by performing the isospin rotation $(\pi^+\pi^0)_{I=2}\to
  (\pi^+\pi^+)_{I=2}$ in Eq.~(\ref{eq:isospin_rotation}). Here the two-pions are in the ground state, i.e. at threshold.
  This value of the matrix element is only about 8\% smaller than an estimate from tree level chiral perturbation theory, where the interaction between the two pions in the I=2 state is neglected.

%% file: WW.tex
\subsection{Evaluation of the matrix element of the bilocal operator for the
\boldmath $W$-$W$ diagrams}\label{subsec:bilocalWW}
   In this section we discuss the evaluation of the matrix element of the
   bilocal operator ${\mathcal B}_{WW}(y)$ defined in Eq.~(\ref{eq:B_WW}). The matrix element $T_{WW}$ 
   for the $W$-$W$ diagrams is given by
   \begin{eqnarray}
   T_{WW}=\int d^4x\,\langle\pi^+\nu\bar{\nu}|T\{O_{u\ell}^{\Delta S=1}(x)
   O_{u\ell}^{\Delta S=0}(y)\}|K^+\rangle - \{u\to c\}\,.
   \end{eqnarray}
   As explained in Ref.~\cite{Christ:2016eae}, $T_{WW}$ can be written in terms of
   the scalar amplitude $F_{WW}(\Delta, s)$ and leptonic spinor product $\bar{u}(p_\nu){\slashed p}_K(1-\gamma_5)v(p_{\bar{\nu}})$:
   \begin{eqnarray}
   \label{eq:F_WW}
   T_{WW}=i~F_{WW}(\Delta, s)\,\left[\bar{u}(p_\nu){\slashed p}_K(1-\gamma_5)v(p_{\bar{\nu}})\right],
   \end{eqnarray}
   where the variables $\Delta$ and $s$ are defined in the paragraph following Eq.~(\ref{eq:Lorentz_invariant}). 
   In practice one can obtain $F_{WW}(\Delta, s)$ through~\cite{Christ:2016eae}
   \begin{equation}\label{eq:FWW4}
   F_{WW}(\Delta,s)=-i\int d^4x\,H_{\alpha\beta}(x,y)\sum_\mu c_\mu\mathrm{Tr}\left[\Gamma_{\alpha\beta}(x,y)\gamma_\mu(1+\gamma_5)\right]\,,
   \end{equation}
   where the coefficient $c_\mu$ is given by
   \begin{equation}
   c_\mu=\frac{1}{8}\frac{b_\mu}{b\cdot p_K}\quad\textrm{where}\quad b_\mu=\frac{1}{4}\mathrm{Tr}\left[\gamma_\mu{\slashed p}_{\bar{\nu}}{\slashed p}_K(1-\gamma_5){\slashed p}_\nu\right].
   \end{equation}
   The hadronic and leptonic parts,
   $H_{\alpha\beta}(x,y)$ and $\Gamma_{\alpha\beta}(x,y)$, are defined by
   \begin{eqnarray}
   H_{\alpha\beta}(x,y)&=&Z_V^2\langle\pi^+(p_\pi)|T[\bar{s}\gamma_\alpha
   (1-\gamma_5)u(x)\,\bar{u}\gamma_\beta(1-\gamma_5)d(y)]|K^+(p_K)\rangle-\{u\to c\}
   \nn\\
   \Gamma_{\alpha\beta}(x,y)&=&\gamma_\alpha
   (1-\gamma_5)S_\ell(x,y)\gamma_\beta(1-\gamma_5)e^{-ip_\nu x}e^{-ip_{\bar{\nu}}y}\,,
   \end{eqnarray}
   where $S_\ell(x,y)$ is the free lepton propagator for $\ell=e,\mu$ or $\tau$.

\subsubsection{Construction of the correlation function}
   Similarly to the calculation of the matrix elements of local operators, we use Coulomb-gauge wall-source interpolating 
   operators to create the kaon in the initial state and the pion in the final state. 
   For the two weak operators $O_{q\ell}^{\Delta S=1}(x)$ and
   $O_{q\ell}^{\Delta S=0}(y)$, one is evaluated at a fixed point which is used as the source for the internal
quark lines connected to that operator. The second operator acts as the sink for
all the propagators joined to it and is summed over the spatial volume.
  To gain a higher precision from the time translation average, we calculate the
  point source propagators at all $T$ time slices.
  We also exchange the source and sink locations between the two weak
  operators and average over both choices.

\subsubsection{Lepton propagator with infinite time extent}
   A subtlety in the calculation of the $W$-$W$ diagrams is the inclusion of the lepton propagators, $S_\ell(x,y)$.
    For the light leptons $\ell=e,\mu$ the round-the-world effects are
    significant in our lattice calculation with temporal extent $T=32$.
   To solve this problem, we first write the lepton propagator
   in the spatial momentum-time mixed representation
   \begin{equation}
   S_\ell^{(T)}({\bf p},t)=\frac{1}{T}\sum_{p_4}S_\ell({\bf
   p},p_4)e^{ip_4t},\quad p_4=\frac{2\pi}{T}n,\quad n=0,1,\cdots T-1,
    \end{equation}
   where $S_\ell({\bf p},p_4)$ is the lepton propagator in momentum space.
   We then construct the propagator with infinite time extent as
   \begin{equation}
   S_\ell^{(\infty)}({\bf
   p},t)\equiv\int_{-\pi}^{\pi}\frac{dp_4}{2\pi}S_\ell({\bf p},p_4)e^{ip_4t}.
   \end{equation}
   Instead of using $S_\ell^{(T)}({\bf p},t)$ with periodic boundary condition we use the time-truncated lepton
   propagator $S_\ell^{[T]}({\bf p},t)$ to avoid round-the-world effects
   \begin{equation}
       S_\ell^{[T]}({\bf p},t)\equiv
    \begin{cases}
        S_\ell^{(\infty)}({\bf p},t) & \mbox{for $-T/2\le t<T/2$} \\
        0 & \mbox{for $t\ge T/2$ or $t<-T/2$}
\end{cases}.
   \end{equation}
   Such a time-truncated lepton propagator is implemented using an overlap
   fermion formulation. The detailed expression of
   $S_\ell^{[T]}({\bf p},t)$ can be found
   in Appendix~\ref{appendix:lepton_prop}.

\subsubsection{Using twisted boundary conditions to insert momenta}

   In the present computation, the kaon is at rest, while the pion, neutrino and anti-neutrino in the final
   state have nonzero momenta as indicated by Eq.~(\ref{eq:mom_setup}). We therefore use twisted boundary conditions
   for the $d$ quark to insert the nonzero momentum
   ${\bf p}_\pi$ for the pion in the final state. Spatial momentum conservation implies that 
   in the process $K^+\to(\ell^+ X)^*\nu\to\pi^+\nu\bar{\nu}$,
   the intermediate state $(\ell^+ X)^*$ has the nonzero momentum ${\bf p}_K-{\bf p_{\nu}}$. 
   Here the superscript $^*$ indicates that the particles are off-shell and $X$ represents 
   hadrons or the vacuum. We use twisted boundary conditions
   for the lepton field and periodic boundary condition for internal up and
   charm quark fields. In this way, the lepton $\ell^+$ has
   momentum ${\bf p}_{\ell}={\bf p}_K-{\bf p}_\nu+\frac{2\pi}{L}{\bf n}$, where ${\bf n}=(n_1,n_2,n_3)$,
   $n_i\in\{0,1,\cdots, L-1\}$, and the 
   hadronic particles $X$ have a total spatial momentum
   ${\bf p}_X=-\frac{2\pi}{L}{\bf n}$. For the intermediate ground state ${\bf
   p}_\ell={\bf p}_K-{\bf p}_\nu$ and ${\bf p}_X={\bf 0}$.

   \subsubsection{Exponentially growing unphysical terms}
\label{subsubsec:exponential}
   In the evaluation of integrals of matrix elements of bilocal operators over a
   large, but finite Euclidean time interval,
   there exist unphysical terms which grow exponentially as the range of the time integration is increased.
   Given the bilocal matrix element $\int d^4x\,\langle\pi^+\nu\bar{\nu}|T[O^{\Delta S=1}(x)O^{\Delta S=0}(0)]|K^+\rangle$, 
   one can insert a complete set of intermediate states between the two interpolating operators, $O^{\Delta S=1}$ and $O^{\Delta S=0}$.
   Integrating over an interval of $-T_a<x_0<T_b$ ($T_a,T_b>0$) gives
   \begin{eqnarray}
   \label{eq:exp_growing}
    \int_{-T_a}^{T_b} dx_0\int d^3\vec{x}\,\langle \pi^+\nu\bar{\nu}|T[O^{\Delta S=1}(x)O^{\Delta S=0}(0)]|K^+   \rangle&=&\nn\\
   &&\hspace{-3.5in}\sum_{n_s}\cfrac{\langle\pi^+\nu\bar{\nu}|O^{\Delta S=1}|n_s\rangle
   \langle n_s|O^{\Delta S=0}|K^+\rangle}{E_{n_s}-E_K}\left(1-e^{(E_K-E_{n_s})T_b}\right)\nn\\  
  &&\hspace{-2.8in}-\sum_n\cfrac{\langle\pi^+\nu\bar{\nu}|O^{\Delta S=0}|n\rangle\langle n|O^{\Delta S=1}|K^+\rangle}{E_K-E_n} \left(1-e^{(E_K-E_n)T_a}\right). \label{eq:Euclidean}
   \end{eqnarray}
   The second and third lines of Eq.~(\ref{eq:exp_growing}) give the second-order weak matrix element 
   together with the unwanted exponential terms.
   For the intermediate states $|n\rangle=|\ell^+\nu\rangle$ and $|\pi^0\ell^+\nu\rangle$,
   the factor $e^{(E_K-E_n)T_a}$ increases exponentially as $T_a$ increases. We have determined the hadronic matrix elements
   $\langle\pi^+|\bar{s}\gamma_\mu\gamma_5 d|0\rangle$ and $\langle0|\bar{s}\gamma_\mu\gamma_5 u|K^+\rangle$ from 2-point correlation functions
   and $\langle\pi^+|\bar{u}\gamma_\mu d|\pi^0\rangle$ and $\langle\pi^0|\bar{s}\gamma_\mu u|K^+\rangle$ from 3-point correlation
   functions (see Table~\ref{tab:lattice_local_matrix} for the results). Therefore we can remove these exponentially growing terms directly.
   At $m_\pi=420$ MeV, the exponential terms from the states $|n\rangle=|\pi\pi\ell^+\nu\rangle$ and $|3\pi\ell^+\nu\rangle$
   vanish at large $T_a$. At the physical pion mass, although the unphysical
   terms from $|\pi\pi\ell^+\nu\rangle$ and $|3\pi\ell^+\nu\rangle$ grow exponentially at large $T_a$, they are significantly
   suppressed by phase space and are expected to be negligible in lattice QCD calculations~\cite{Christ:2016eae}.

\subsubsection{Double integration method}\label{subsubsec:double}
   Since the point-source propagators are placed on each time slice, we can adopt the method proposed in Ref.~\cite{Christ:2012se} and
   perform the time integral over the time locations of both $O_{q\ell}^{\Delta S=1}$ and $O_{q\ell}^{\Delta S=0}$
   \begin{eqnarray}
   \label{eq:double_int}
   &&\sum_{t_1=t_a}^{t_b} \sum_{t_2=t_a}^{t_b}\int d^3{\bf x}\,\langle \pi^+\nu\bar{\nu}|T[O_{q\ell}^{\Delta S=1}({\bf x},t_1)O_{q\ell}^{\Delta S=0}({\bf 0},t_2)]|K^+\rangle
   \nn\\
   &=&T_{\mathrm{box}}\int d^3{\bf x}\,\langle\pi^+\nu\bar{\nu}|O_{q\ell}^{\Delta S=1}({\bf x},0)O_{q\ell}^{\Delta S=0}({\bf 0},0)|K^+\rangle
   \nn\\
   &+&\sum_{n_s}\frac{\langle\pi^+\nu\bar{\nu}|O_{q\ell}^{\Delta S=1}(0)|n_s\rangle\langle n_s|O_{q\ell}^{\Delta S=0}(0)|K^+\rangle}{E_{n_s}-E_K}\left(T_{\mathrm{box}}+\frac{e^{(E_K-E_{n_s})T_{\mathrm{box}}}-1}{E_{n_s}-E_K}\right)
   \nn\\
   &+&\sum_n\frac{\langle\pi^+\nu\bar{\nu}|O_{q\ell}^{\Delta
   S=0}(0)|n\rangle\langle n|O_{q\ell}^{\Delta
   S=1}(0)|K^+\rangle}{E_n-E_K}\left(T_{\mathrm{box}}+\frac{e^{(E_K-E_n)T_{\mathrm{box}}}-1}{E_n-E_K}\right)\,,
    \end{eqnarray}
   where the interval size $T_{\mathrm{box}}=t_b-t_a+1$.
   Given the time locations $t_K$ for the kaon interpolating operator and
   $t_\pi$ for the pion operator, $t_a$ and $t_b$ are required
   to satisfy $t_K\ll t_a$ and $t_\pi\gg t_b$ to guarantee ground-state dominance.
   In practice, we find that for $t_a-t_K\ge 6$ and $t_\pi-t_b\ge6$, the excited-state effects can safely be neglected.
   Therefore, given $t_\pi$ and $t_K$, we can change $T_{\mathrm{box}}$ in a
   range of $[1,t_\pi-t_K-11]$. We can also increase the separation between
   $t_\pi$ and $t_K$ to increase the upper bound for $T_{\mathrm{box}}$.
   On the other hand, $t_\pi-t_K$ should not be too large in order to suppress the
   around-of-world effects. In our calculation, the time extent of the lattice is
   $T=32$.
   We compute propagators for both periodic and anti-periodic boundary conditions in the temporal direction
   and use their average in the calculation.
   This trick effectively doubles the temporal extent of the lattice
   and suppresses round-the-world effects to a negligible level when we choose the maximal value of $t_\pi-t_K=30$.
   For each $t_\pi-t_K$ separation, we shift the whole system in the temporal
   direction and perform the average over all time slices by using
   time translation invariance. We find that such an averaging effectively reduces the statistical
   uncertainty by a factor of about $1/\sqrt{T}$.

   After we obtain the matrix element using the double-integration method for various values of $T_{\mathrm{box}}$,
   we remove the unphysical terms associated with the $|\ell^+\nu\rangle$ and $|\pi^0\ell^+\nu\rangle$ intermediate states. We then
   fit the $T_{\mathrm{box}}$ dependence of the double-integrated
   matrix element to a linear function $b_0+b_1T_{\mathrm{box}}$.
   The slope $b_1$ yields the physical bilocal matrix element.

\subsubsection{Lattice results for the $W$-$W$ diagrams}
   To show the time dependence of the $W$-$W$ diagrams explicitly,
   we define the unintegrated scalar amplitude $F_{WW}(t)$ as a function of the variable $t=t_{\Delta S=1}-t_{\Delta S=0}$ ,
   where $t_{\Delta S=1}$ is the time at which the operator $O_{q\ell}^{\Delta S=1}$ is inserted and
   $t_{\Delta S=0}$ is the time of the insertion of $O_{q\ell}^{\Delta S=0}$:
     \begin{equation}
       F_{WW}(t)=-i\int d^3{\bf x}\,H_{\alpha\beta}(x,y)\sum_\mu c_\mu\mathrm{Tr}\left[\Gamma_{\alpha\beta}(x,y)\gamma_\mu(1+\gamma_5)\right]\,,
   \end{equation}
where $x=({\bf x},t_{\Delta S=1})$ and $y=({\bf y},t_{\Delta S=0})$.
Recalling Eq.\,(\ref{eq:FWW4}), the scalar amplitude $F_{WW}(\Delta, s)$ is obtained by integrating $F_{WW}(t)$ over the time separation
$t$.

   \begin{figure}
   \centering
   \includegraphics[width=.8\textwidth]{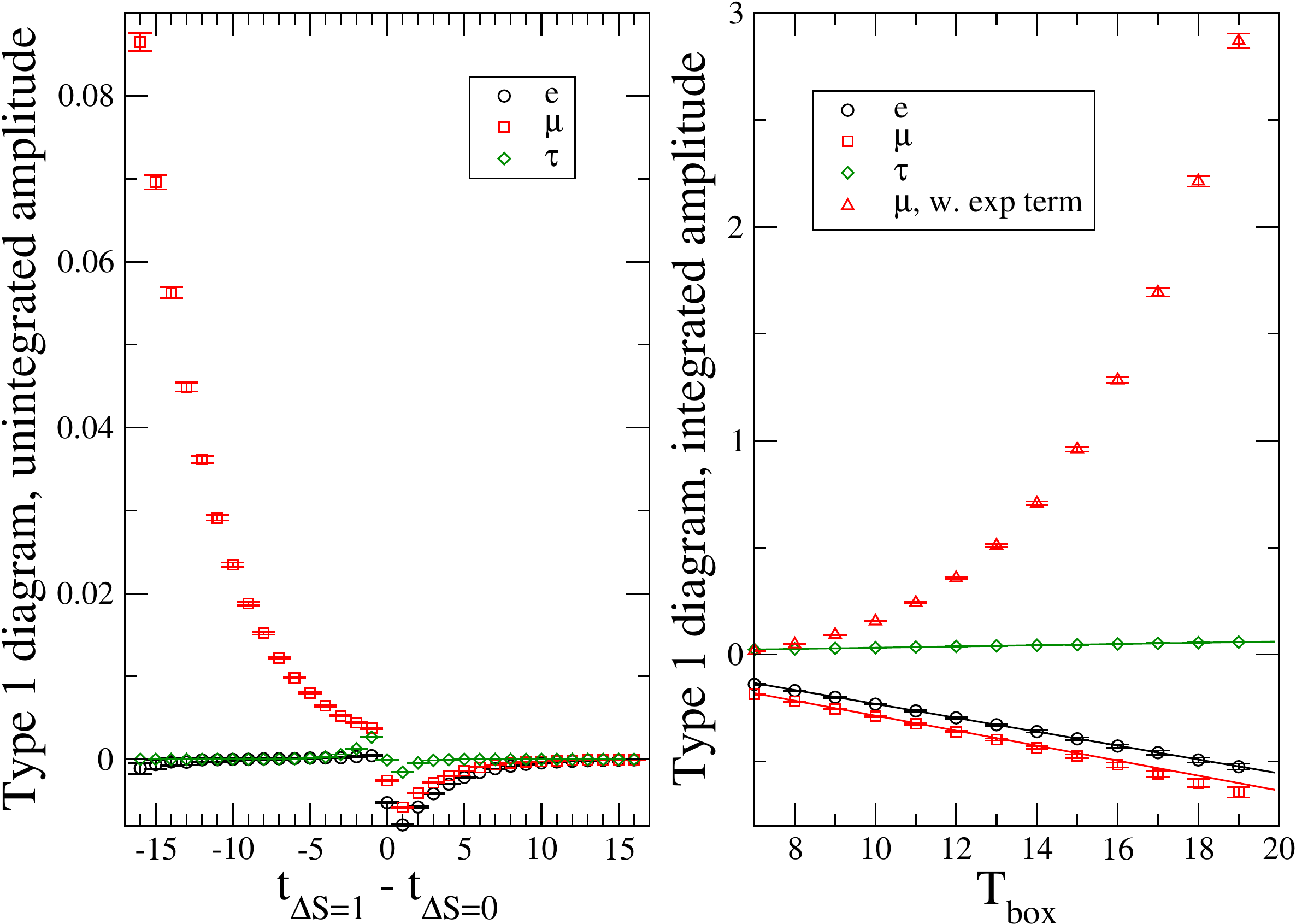}
   \caption{The scalar amplitude for the Type 1 diagram. In the left panel the unintegrated scalar amplitude 
   $F_{WW}(t)$ is shown as a function
   of $t=t_{\Delta S=1}-t_{\Delta S=0}$. The black circles, red squares and green diamonds show the contributions 
   from each of the three leptons
   $e$, $\mu$ and $\tau$ respectively. In the right panel, the integrated scalar amplitude is shown 
   as a function of $T_{\mathrm{box}}$. 
   The exponentially growing term has been removed. For comparison, 
   we also show the results for the muon before the subtraction of the unphysical 
   exponentially growing term (red triangles).}
   \label{fig:type1}
   \end{figure}

   For the Type 1 diagram shown in Fig.~\ref{fig:contraction}, the corresponding unintegrated scalar amplitude
   is shown in the left panel of Fig.~\ref{fig:type1}. For the time region in which $t_{\Delta S=1}\ll t_{\Delta S=0}$,
   this amplitude is dominated by the contribution from ground state, i.e. the
   $|\ell^+\nu\rangle$ state. From among the three lepton flavors
   $\ell=e,\mu,\tau$, we observe the exponentially
   growing time dependence for the muon. This is to be expected since the muon mass is lighter
   than the initial kaon mass. For the electron $e$, the exponentially growing behavior does not appear due to
   the helicity suppression in the process of $K^+\to e^+\nu \to\pi^+\nu\bar{\nu}$. For the $\tau$ flavor, since the intermediate
   states are much heavier than the initial state, there are no exponentially growing contributions.

   We perform the double integration and show the matrix element as a function
   of $T_{\mathrm{box}}$ in the right panel of Fig.~\ref{fig:type1}. The data points marked by the red triangles 
   show the amplitude for the muon, which
   contains the exponentially growing term. The red square points show the same
   amplitude after the subtraction of the unphysical exponentially growing terms. After removing
   the unphysical term, the data is well described by a linear function and by performing a fit 
   we determine the scalar amplitude $F_{WW}(\Delta,s)$ for the three
   lepton flavors. The corresponding results are shown in Table~\ref{table:WW}.
   For comparison, we also calculate the scalar amplitude including only the contributions
   from the ground $|n\rangle$ and $|n_s\rangle$ states, 
   $|\ell^+\nu\rangle$ \& $|K^+\pi^+\ell^-\bar{\nu}\rangle$ respectively. This contribution to $F_{WW}$ is\,\cite{Christ:2016eae}
   \begin{equation}
   -f_Kf_\pi\frac{2q^2}{q^2+m_\ell^2},\quad\textrm{where}\quad q^2=(p_K-p_\nu)^2,
   \end{equation}
   and $f_\pi$ and $f_K$ are the pion and kaon decay constants. As shown in
   Table~\ref{table:WW}, the ground-state dominates the contributions to the Type 1 diagram,
   and the effects of excited intermediate states are very small ($\lesssim3\%$).

\begin{table}[t]
\centering
\begin{tabular}{c|cc|cc}
\hline\hline
$F_{WW}$ & Type 1 & $|\ell^+\nu\rangle$ \& $|K^+\pi^+\ell^-\bar{\nu}\rangle$ & Type 2 & $|\pi^0\ell^+\nu\rangle$ \\
\hline
$e$& $-1.685(47)\times10^{-2}$ & $-1.740(6)\times10^{-2}$ &
$1.123(17)\times10^{-1}$ & $-$\\
$\mu$ & $-1.818(40)\times10^{-2}$ & $-1.822(6)\times10^{-2}$ &
$1.194(18)\times10^{-1}$ & $1.869(14)\times10^{-2}$\\
$\tau$ & $1.491(36)\times10^{-3}$ & $1.471(5)\times10^{-3}$ &
$4.690(77)\times10^{-2}$ & $1.026(3)\times10^{-3}$ \\
\hline
\end{tabular}
\caption{Lattice results, in lattice units, for the scalar amplitude from the $W$-$W$ diagrams. The third and fifth columns show the contributions from the ground states as explained in the text.}
\label{table:WW}
\end{table}

   \begin{figure}
   \centering
   \includegraphics[width=.8\textwidth]{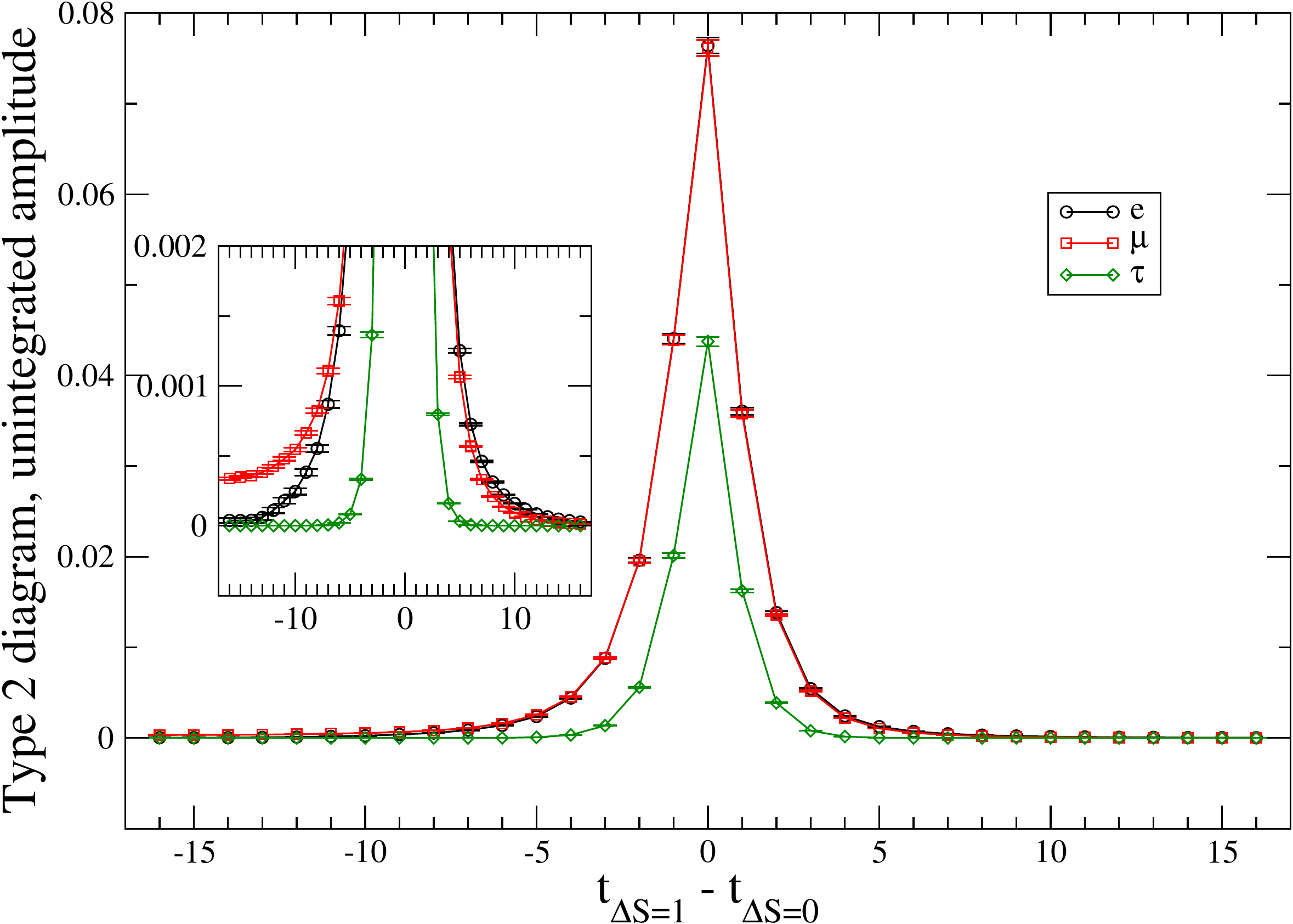}
   \caption{Unintegrated scalar amplitude for the Type 2 diagram.}
   \label{fig:type2}
   \end{figure}

   \begin{figure}
   \centering
   \includegraphics[width=.8\textwidth]{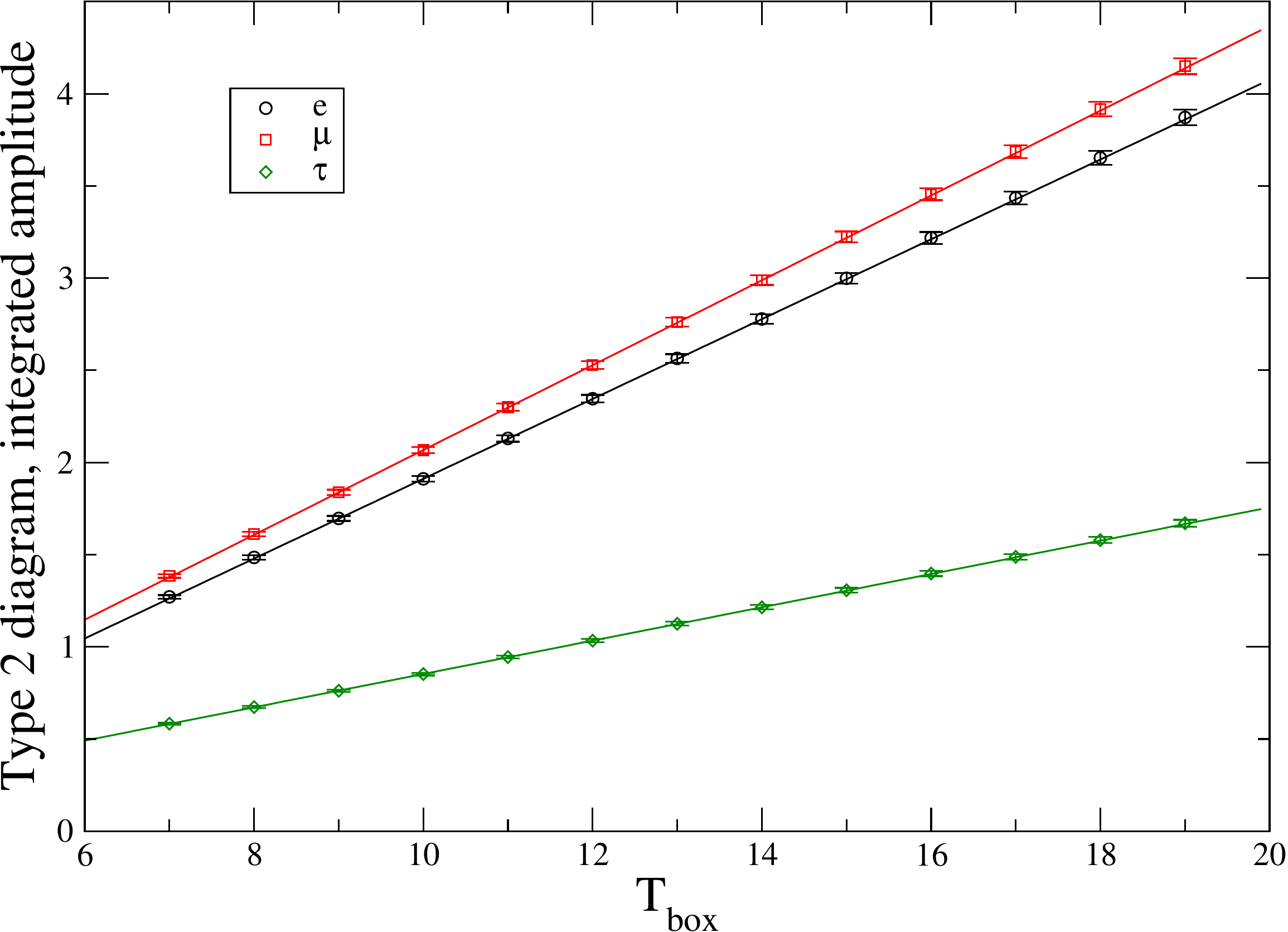}
   \caption{Integrated scalar amplitude for the Type 2 diagram. The unphysical exponentially growing terms for the muon
   have been subtracted.}
   \label{fig:type2_integrated}
   \end{figure}

   In contrast to the Type 1 diagram, even after the GIM subtraction, 
   the Type 2 diagram contains a logarithmic SD (ultraviolet) divergence 
   which needs to be removed as explained in detail in Sec.\,\ref{sec:SD_NPR}.
   The unintegrated scalar amplitude is shown in Fig.~\ref{fig:type2} as a function of $t_{\Delta S=1}-t_{\Delta S=0}$.
   By zooming into the plots, we can observe the exponentially growing time dependence
   for the muon. This exponential behavior is not very significant however, since now
   the intermediate ground state is $|\pi^0\ell^+\nu\rangle$ and
   its energy is similar to $m_K$. Nevertheless this 
   unphysical term still
   contributes a sizeable systematic effect and needs to be subtracted. We therefore calculate the matrix elements 
   $\langle\pi^+\nu\bar{\nu}|O_{u\ell}^{\Delta S=0}(0)|\pi^0\ell^+\nu\rangle$ and 
   $\langle\pi^0\ell^+\nu|O_{u\ell}^{\Delta S=1}(0)|K^+\rangle$
   to remove this unphysical term. For the Type 2 diagram, we do not observe the 
   exponentially growing behavior for the electron.
   In general we would expect there to be no
   helicity suppression in this case, since the intermediate ground state is now 
   semi-leptonic, rather than the leptonic one for the Type I diagram.
   In our calculation, we use the discrete
   lattice momenta $-(2\pi/L){\bf n}$ for the intermediate hadronic particles
   and momenta ${\bf p}_K-{\bf p}_\nu+(2\pi/L){\bf n}$ for the intermediate
   lepton $\ell^+$.
   With such assignments, in the intermediate ground state,
   the neutral pion carries zero momentum
   and the helicity suppression still holds for the electron.  This is
   the reason why we don't observe an exponentially growing term  
   for the electron. The assignment of the spatial momenta for the
   intermediate-state particles is clearly not unique. Different assignments will introduce
   different finite-size effects~\cite{Christ:2015pwa} and we will discuss this
   topic later.

   The integrated scalar amplitude for the Type 2 diagram is shown in Fig.~\ref{fig:type2_integrated}.
   After removing the exponential unphysical contributions and fitting the lattice data to a linear function of $T_{\mathrm{box}}$, we
   determine the values of $F_{WW}$ and include them in Table~\ref{table:WW}. We
   also compute the contributions from the lowest $|\pi^0\ell^+\nu\rangle$
   intermediate state and compare them with the total result for the Type 2
   diagram. For the muon the contribution from $|\pi^0\mu^+\nu\rangle$ 
   is only 16\% of the total contribution. Significant contributions come from
   the excited states, suggesting that the amplitude for Type 2 diagram contains
   a large SD contribution. This SD contribution is cut off by the unphysical
   lattice scale $1/a$. We must introduce a counter term to obtain the physical amplitude, as explained 
   in Sec.\,\ref{sec:SD_NPR} below.

%% file: Z_exchange.tex
\subsection{The matrix element of the bilocal operator for the \boldmath $Z$-exchange diagrams}
\label{subsec:bilocalZ}
   Examples of $Z$-exchange diagrams are given in Fig.~\ref{fig:contraction}.
   We write the bilocal matrix element in the form
   \begin{eqnarray}
   T_Z({\bf p}_K,{\bf p}_\pi)&=&\int d^4x\,\langle\pi^+\nu\bar{\nu}|T[O_u^W(x)O_\ell^Z(0)]|K^+\rangle-\{u\to c\}
   \nn\\
   &=&T^Z_\mu({\bf p}_K,{\bf p}_\pi)\left[\bar{u}(p_\nu)\gamma_\mu(1-\gamma_5)v(p_{\bar{\nu}})\right],
   \end{eqnarray}
   where $O_q^W$ and $O_\ell^Z$ are defined in Eq.~(\ref{eq:operator_Z}).
   The hadronic part of $T_Z$ is given by
   \begin{eqnarray}
   \label{eq:Z_hadronic}
   T^Z_\mu({\bf p}_K,{\bf p}_\pi)=\int d^4x\,\langle\pi^+|T[O_u^W(x) J_\mu^Z(0)]|K^+\rangle-\{u\to c\}.
   \end{eqnarray}
   We separate $T^Z_\mu$ into two parts: $T^Z_\mu=T^{Z,V}_\mu+T^{Z,A}_\mu$, corresponding to the vector
   ($V$) and axial vector ($A$) components of $J_\mu^Z$.
   The $K\to\pi Z^*$ form factors are conventionally defined by
   \begin{eqnarray}
   \label{eq:Z_form_factor}
   T_\mu^{Z,i}({\bf p}_K,{\bf
   p}_\pi)=i\,\left(F_+^{Z,i}(s)(p_K+p_\pi)_\mu+F_-^{Z,i}(s)(p_K-p_\pi)_\mu\right),\quad i=V,A,
   \end{eqnarray}
   where $s=-(p_K-p_\pi)^2$.

   Since the spinor product $\bar{u}(p_\nu){\slashed q}(1-\gamma_5)v(p_{\bar{\nu}})$ vanishes
   for massless neutrinos, only the form factors $F_+^{Z,i}(q^2)$ contribute to the decay amplitude.
   For the vector current, the Ward-Takahashi identity guarantees
   \ba
   \label{eq:Ward_identity}
   (m_K^2-m_\pi^2)F_+^{Z,V}(s)=-sF_-^{Z,V}(s).
   \ea 
   For the axial vector current, in order to determine $F_+^{Z,A}(s)$ from
   $T_\mu^{Z,A}({\bf p}_K,{\bf
   p}_\pi)$, we need to compute the amplitude $T_\mu^{Z,A}({\bf p}_K,{\bf
   p}_\pi)$ for different choices of the polarization $\mu$.
   This requires that either the kaon in the initial state
   or the pion in the final state (or both) carries a non-zero spatial momentum.

   Although we cannot determine $F_+^{Z,i}(s)$ directly from
   $T_\mu^{Z,i}({\bf 0},{\bf 0})$, where both kaon and pion are at rest,
   we still calculate such matrix element for two reasons. 
   Firstly, in our calculation we have used the local
   vector current rather than the conserved vector current. Due to the violation
   of the Ward-Takahashi identity, there will be a SD singularity when the operator
   $J_\mu^{Z,V}$ approaches the operator $O_q^W$. This SD contribution is
   independent of the kaon and pion momenta ${\bf p}_K$ and ${\bf p}_\pi$.
   As a result, we can use $T_\mu^{Z,V}({\bf 0},{\bf 0})$ to remove the SD
   divergence in $T_\mu^{Z,V}({\bf p}_K,{\bf p}_\pi)$.
   Secondly, for the insertion of the axial vector current ($i=A$), the matrix element
   $T_\mu^{Z,A}({\bf 0},{\bf 0})$ provides the most
   accurate data we can obtain for the $Z$-exchange diagrams.
   We define the scalar function $F_0^{Z,A}(s)$ by
   \begin{equation}\label{eq:F0F+F-}
   F_0^{Z,A}(s)\equiv F_+^{Z,A}(s)+\frac{s}{m_K^2-m_\pi^2}F_-^{Z,A}(s).
   \end{equation}
   At ${\bf p}_K=0$ and ${\bf p}_\pi=0$, we obtain
   from $T_\mu^{Z,A}({\bf 0},{\bf 0})$ the scalar function of
   $F_0^{Z,A}(s_{\mathrm{max}})$, where the variable $s$ takes its 
   maximal value of $s_{\mathrm{max}}=(m_K-m_\pi)^2$.
   As we will argue later, $F_0^{Z,A}(s_{\mathrm{max}})$ gives a good
   approximation to $F_+^{Z,i}(s)$ at $s=0$ (for the momentum choice
   in Eq.~(\ref{eq:mom_setup})).

   \subsubsection{Quark loops and disconnected diagrams}
   The operators $O_q^W$ defined in Eq.~(\ref{eq:operator_Z}) 
   can induce closed quark loops through the contraction of $u$ and $c$-quark loops.
   Given each gauge configuration, the $N_r$ components of the random volume-source
   light and charm quark propagators, which have already been used for the 3-point
   correlator,
   can also be used for the 4-point correlator. In addition, in order to be able to evaluate the disconnected
   diagrams in which $\pi\,O_q^W\,K$ and $J_\mu^Z$ form two separate loops,
   we have also calculated 32 random volume-source propagators for the strange quark.
   Thus we can perform a full calculation, which includes all connected,
   self-loop and disconnected
   diagrams.

   \subsubsection{Using chiral ward identities to remove the unphysical terms}

   For the $Z$-exchange diagrams, we start by inserting a complete set of intermediate states between the operators
   $O_u^W$ and $J_\mu^Z$ in Eq.~(\ref{eq:Z_hadronic}). In order to obtain the physical result 
   we need to remove the exponentially growing terms arising from the intermediate states whose energies
   are smaller than the mass of the initial kaon. For the vector current component of $J_\mu^Z$,
   the odd-parity intermediate states
   $|\pi^+\rangle$ and $|3\pi\rangle$ contain
   exponentially growing contributions\,\cite{Christ:2015aha}. The exponentially growing contribution from the three-pion state
   can safely be neglected because of phase space suppression (and in the present calculation it is absent 
   since $m_K<3m_\pi$).
   The unphysical contribution from the single-pion state can be removed by adding to
   the weak Hamiltonian $H_W=O_u^W-O_c^W$ a term proportional to the scalar density: 
   $H_W'=H_W-c_s\,\bar{s}d$.
   The chiral Ward identities imply that the addition of the term proportional to the scalar
   density does not change the on-shell matrix element~\cite{Christ:2012se,Bai:2014cva,Christ:2015aha}. 
   The coefficient $c_s$ can be determined by requiring that
   \begin{equation}
   \langle\pi^+({\bf 0})|H_W(0)-c_s\,\bar{s}d(0)|K^+({\bf 0})\rangle=0.
   \end{equation}
   and our lattice results for $c_s$ are listed in
   Table~\ref{tab:lattice_local_matrix}.

   For the axial-vector current component of $J_\mu^Z$, the parity-even state
   $|2\pi\rangle$ can produce an exponentially growing unphysical term. 
   In this case it is not possible to add a term proportional to the pseudoscalar density 
   ($H_W\to H_W'=H_W-c_p\,\bar{s}\gamma_5d$)
   in such a way as to remove the $I=2$ two-pion  contribution. This is because
   the combination of initial $K^+$ state and the pseudoscalar density $\bar{s}\gamma_5d$ cannot create an $I=2$ $\pi\pi$ state.
   Instead, as shown in
   Table~\ref{tab:lattice_local_matrix},
   we have explicitly calculated the matrix elements
   $\langle(\pi^+\pi^0)_{I=2}({\bf 0})|Q_{i,q}|K^+({\bf 0})\rangle$ and
   $\langle\pi^+({\bf 0})|\bar{u}\gamma_t\gamma_5u-\bar{d}\gamma_t\gamma_5d|(\pi^+\pi^0)_{I=2}({\bf 0})\rangle$
   and are therefore able to remove the unphysical term from the $|2\pi\rangle$ intermediate state (if it
   exists). For the current lattice calculation, since $m_K<2m_\pi$, no removal of
   such an unphysical term is required. Nevertheless the evaluation of
   these matrix elements of local operators allows
   us to determine the contribution to the
   $Z$-exchange diagrams from the $\pi\pi$ intermediate ground state in preparation for future simulations 
   at physical light-quark masses.

   \subsubsection{The local vector current and the short-distance divergence}
   \label{sec:vector_matrix_SD}

   If one uses the conserved vector current, then gauge invariance implies that one
   can write $T_\mu^{Z,V}({\bf p}_K,{\bf p}_\pi)$ as
   \ba\label{eq:gaugeinv}
   T_\mu^{Z,V}({\bf p}_K,{\bf
   p}_\pi)=i\left(-\frac{s}{m_K^2-m_\pi^2}(p_K+p_\pi)_\mu+(p_K-p_\pi)_\mu\right)F_-^{Z,V}(s).
   \ea
   The simplest choice of momenta for the $K\to\pi Z^*$ transition is ${\bf
   p}_K={\bf p}_\pi={\bf 0}$, where
   ${\bf p}_K$ and ${\bf p}_\pi$ are the spatial momenta of the kaon in the initial state and 
   the pion in the final state. Such a choice of momenta 
   is not very useful however, since the kinematic factor
   $-\frac{s}{m_K^2-m_\pi^2}(p_K+p_\pi)_\mu+(p_K-p_\pi)_\mu$ is then equal to $0$. As a consequence, the transition amplitude
   $T_\mu^{Z,V}({\bf 0},{\bf 0})$ vanishes. However, by using the 
   local vector current instead of the conserved one, this simple choice of momenta proves to be useful in making
   a SD correction as we now explain. 

   With the local vector current we can no longer use the Ward-Takahashi identity to obtain (\ref{eq:gaugeinv}).
   The operator product expansion of
   $Q_{i,q}(x)J_\mu^{V_\textrm{loc}}(0)$ can be written in the form
   \ba
   \label{eq:OPE}
   Q_{i,q}(x)J_\mu^{V_\textrm{loc}}(0)\simeq
   c_1\bar{s}\gamma_\mu^Ld+
   c_2\bar{s}\gamma_\nu^L(\partial^2\delta_{\mu\nu}-\partial_\mu\partial_\nu)d+
   c_3\bar{s}\gamma_\nu^L\partial_\mu\partial_\nu d+\cdots
   \ea
   where $\gamma_\mu^L\equiv \gamma_\mu(1-\gamma_5)$ and for compactness of notation we 
   have suppressed the label $i$ on the right-hand side. Dimensional analysis shows that the coefficient $c_1\sim 1/x^6$ 
   at small distances, leading to a $1/a^2$ quadratic divergence after integration over $x$, while
   $c_2$ and $c_3$ both $\sim 1/x^4$ corresponding to a $\log a^2$ logarithmic divergence. All the
   higher-dimension terms are accounted for by the ellipsis in Eq.~(\ref{eq:OPE}).
   It is the $c_2$-term which is physical and the terms with coefficients
   $c_1$ and $c_3$ appear because of the use of the local vector current. By applying the GIM
   mechanism, i.e. subtracting the charm quark contribution ($i=c$) from that of the up quark ($i=u$) 
   we reduce the divergence in the integrated correlation function from the term proportional to $c_1$ to a logarithmic one and 
   remove the divergences from the terms 
   proportional to  $c_{2,3}$, leaving them finite. The logarithmic divergence in the term proportional to $c_1$ arises from 
   the contact term 
   as $x$ approaches $0$ in 
   $Q_{i,q}(x)J_\mu^{V_\textrm{loc}}(0)$. In order to subtract this divergence we introduce a counter term 
   $X_V\,\bar{s}\gamma_\mu^Ld$ writing 
   \begin{equation}
   T_\mu^{Z,V}({\bf p}_K,{\bf p}_\pi)=Z_V\left(T_\mu^{Z,V_\textrm{loc}}({\bf p}_K,{\bf
   p}_\pi)-X_V\,\langle\pi^+({\bf p}_\pi)|\bar{s}\gamma_\mu^Ld(0)|K^+({\bf p}_K\rangle\right)\,,
   \end{equation}
   where 
   \begin{equation}
   T_\mu^{Z,V_\textrm{loc}}({\bf p}_K,{\bf p}_\pi)=\int
   d^4x\,\langle
   \pi^+({\bf p}_\pi)|T[O_u^W(x)J_\mu^{V_\textrm{loc}}(0)]|K^+({\bf p}_K)\rangle-\{u\to c\},
   \end{equation}
   and the superscript $V_\textrm{loc}$ indicates the insertion of the local vector current. A natural condition which 
   can be used to define and determine the coefficient $X_V$ is $T_\mu^{Z,V}({\bf
   0},{\bf 0})=0$, i.e.
   \begin{equation}
   \label{eq:X_V}
   T_\mu^{Z,V_\textrm{loc}}({\bf 0},{\bf 0})-X_V\,\langle\pi^+({\bf 0})|\bar{s}\gamma_\mu
   d(0)|K^+({\bf 0})\rangle=0.
   \end{equation}
   Once $X_V$ is determined, we obtain the form factor $F_+^{Z,V}(s)$ 
   for the choice of momenta in\,(\ref{eq:mom_setup}) with the contact term removed
   using
   \begin{equation}
   F_+^{Z,V}(s)=F_+^{Z,V_{\textrm{loc}}}(s)-X_V\, f_+(s),
   \end{equation}
   where $F_+^{Z,V_\textrm{loc}}(s)$ is obtained from 
   \ba
   Z_VT_\mu^{Z,V_{\textrm{loc}}}({\bf
   p}_K,{\bf
   p}_\pi)=i\,\left(F_+^{Z,V_{\textrm{loc}}}(s)(p_K+p_\pi)_\mu+F_-^{Z,V_{\textrm{loc}}}(s)(p_K-p_\pi)_\mu\right)
   \ea 
   and $f_+(s)$ is defined in Eq.~(\ref{eq:local_form_factor}).
   For the particular choice of momenta given in Eq.~(\ref{eq:mom_setup}) 
   $s=0$ and the Ward-Takahashi identity\,(\ref{eq:Ward_identity}) implies that $F_+^{Z,V}(s)=0$ at $s=0$. We will show later 
   that our lattice result for $F_+^{Z,V}(s)$ is indeed
   consistent with $0$ within the statistical errors. For other values of $s$, $F_+^{Z,V}(s)$ does not vanish and the procedure 
   described in this section allows for its determination.
   
   Note that the term proportional to $c_3$ vanishes in the continuum limit. Having used the GIM mechanism to 
   reduce the degree of divergence and subtracted the remaining contact term by introducing the counterterm, we can relate the
   conserved and local vector currents ($J_\mu^{V_\textrm{con}}$ and $J_\mu^{V_\textrm{loc}}$ respectively)  by 
   $J_\mu^{V_\textrm{con}}=Z_VJ_\mu^{V_\textrm{loc}}$ up to lattice artifacts. Since the artifacts vanish in the continuum limit, 
   so does $c_3$.

   \subsubsection{Single integration method}
   As explained in Sec.\,\ref{subsubsec:double}, when calculating the matrix element for the $W$-$W$ diagrams we have
   used the double integration method. At large $T_{\mathrm{box}}$ the method requires the 
   lattice data to be fit using a simple linear function. However, the drawback of this method is that the lattice data for small
   separations $t_2-t_1$ of the two weak operators are included only when the source-sink separation 
   $t_\pi-t_K \gg T_{\mathrm{box}}$. In fact, this data will accurately contribute to the bilocal matrix element provided 
   $t_\pi-t_K \gg |t_2-t_1|$. The smaller values of $t_\pi-t_K$ allowed by this less stringent condition will give data with 
   smaller errors. The single integration method described in this section makes use of this more accurate data, and are 
   able to significantly improve the precision for the $Z$-exchange diagrams. For the $W$-$W$ diagrams the
   lepton in the intermediate state is not affected by the gauge noise and there would be no improvement.
   
   For the $Z$-exchange diagrams we adopt the single integration method. Given the time locations of the kaon and pion 
   interpolating operators, $t_K$ and $t_\pi$ respectively, we determine the unintegrated matrix element
   using
   \begin{equation}
   T_\mu^{Z,i}(t_\pi,t_H,t_J,t_K)=\frac{4E_\pi E_K}{N_\pi N_K}\langle\phi_\pi(t_\pi)O_q^W(t_H)J_\mu^{Z,i}(t_J)\phi^\dagger_K(t_K)\rangle
   e^{E_\pi(t_\pi-t_J)}e^{E_K(t_J-t_K)}\,.
   \end{equation}
   By examining  the numerical results for $T_\mu^{Z,i}(t_\pi,t_H,t_J,t_K)$ as functions of $t_H$ and $t_J$, we conclude that
   for $t_\pi-t_{H,J}\ge6$ and $t_{H,J}-t_K\ge6$, the effects from excited states 
   can be safely neglected (this is consistent with the corresponding observations for the $W$-$W$ diagrams). 
   For such time separations, by using time-translation invariance $T_\mu^{Z,i}(t_\pi,t_H,t_J,t_K)$ only depends on the time
   difference between $t_H$ and $t_J$. For fixed time separations $t=t_H-t_J$ (but different locations of $t_J$) we fit 
   the matrix elements $T_\mu^{Z,i}(t_\pi,t_H,t_J,t_K)$ to a constant and obtain the average value
   $\bar{T}_\mu^{Z,i}(t_\pi,t,t_K)$.
   We then use these results for $\bar{T}_\mu^{Z,i}(t_\pi,t,t_K)$, to perform a second fit, 
   this time over $t_\pi$ and $t_K$ for each value of $t$. In
   this way, we obtain the matrix element
   $\bar{\bar{T}}_\mu^{Z,i}(t)$, which contains the information from all the
   lattice data constrained by
   $\{t_\pi,t_H,t_J,t_K|\,t_H-t_J=t,t_{H,J}-t_K\ge6,t_\pi-t_{H,J}\ge6\}$. 
   We then perform a single integration of $\bar{\bar{T}}_\mu^{Z,i}(t)$ over 
   the variable $t$ in the range $-T_\textrm{box}\le t\le T_\textrm{box}$ and find the plateau for large $T_\textrm{box}$, 
   once the unphysical terms growing exponentially with $T_\textrm{box}$ have been removed. 
   Since all the possible data for $t_H-t_J=t$ have been used, 
   the single integration method decreases the
   statistical error for the $Z$-exchange diagrams by 30\%-40\% when compared to the
   double integration method.

   \subsubsection{Lattice results}

   \begin{figure}
   \centering
   \includegraphics[width=.8\textwidth]{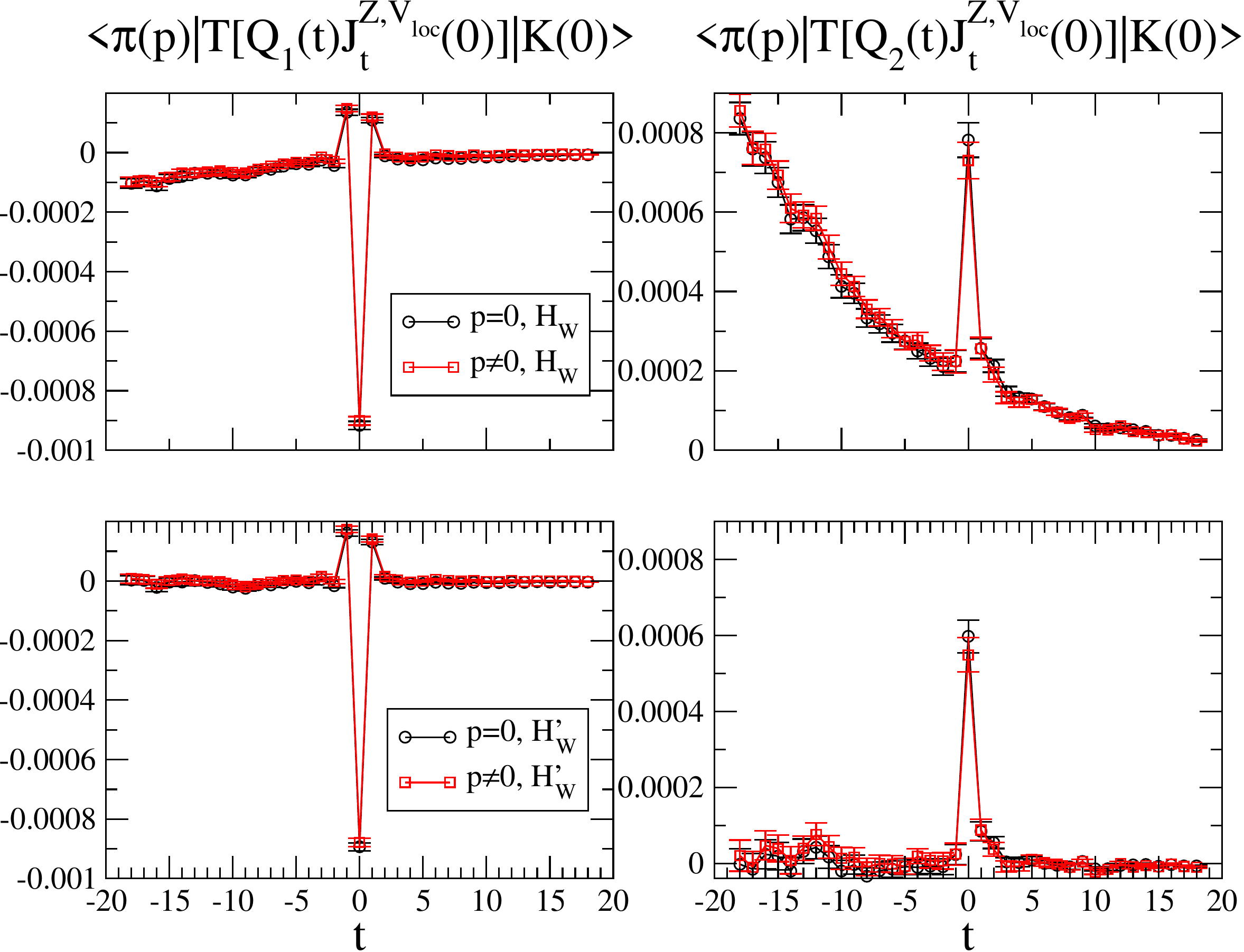}
   \caption{The unintegrated matrix elements for the $Z$-exchange diagrams with the vector current component of $J_\mu^Z$. 
   The vector current polarization direction
   is chosen to be $\mu=t$. In the upper panel, the matrix elements 
   $\int d^3\vec{x}\,\langle\pi^+(p_\pi)|T[H_W(\vec{x},t)J_{\mu=t}^{V_{\textrm{loc}}}(0)|K^+(0)\rangle$
   (for the $Q_{1,q}$ and $Q_{2,q}$ components) are shown as functions of $t=t_H-t_J$. 
   The black circle data points show the lattice results for the momentum mode
   ${\bf p}_K={\bf p}_\pi={\bf 0}$; the red square points show the results for
   ${\bf p}_K={\bf 0}$ and with ${\bf p}_\pi\neq{\bf 0}$ and taking the value in Eq.\,(\ref{eq:mom_setup}). The exponentially growing
   time dependence can be seen at $t\ll0$. In the lower panel, the matrix elements are calculated using the modified Hamiltonian
   $H_W'=H_W-c_s\,\bar{s}d$, so that the exponentially growing terms have been removed.}
   \label{fig:unintegrated_V_insertion_mu3}
   \end{figure}

   We start by presenting the numerical results for the vector current component of $J_\mu^Z$.
   The unintegrated matrix elements $\bar{\bar{T}}_\mu^{Z,V_{\textrm{loc}}}(t)$ as a function of 
   the time separation $t=t_H-t_J$ are shown in the upper panel of
   Fig.~\ref{fig:unintegrated_V_insertion_mu3}.
   Since the four-fermion operator $O_q^W$ is a linear combination 
   of $Q_{1,q}$ and $Q_{2,q}$, we show the numerical results for each operator. When the polarization 
   index of the vector current $J_\mu^{V_{\textrm{loc}}}$ is a
   spacial one, i.e. when $\mu=i=x,\,y$ or $z$, the matrix element is supressed by a factor of ${\bf p}_{\pi,i}/m_K$ 
   as shown in Eq.\,(\ref{eq:gaugeinv}). For this reason and in order to facilitate the comparison of the matrix 
   element at zero and non-zero ${\bf p}_\pi$ we plot the matrix element with 
   $\mu=t$. The black circle data points show the lattice results for the momentum
   ${\bf p}_K={\bf p}_\pi={\bf 0}$; the red square points show the results for
   ${\bf p}_K={\bf 0}$ and with ${\bf p}_\pi$ taking the non-zero value given in
   Eq.~(\ref{eq:mom_setup}).
   As ${\bf p}_\pi$ is small, it is not surprising that the black circle and red square data points
   are very close to each other.

   In the time region $t\ll0$, the dominant intermediate state is the $| \pi^+\rangle$.
   Since this state is lighter than the initial kaon there is an
   exponentially growing contribution as shown in the upper panel of Fig.~\ref{fig:unintegrated_V_insertion_mu3}. 
   We remove this unphysical contribution by adding to the weak Hamiltonian a term proportional to the scalar density 
   $c_s\,\bar{s}d$, with the value of $c_s$ given in Table~\ref{tab:lattice_local_matrix} and show
   in the lower panel of
   Fig.~\ref{fig:unintegrated_V_insertion_mu3} that after correction the lattice data
   does indeed converge to a constant at $t\ll0$.
   
\begin{figure}
   \centering
   \includegraphics[width=.8\textwidth]{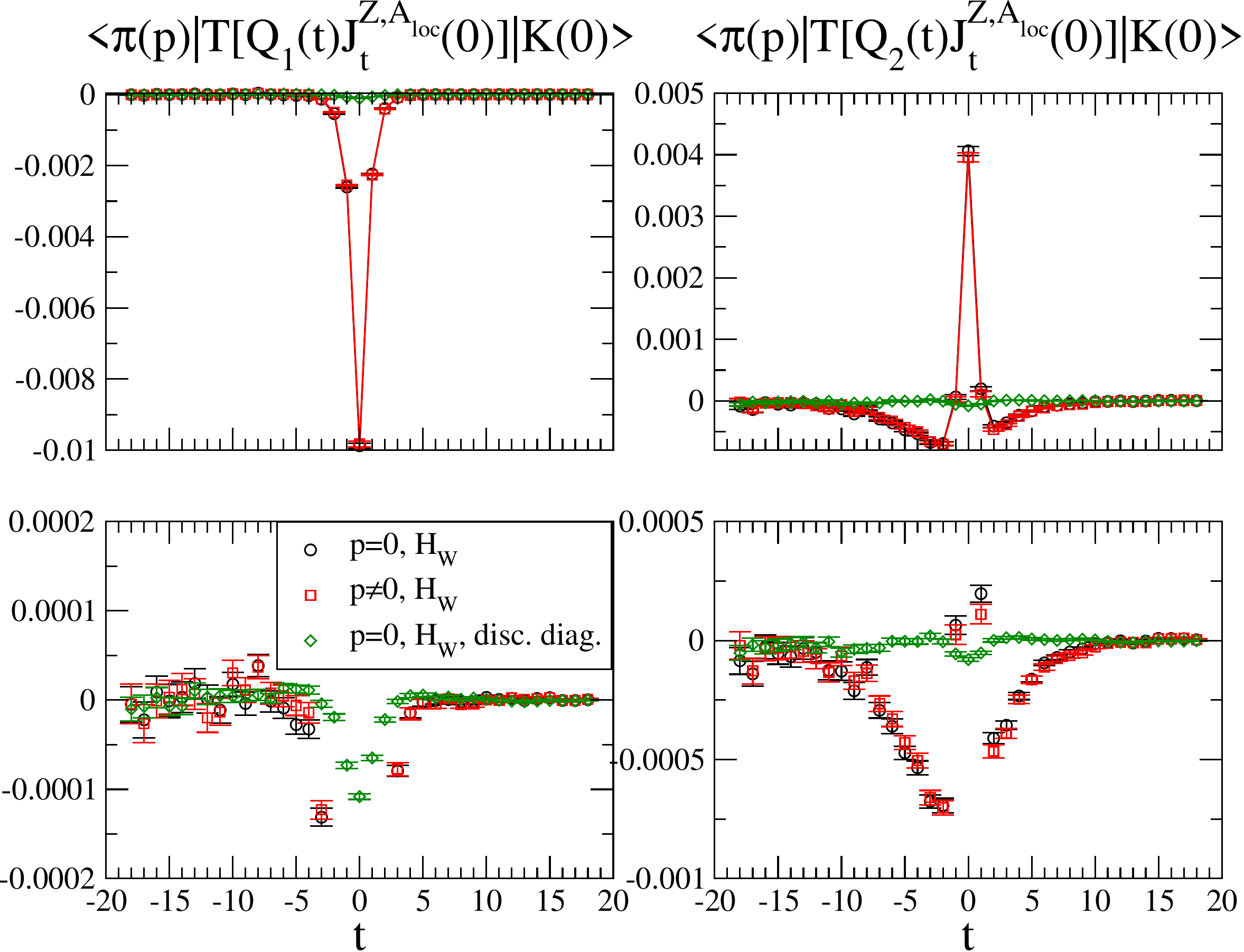}
   \caption{Unintegrated matrix elements for the $Z$-exchange diagrams with the axial vector current component. The axial vector 
   current polarization direction
   is chosen to be $\mu=t$. At $m_\pi=420$\,MeV, no exponentially growing term is observed at $t\ll0$. The black circle data points show the lattice results for the momentum
   ${\bf p}_K={\bf p}_\pi={\bf 0}$; the red square points show the results for
   ${\bf p}_K={\bf 0}$ and with ${\bf p}_\pi\neq{\bf 0}$ and taking the value in Eq.\,(\ref{eq:mom_setup}). These results include only
   the connected and self-loop diagrams. For the disconnected diagrams, the corresponding results are shown by the green diamond 
   symbol. Although noisy, the disconnected contributions are much smaller than the connected ones.}
   \label{fig:unintegrated_axial}
   \end{figure}

   For both the vector and axial-vector components of the weak current $J_\mu^Z$ we have only calculated the 
   contribution of the disconnected diagrams with ${\bf p}_K={\bf p}_\pi={\bf 0}$. For the vector current, the 
   Ward identity implies that the amplitude is zero in this case (i.e. the numerical results are simply gauge noise) and 
   so we do not include the contribution from the disconnected diagrams in Fig.\,\ref{fig:unintegrated_V_insertion_mu3}. 
   For the axial current the amplitude does not vanish for ${\bf p}_K={\bf p}_\pi={\bf 0}$ and below we 
   do include the contribution
   from the disconnected diagrams in Fig.\,\ref{fig:unintegrated_axial} and the corresponding text.

   For the axial-vector current component of $J_\mu^Z$ it is not possible to use the (partially) conserved 
   current to avoid having to make a 
   subtraction of the short-distance divergence, as was done for
   the vector current in Sec.\,\ref{sec:vector_matrix_SD}.   We therefore use the local axial-vector current and follow
   the general procedure for the subtraction of the SD divergence using the RI/SMOM intermediate scheme, 
   as explained in detail in Sec.\,\ref{sec:SD_NPR}.
   The unintegrated matrix elements 
   are shown in Fig.~\ref{fig:unintegrated_axial}. At $t\ll0$ the time
   dependence is dominated by the two-pion state, whose energy
   $E_{\pi\pi}\approx 2m_\pi$ with, in this simulation, $m_\pi=420$ MeV which is larger than the initial kaon mass. 
   Thus we do not observe the exponentially growing $t$ dependence.
 
  In addition to the connected diagrams in Fig.~\ref{fig:contraction}, we also calculate the disconnected diagrams 
  and produce results including all quark contractions.
  The summation of up, down and strange quark loops vanish in the flavor
  $SU(3)$ limit. The remaining charm quark loop is suppressed due to the
  heavy charm quark mass.
  So we expect that the absolute size of the disconnected diagrams is small.
   This expectation is confirmed by a comparison between disconnected data
   points (the green diamond symbol in
   Fig.~\ref{fig:unintegrated_axial}) and the connected and self-loop ones (the
   black circle symbol). 
   Due their small size, although the disconnected diagrams
   have much larger relative statistical errors, they do not contribute a large uncertainty in the total decay amplitude. 
   Thus a complete lattice QCD
   calculation including all the diagrams is practical.

   The lattice results for the matrix elements of the bilocal operators from the $Z$-exchange
   diagrams are summarized in Table~\ref{tab:lattice_results_Z_exchange}.
   The lattice data are shown in three columns for the $Q_{1,q}^{\lat}$ and
   $Q_{2,q}^{\lat}$
   operators and also for the combination $C_1^{\lat}Q_{1,q}+C_2^{\lat}Q_{2,q}$. Here
   $C_1^{\lat}=-0.2186$ and $C_2^{\lat}=0.6424$
   are Wilson coefficients in the lattice regularization. They can be
   related to the Wilson coefficients in the $\MS$ scheme by a $2\times2$
   conversion matrix
   $Z^{\lat\to\MS}$. The details
   will be discussed in Sec.~\ref{sec:SD_NPR}.
   In Table~\ref{tab:lattice_results_Z_exchange}, starting at the top we first
   show the matrix elements for the $K^+({\bf 0})\to\pi^+(\bf{0})$ transition. 
   For the vector-current component,
   these matrix elements can be used to
   determine the coefficient $X_V$ of the counter-term and to correct
   the SD divergence for the $Q_{i,q}(x)J_\mu^{Z,V_\textrm{loc}}(0)$ bilocal operator.
   For the axial vector-current component, we can use these
   matrix elements to determine $F_0^{Z,A}(s_{\mathrm{max}})$.
   The calculation of 
   $F_0^{Z,A}(s_{\mathrm{max}})$ proves to be useful 
   for our exploratory study as it provides approximate information about 
   $F_+^{Z,A}(s_{\mathrm{max}})$,
   see Eq.\,(\ref{eq:F0F+F-}).  
   For the $K^+({\bf 0})\to\pi^+({\bf 0})$ transition, we also include the contributions from the disconnected 
   diagrams in the calculation of $F_0^{Z,A}(s_{\mathrm{max}})$; these data are labeled by the subscript {\em disc}. 
   This is not currently possible for the direct evaluation of $F_+^{Z,A,disc}(0)$, for which we need to 
   use twisted boundary conditions.
   Next in Table\,\ref{tab:lattice_results_Z_exchange} we show the matrix elements 
   for the $K^+({\bf 0})\to\pi^+({\bf p})$ transition, where
   the spatial momentum of the pion is given by Eq.~(\ref{eq:mom_setup}). Due to
   the non-zero momentum of the pion, we are able to obtain the scalar
   function $F_+^{Z,i}(0)$ from these data. From
   Table~\ref{tab:lattice_results_Z_exchange} we obtain the following
   information.
   \bi
   \item The contribution from the vector current $F_+^{Z,V}(s)$ (which is proportional to $s$) is
   expected to be much smaller than that from the axial vector
   current $F_+^{Z,A}(s)$ (which is proportional to $m_c^2$). This is confirmed by our
   lattice data.
   \item At the special momentum transfer $s=0$ we expect that $F_+^{Z,V}(0)=0$
   because of the Ward-Takahashi identity~(\ref{eq:Ward_identity}). This holds for the conserved vector current or, 
   as in the present case,
   by using the local vector current and subtracting the SD counterterm. We see from the table that after subtracting the counter-term, 
   $F_+^{Z,V_\textrm{loc}}(0)-X_Vf_+(0)$ is consistent with zero within 1$\sigma$.
   We also see that $F_+^{Z,V_\textrm{loc}}(0)$ 
   itself is significantly different from 0.
   \item For the axial vector current, we observe that 
   $F_0^{Z,A}(s_{\mathrm{max}})\approx F_0^{Z,A}(0)=F_+^{Z,A}(0)$. 
   Although we are interested in $F_+^{Z,A}(s)$, we conclude that the lattice
   determination of $F_0^{Z,A}(s)$ can be used as a good approximation for
   $F_+^{Z,A}(s)$ for small values of $s$ since $F_-^{Z,A}(s)$ is much smaller than $F_+^{Z,A}(s)$. 
   \item The disconnected diagrams have been evaluated for the
   transition $K^+({\bf 0})\to\pi^+({\bf 0})$. The contribution from these diagrams
   $F_0^{Z,A,disc}(s_{\mathrm{max}})$ is
   about 3\% of that from the connected diagrams
   $F_0^{Z,A}(s_{\mathrm{max}})$. If we accept that $F_0^{Z,A}(s)$
   approximates $F_+^{Z,A}(s)$, then the disconnected diagrams only make a small contribution to the $Z$-exchange diagrams.
   \ei

\begin{table}
\begin{tabular}{l c c c c c c c}
\hline\hline
$Z$-exchange diagrams\\
\hline\hline
$K^+({\bf 0})\to\pi^+({\bf 0})$ & $Q_1$ & $Q_2$ & $C_1^{\lat}Q_1+C_2^{\lat}Q_2$ \\
\hline
$\langle\pi^+({\bf 0})|H_W\,J_t^{V_\textrm{loc}}|K^+({\bf 0})\rangle_{conn}$ 
 & $31.5(1.8)\cdot10^{-4}$ & $13.5(2.0)\cdot10^{-4}$ & $1.8(1.5)\cdot10^{-4}$ \\
$X_V$ defined by Eq.~(\ref{eq:X_V})
& $-39.4(2.2)\cdot10^{-4}$ & $-16.9(2.5)\cdot10^{-4}$ &
$-2.2(1.9)\cdot10^{-4}$ \\
\hline
$\langle\pi^+({\bf 0})|H_W\,J_t^{A_\textrm{loc}}|K^+({\bf 0})\rangle_{conn}$ 
& $7.313(41)\cdot10^{-2}$ & $-0.121(22)\cdot10^{-2}$ & $-1.676(19)\cdot10^{-2}$ \\
$F_0^{Z,A}(s_{\mathrm{max}})$ & $-9.202(61)\cdot10^{-2}$ &
$0.152(28)\cdot10^{-2}$ & $2.109(25)\cdot10^{-2}$ \\
\hline
$\langle\pi^+({\bf 0})|H_W\,J_t^{A_\textrm{loc}}|K^+({\bf 0})\rangle_{disc}$ 
& $11.1(1.3)\cdot10^{-4}$ & $-3.7(1.1)\cdot10^{-4}$ & $-4.8(0.9)\cdot10^{-4}$ \\
$F_0^{Z,A,disc}(s_{\mathrm{max}})$ & $-13.9(1.7)\cdot10^{-4}$ &
$4.7(1.4)\cdot10^{-4}$ & $6.0(1.2)\cdot10^{-4}$  \\
\hline
\hline
$K^+({\bf 0})\to\pi^+({\bf p_\pi})$ \\
\hline
$\langle\pi^+({\bf p_\pi})|H_W\,J_t^{V_\textrm{loc}}|K^+({\bf 0})\rangle_{conn}$ &
$27.9(1.8)\cdot10^{-4}$ & $15.0(2.0)\cdot10^{-4}$ & $3.5(1.6)\cdot10^{-4}$ \\
$\langle\pi^+({\bf p_\pi})|H_W\,J_i^{V_\textrm{loc}}|K^+({\bf 0})\rangle_{conn}$ & 
$i\cdot3.4(0.8)\cdot10^{-4}$ & $i\cdot1.5(0.8)\cdot10^{-4}$ & $i\cdot0.2(0.7)\cdot10^{-4}$ \\
$F_+^{Z,V_\textrm{loc}}(0)$ & $-37.3(2.5)\cdot10^{-4}$ & $-19.1(2.7)\cdot10^{-4}$
& $-4.1(2.1)\cdot10^{-4}$\\
$F_+^{Z,V_\textrm{loc}}(0)-X_Vf_+(0)$ & $-1.8(1.7)\cdot10^{-4}$ & $-2.4(1.8)\cdot10^{-4}$ &
$-1.9(1.4)\cdot10^{-4}$ \\
\hline
$\langle\pi^+({\bf p_\pi})|H_W\,J_t^{A_\textrm{loc}}|K^+({\bf 0})\rangle_{conn}$ & 
$7.276(44)\cdot10^{-2}$ & $-0.141(24)\cdot10^{-2}$ & $-1.681(20)\cdot10^{-2}$\\
$\langle\pi^+({\bf p_\pi})|H_W\,J_i^{A_\textrm{loc}}|K^+({\bf 0})\rangle_{conn}$ & 
$i\cdot0.600(17)\cdot10^{-2}$ & $-i\cdot0.026(16)\cdot10^{-2}$ &
$-i\cdot0.148(12)\cdot10^{-2}$\\
$F_+^{Z,A_\textrm{loc}}(0)=F_0^{Z,A_\textrm{loc}}(0)$ & $-9.158(64)\cdot10^{-2}$ & $0.204(41)\cdot10^{-2}$ &
$2.133(32)\cdot10^{-2}$\\
$F_-^{Z,A_\textrm{loc}}(0)$ & $1.22(27)\cdot10^{-2}$ & $-0.24(24)\cdot10^{-2}$ & $-0.42(18)\cdot10^{-2}$ \\
\hline
\end{tabular}
\caption{Summary of the matrix elements of the bilocal operators and the form factors for the
$Z$-exchange diagrams. The momentum transfer $s$ is given by $s=s_{\mathrm{max}}=(m_K-m_\pi)^2$ for the
$K^+({\bf 0})\to\pi^+({\bf 0})$ transition and $s=0$ for the $K^+({\bf 0})\to\pi^+({\bf p_\pi})$ with ${\bf p_\pi}$ given in Eq.\,(\ref{eq:mom_setup}).}
\label{tab:lattice_results_Z_exchange}
\end{table}

    We end this section by estimating the contribution from the lowest energy
    $|(\pi\pi)_{I=2}\rangle$ state to the $Z$-exchange diagrams. 
    Using the computed matrix elements
    $A_{\pi\pi\to\pi}\equiv\langle\pi^+(0)|\bar{u}\gamma_t\gamma_5u-\bar{d}\gamma_t\gamma_5
    d|(\pi^+\pi^0)_{I=2}(0)\rangle$ and
    $A_{K\to\pi\pi}\equiv\langle(\pi^+\pi^0)_{I=2}(0)|Q_{i,q}|K^+(0)\rangle$
    given in Table\,\ref{tab:lattice_local_matrix} we
    construct the $\pi\pi$ contribution as
    \ba
    T_{\mu=t}^{Z,A,\pi\pi}(s_{\mathrm{max}})=Z_A (-T_3^u) A_{\pi\pi\to\pi}\frac{1}{2E_{\pi\pi}}\frac{1}{E_{\pi\pi}-m_K}(C_1^\lat+C_2^\lat)A_{K\to\pi\pi}\,,
    \ea
    where $Z_A=Z_V$ is the (axial) vector current renormalization factor and
    $T_3^u=\frac{1}{2}$ is the weak isospin associated with the axial vector
    current. The minus sign corresponds to that in the $V-A$
    structure of the weak Hamiltonian. We finally determine the $\pi\pi$ contribution to the form factor
    using
    $F_0^{Z,A,\pi\pi}(s_{\mathrm{max}})=T_{\mu=t}^{Z,A,\pi\pi}(s_{\mathrm{max}})/(-(m_K+m_\pi))=1.526(10)\cdot
    10^{-3}$, which is only 7\% of the $F_0^{Z,A}(s_{\mathrm{max}})$ given in
    Table~\ref{tab:lattice_results_Z_exchange}, suggesting that the dominant
    contribution to the $Z$-exchange diagrams comes from higher excited states and
    SD physics. Once simulations at physical quark masses are performed, when the two-pion state contributes exponentially growing 
    contributions in $T_{\textrm{box}}$ which will need to be subtracted, its contribution to $F_+^{Z,A}$ will have to be studied again.



%% file: NPR.tex
In this section we discuss the subtraction of the additional ultraviolet divergences which appear when the two local
operators which are the components of a bilocal operator approach each other. In Sec.\,\ref{subsec:NPR} we review the 
theoretical background and in Sec.\,\ref{subsec:NPRnumerical} we present the numerical results for the subtraction constants.

\subsection{Non-perturbative renormalization using RI/SMOM scheme}
    \label{subsec:NPR}

    In Sec.~\ref{sec:vector_matrix_SD}, for the vector current insertion we have used the matrix element
    of the transition $K({\bf 0})\to\pi({\bf 0})$ to remove the
    SD divergence in the matrix element of the bilocal operators. Here we describe 
    a more general method to remove the SD divergence, following the procedures
    developed in Ref.~\cite{Christ:2016eae}.

\begin{figure}
   \centering
   \includegraphics[width=.55\textwidth]{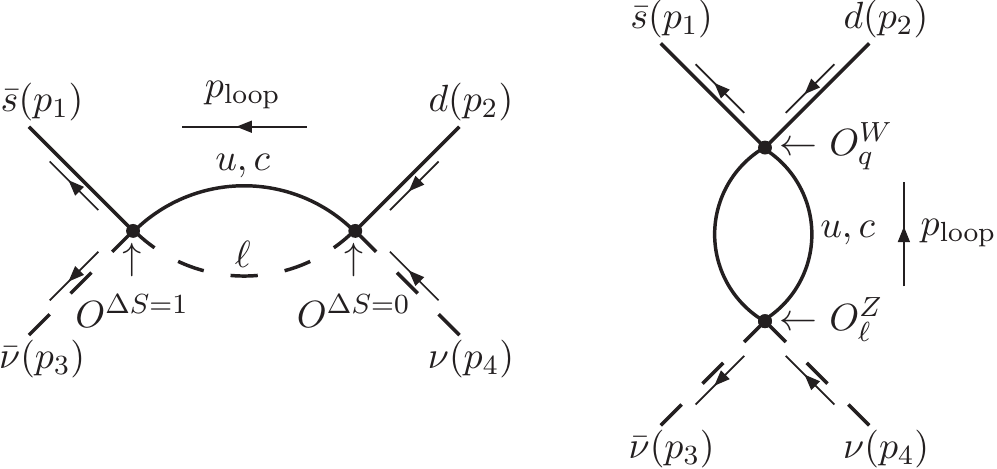}
   \caption{Left: SD divergent loop in $W$-$W$ diagrams. Right: SD divergent loop in $Z$-exchange diagrams.}
   \label{fig:SD_loop}
   \end{figure}

    Given a bare lattice bilocal operator $\{Q_A^\lat Q_B^\lat\}^\lat_a$, in order to define and 
    determine its SD component, we construct an off-shell Green's function  
   \begin{eqnarray}
   \label{eq:Green_function}
   G_{\alpha\beta\rho\sigma}^{AB}=\langle s_\alpha(p_1)\nu_\rho(p_3)\left[
   \int d^4x\,Q_A(x)Q_B(y)\right]\bar{d}_\beta(p_2)\bar{\nu}_\sigma(p_4)\rangle
   \end{eqnarray}
   where the fermionic fields $s$, $\bar{d}$, $\nu$ and $\bar{\nu}$ carry the non-exceptional Euclidean 4-momenta
   \ba
   \label{eq:extern_mom}
   p_1=(\xi,\xi,0,0),\quad p_2=(\xi,0,\xi,0),\quad p_3=(0,-\xi,0,-\xi),\quad
   p_4=(0,0,-\xi,-\xi).
   \ea
   The quark and lepton contractions contributing to the SD divergence are shown in
   Fig.~\ref{fig:SD_loop}.
   We choose the external momenta $p_i$ to satisfy $p_i^2\equiv\mu_0^2=2\xi^2$.
   The momentum $p_{\mathrm{loop}}$ flowing into the internal loop is given by
   $p_{\mathrm{loop}}=(\xi,0,0,-\xi)$ for $W$-$W$ diagrams
   and $p_{\mathrm{loop}}=(0,\xi,-\xi,0)$ for $Z$-exchange diagrams. 

   \label{sec:SD_correction}

   For the $Z$-exchange diagrams the weak Hamiltonian is a linear combination of two operators $O_{1,q}$ 
   and $O_{2,q}$ 
   which mix under renormalization. The second operator however, is either the local vector or 
   axial vector current with a multiplicative renormalization constant $Z_V$. For the $W$-$W$ diagrams 
   both the operators $Q_A$ and $Q_B$, i.e. $Q_{q\ell}^{\Delta S=1}$ and $Q_{q\ell}^{\Delta S=0}$, 
   renormalise multiplicatively. Nevertheless, in this section we 
   present a general discussion in which both $Q_A$ and $Q_B$ mix with other operators and 
   in the absence of such mixing the corresponding renormalization matrices in the formulae become 
   numerical constants. In order to allow the RI/SMOM normalization to be
   imposed at four-momenta that can be held fixed in physical units in both magnitude and direction when we later perform a continuum 
   extrapolation, we will use twisted external momenta whose components are nonecessarily integer multiples of 
   $2\pi/ L$~\cite{Arthur:2010ht}.

   We perform the calculation in the Landau gauge. Imposing the twisted boundary condition on the quark field, 
   $q(x+L\hat{\mu})=e^{i\theta_\mu}q(x)$, is equivalent to multiplying the gauge field by a factor of 
   $e^{ip_\mu}$: $U_\mu\to U_\mu^\prime = e^{ip_\mu}U_\mu $, with $\theta_\mu = p_\mu L (\bmod 2\pi)$.
   We can consider this multiplication as a global $U(1)$ rotation. Since $p_1\neq p_2$, we multiply the gauge field by a  
   different factor $e^{i p_{i,\mu}}$ when calculating the corresponding quark propagator.
   Calculating a zero-momentum  volume-source quark propagator on
   the rotated gauge fields $U_\mu'$ naturally assigns the non-zero external momentum $p_i$
   for the external quark propagator. For the $Z$-exchange diagram, we rotate the gauge fields
   with a phase factor of $e^{ip_{{\mathrm{loop},\mu}}}$. Combining the
   point-source quark
   propagators with and without this gauge rotation, we can arrange that the internal loop can carry an appropriately twisted momentum
   $p_{\mathrm{loop}}$.  For the $W$-$W$ diagram, the momentum
   $p_{\mathrm{loop}}$ is carried by the internal lepton field while the internal
   quark propagators are calculated with unrotated gauge fields.
   We treat the position $x$ of one operator as the source and the position $y$ of the other operator as the sink.  The source $x$ is treated as a fixed, point source while the sink $y$ is summed over the full space-time volume after the other propagators connected to $y$ have been included.
   To improve the precision, we place the point source at $32$
   different positions and then exploit translation-invariance to average over
   these source locations.
   
When implementing the non-perturbative renormalization as described above, we impose different (twisted) boundary conditions within the same diagram for different fermion propagators of the same flavor. We argue below that this can be done consistently for connected diagrams evaluated in the perturbative regime.
This is in contrast to the use of different boundary conditions for different portions of an amplitude at low energies.  For example, the effects of using different boundary condition for the valence and sea quarks require the study of an effective field theory and careful consideration of possible on-shell intermediate states~\cite{Sachrajda:2004mi}.   Our use of multiple boundary conditions is introduced to allow specific external momenta and we now show that the errors introduced by this approach fall exponentially with the volume.
 
Because the usual RI/SMOM conditions are applied for large non-exceptional Euclidean external momenta, the amplitudes being studied are infrared safe and may be represented by a standard, all-orders perturbative sum.  Further, we assume that the twist angles $\theta_\mu$ are rational multiples of $2\pi$,  $\theta_\mu = 2\pi(r_\mu/r)$ for five integers $r_\mu$, $0 \le \mu \le 3$ and $r$.  For a quark-line-connected diagram of the sort described above a sequence of twisted quark propagators is introduced connecting the vertex at which the twisted momentum enters to the vertex at which it exits so that momentum will be conserved at each vertex of the graph.  If this same Green's function were evaluated in a much larger volume of side
$L^\prime= r L$, all of the momenta would be integral multiples of $2\pi/L^\prime$ with no twisting needed. 
 
We now use the Poisson summation formula to argue that these two Green's functions must differ by terms which vanish exponentially in the length $L$.  In both cases we can use momentum conservation to route the twisted external momenta on the same path through the graph.  The internal momentum sums for both volumes then involve momenta that are added to the twisted momentum, when present, carried by each quark line.  For the original volume $L^4$, the result depends on the arbitrary routing of the twisted momentum.  For the larger volume $(rL)^4$ the loop momenta can be redefined to move the path followed
within the graph by the external momentum.  Since there are no nearby singularities for such an off-shell Euclidean amplitude, the Poisson summation formula guarantees that these two sums over discrete internal momenta, one with $r^4$ more terms than the other, will differ by terms which vanish exponentially in the distance $L$~\footnote{The authors thank Chulwoo Jung and William Detmold for independently explaining this argument to us.}.

   In the next step we calculate the amputated vertex
   $\Gamma_{\alpha\beta\rho\sigma}^{\mathrm{AB}}$
   from the Green's function through
   \begin{eqnarray}
    \label{eq:amputated_Green_F}
   \Gamma_{\alpha\beta\rho\sigma}^{\mathrm{AB}}=\langle S^{-1}_s(p_1)\rangle_{\alpha\alpha'} \langle S^{-1}_d(p_2)\rangle_{\beta\beta'}
   \langle S^{-1}_\nu(p_3)\rangle_{\rho\rho'}  \langle S^{-1}_{\bar{\nu}}(p_4)\rangle_{\sigma\sigma'}
   G_{\alpha'\beta'\rho'\sigma'}^{\mathrm{AB}}
   \end{eqnarray}
   where $S^{-1}_{s,d}(p_i)$ stand for the inverse of the full strange and down quark propagators 
   and $S^{-1}_{\nu,\bar{\nu}}(p_i)$
   for the inverse of free neutrino propagators.
   Another amputated vertex $\Gamma^{0}_{\alpha\beta\rho\sigma}$ can be obtained
   from the Green's function in Eq.~(\ref{eq:Green_function}) if the bilocal
   operator product $\int d^4x\,Q_A^\lat(x)Q_B^\lat(y)$ is 
   replaced by a bare local operator $Q_0^\lat(y)$. At tree level,
   $\Gamma^0$ is simply given by $\hat{\Gamma}=[\gamma_\mu(1-\gamma_5)]_{q}\otimes [\gamma_\mu(1-\gamma_5)]_\nu$,
   where the subscript $q$ indicates the quark flavor space and $\nu$ the neutrino flavor space. 
   The color structure is not shown explicitly in $[\gamma_\mu(1-\gamma_5)]_{q}$
   since at tree level it is trivial.
   We use $\hat{\Gamma}$ to construct the projector
   \begin{eqnarray}
   \label{eq:projector}
   \mathrm{P}=\frac{\hat{\Gamma}^{\dagger}}{\textmd{Tr}_{s,c}[\hat{\Gamma}^{\dagger}\hat{\Gamma}]}\,,
   \end{eqnarray}
   where $\textmd{Tr}_{s,c}$ requires the trace over both the spin and color indices.
   When the projector acts on $\hat{\Gamma}$ it yields
   $\mathrm{Tr}_{s,c}[\mathrm{P}\hat{\Gamma}]=1$.

   We use the large external momenta
   $p_i^2=\mu_0^2\gg\Lambda_{\mathrm{QCD}}^2$ to capture the SD contribution to the
   bilocal operator product $\int d^4x\,Q_A^{\lat}(x) Q_B^{\lat}(y)$ and then relate this contribution to the projection of 
   the amputated Green function $\Gamma_0$ of the local operator $Q_0^{\lat}(y)$, where 
   with the same external momenta we require:
   \ba
   \label{eq:SD_projection}
   \textmd{Tr}_{s,c}[\mathrm{P}\Gamma^{\mathrm{AB}}]=X_{AB}^{\lat}(\mu_0,a)\,\textmd{Tr}_{s,c}[\mathrm{P}\Gamma^{0}].
   \ea
   Recall that the local operator is $Q_0=(\bar{s}d)_{V-A}(\bar{\nu}\nu)_{V-A}$. 
   Using the coefficient $X_{AB}^{\lat}(a,\mu_0)$, we remove the SD
   divergence by constructing the subtraction $\int d^4x\,Q_A^{\lat}(x)
   Q_B^{\lat}(y)-X_{AB}^{\lat}(a,\mu_0)Q_0^{\lat}(y)$.

   Following Ref.~\cite{Christ:2016eae}, we adopt the renormalization condition
   \ba
   \label{eq:renorm_cond}
   \langle
   \{Q_A^{\RI}Q_B^{\RI}\}^{\RI}_{\mu_0}\rangle_{p_i^2=\mu_0^2}=\langle\{Q_A^{\RI}Q_B^{\RI}\}_a^\lat\rangle_{p_i^2=\mu_0^2}-X_{AB}(\mu_0,a)\,\langle
   Q_0^{\RI}(\mu_0)\rangle_{p_i^2=\mu_0^2}=0,
   \ea 
   to define the bilocal operator in the RI/SMOM scheme
   \ba\label{eq:bilocalRI}
   \{Q_A^{\RI}Q_B^{\RI}\}^{\RI}_{\mu_0}\equiv \{Q_A^{\RI}Q_B^{\RI}\}_a^\lat -
   X_{AB}(\mu_0,a) Q_0^{\RI}(\mu_0)\,.
   \ea
   The local operators in the RI/SMOM scheme $Q_i^{\RI}(\mu_0)$
   are related to the bare lattice operators $Q_i^{\lat}(a)$ through the
   renormalization relation $Q_i^{\RI}(\mu_0)=Z_{ij}^{\lat\to\RI}(a\mu_0)Q_j^{\lat}(a)$. 
   The angled brackets $\langle\cdots\rangle_{p_i^2=\mu_0^2}$ in Eq.\,(\ref{eq:renorm_cond}) indicate the
   amputated Green's function with the momentum assignments in Eq.\,(\ref{eq:extern_mom}).
   Given the external momenta $p_i$, we impose the standard RI/SMOM renormalization condition for local operators.
   Specifically, the amputated Green's function of the renormalized operator in the RI/SMOM scheme $Q_i^{\RI}(\mu_0)$ 
   is required to be equal to the tree-level amputated Green's function at the scale $\mu_0$ and this determines the 
   matrix of renormalization constants $Z_{ij}^{\lat\to\RI}(a\mu_0)$. 
   $X_{AB}(\mu_0,a)$ defined in Eq.\,(\ref{eq:renorm_cond}) is related 
   to $X_{AB}^{\lat}(\mu_0,a)$ defined in
   Eq.\,(\ref{eq:SD_projection}) by 
   \ba
   X_{AB}(\mu_0,a)=\frac{Z_{AC}^{\lat\to\RI}(a\mu_0)
   Z_{BD}^{\lat\to\RI}(a\mu_0)}{Z_{Q_0}^{\lat\to\RI}(a\mu_0)}\,
   X_{CD}^{\lat}(\mu_0,a)\,,
   \ea
   where it is understood that a sum is to be performed over the operator types $C$ and $D$ which mix with $A$ and $B$ respectively.

   Once the renormalization condition~(\ref{eq:renorm_cond}) has been specified, the bilocal operator
   $\{Q_A^{\RI}Q_B^{\RI}\}^{\RI}_{\mu_0}$ is defined with no ambuiguity.
   The bilocal operator in the $\MS$ scheme,
   $\{Q_A^{\MS}Q_B^{\MS}\}^{\MS}_{\mu}$, is given in terms of
   bilocal and local RI operators
   as shown in Eq.~(\ref{eq:bilocal_MS_RI}).
   By multiplying the Wilson coefficient
   $C_A^{\MS}(\mu)C_B^{\MS}(\mu)$, we have
   \begin{eqnarray}
   C_A^{\MS}(\mu)C_B^{\MS}(\mu)\{Q_A^{\MS}Q_B^{\MS}\}^{\MS}_{\mu}&=&\nonumber\\
   \label{eq:bilocal_MSbar}
   &&\hspace{-1.5in}C_A^{\RI}(\mu_0)C_B^{\RI}(\mu_0)\{Q_A^{\RI}Q_B^{\RI}\}^{\RI}_{\mu_0}
   +C_A^{\MS}(\mu)C_B^{\MS}(\mu)Y_{AB}(\mu,\mu_0)Q_0^{\RI}(\mu_0).
   \end{eqnarray}
   Here,  for example,
   $C_A^{\RI}(\mu_0)Q_A^{\RI}(\mu_0)=C_A^{\MS}(\mu)Z_{AC}^{\RI\to\MS}(\mu/\mu_0)Q_C^{\RI}(\mu_0)$
   where $Z^{\RI\to\MS}(\mu/\mu_0)$ is the $\RI\to\MS$ conversion matrix and we sum over all 
   operators $C$ which mix with $A$. There is a similar expression for $Q_B$ and all the operators 
   which mix with it. The parameter 
   $Y_{AB}(\mu,\mu_0)$, which is determined perturbatively, accounts for the difference between the bilocal
   operators in the $\MS$ and RI schemes. We will discuss the determination
   of $Y_{AB}(\mu,\mu_0)$ in Sec.~\ref{sec:PT_YAB}. 
   
   It is useful to write the
   $\MS$ bilocal operators in terms of the bare lattice operators whose matrix elements are 
   computed non-perturbatively
   \ba
   C_A^{\MS}(\mu)C_B^{\MS}(\mu)\{Q_A^{\MS}Q_B^{\MS}\}^{\MS}_{\mu}&=&
   C_A^{\lat}(a)C_B^{\lat}(a)\left(\{Q_A^{\lat}Q_B^{\lat}\}_a^{\lat}-X_{AB}^{\lat}(\mu_0,a)Q_0^{\lat}(a)\right)
   \nn\\
   &&\hspace{1cm}+C_A^{\MS}(\mu)C_B^{\MS}(\mu)Y_{AB}(\mu,\mu_0)Z_{Q_0}^{\lat\to
   RI}Q_0^{\lat}(a).
   \ea
   where 
   \ba
   C_A^{\lat}(a)Q_A^\lat(a)=C_A^{\MS}(\mu)(Z^{\RI\to\MS}(\mu/\mu_0)Z^{\lat\to\RI}(a\mu_0))_{AC}\,Q_C^{\lat}(a)\,,
   \ea
   and again there is a summation over all operators which mix with $A$; a similar 
   expression holds for $Q_B$.
   
   We now consider the specific case of the Z-exchange diagrams where $Q_B$ is a vector or axial-vector 
   current  and  for $Q_A$ we
   consider each of the two operators $Q_{1q}$ and $Q_{2q}$ which mix under renormalization. (Here we use the conventional 
   operators $Q_1$ and $Q_2$ rather than the combinations $Q_\pm = Q_1\pm Q_2$ which belong to different representations of 
   $SU_L(4)$ and do not mix under renormalization.)
   The conversion matrix for these two operators,
   $Z^{\RI\to\MS}(\mu/\mu_0)=I+\Delta
   r^{\RI\to\MS}$, has been given by Ref.~\cite{Lehner:2011fz} at the
   scale $\mu=\mu_0$. For the entries of the renormalization matrix
   $Z^{\lat\to\RI}(a\mu_0)$ we take the values from
   Ref.~\cite{Christ:2012se}.
   At the scale $\mu=\mu_0=2.15$ GeV, the parameters used to determine
   $C_1^{\lat}$ and $C_2^{\lat}$ are given in Table~\ref{tab:npr_local}.
   These are given by
   \begin{equation}\label{eq:lat_to_RI_matching}
   C_{i}^{\lat}(a)=\sum_{k,l=1,2}C_{k}^{\MS}(\mu)Z^{\RI\to\MS}_{kl}(\mu/mu_0)
   Z^{\lat\to\RI}_{li}(a\mu_0)\,\qquad (i=1,2)\,.
   \end{equation}
   The values for $C_i^{\lat}$ quoted here are about 1.4\% different from the values used in
   Ref.~\cite{Christ:2012se}, as in this paper we use a 3-loop formula for the strong
   coupling evolution while Ref.~\cite{Christ:2012se} used a 2-loop formula.
  
\begin{table}
\begin{tabular}{cc|cc|cc|cc}
\hline
\hline
$C_1^{\MS}$ & $C_2^{\MS}$ & $\Delta r_{11} = \Delta
r_{22}$ &  $\Delta r_{12} =
\Delta r_{21}$ & $Z_{11}^{\lat\to\RI}=Z_{22}^{\lat\to\RI}$ &
$Z_{12}^{\lat\to\RI}=Z_{21}^{\lat\to\RI}$ & $C_1^{\lat}$ & $C_2^{\lat}$ \\
\hline
$-0.2911$ & $1.1353$ & $-6.482\cdot 10^{-2}$ & $7.429\cdot
10^{-3}$ & $0.5916$ &
$-0.05901$ & $-0.2186$ & $0.6424$ \\
\hline
\end{tabular}
\caption{Parameters relevant for the $Z$-exchange diagram. The Wilson coefficients in the $\MS$ 
scheme $C_{1,2}^{\MS}(\mu)$, the entries of the 
$\RI\to\MS$ matching matrix $\Delta r^{\RI\to\MS}(\mu,\mu_0)$, the entries of the 
non-perturbative $\lat\to\RI$ operator renormalization matrix
$Z^{\lat\to\RI}(a\mu_0)$ and the Wilson
coefficients $C_{1,2}^{\lat}(a)$, defined in Eq.~(\ref{eq:lat_to_RI_matching}), are evaluated at the
scale $\mu=\mu_0=2.15$ GeV.}
\label{tab:npr_local}
\end{table}

\subsection{Lattice results for the renormalization of bilocal operators}\label{subsec:NPRnumerical}

The coefficients $X^{\lat}_{AB}(\mu_0,a)$ have been determined using
Eq.~(\ref{eq:SD_projection}).
From the full ensemble of $800$ configurations, we use one from every ten configurations to
calculate the off-shell Green's function for both bilocal and local operators.
To study the scale dependence, we vary $\mu_0$ from 1 GeV to 4 GeV in steps of 0.25 GeV and the results
are presented in Table~\ref{tab:result_npr}. For the $Z$-exchange diagram, we give
the results for $Q_{1,q}$ and $Q_{2,q}$ separately and also for the combination $C_1^{\lat}Q_{1,q}+C_2^{\lat}Q_{2,q}$. For the $W$-$W$ diagrams, we write the results
for the three lepton flavors $\ell=e,\mu,\tau$ respectively.

\begin{table}
\centering
\begin{tabular}{c | c c c | c c c}
\hline\hline
 & \multicolumn{3}{c|}{$X_{AB}^{\lat}(\mu_0)$ from the $Z$-exchange diagrams} &
 \multicolumn{3}{c}{$X_{AB}^{\lat}(\mu_0)$ from the $W$-$W$ diagrams}   \\
\hline
$\mu_0$ [GeV] & $Q_1$ & $Q_2$ & $C_1^{\lat}Q_1+C_2^{\lat}Q_2$ & $e$ & $\mu$ & $\tau$ \\
\hline
1.00 & $-6.659(39)$ & $-1.671(18)$ & $0.382(12)$ & $4.958(140)$ & $5.481(155)$ & $2.866(80)$ \\
1.25 & $-6.019(32)$ & $-1.516(14)$ & $0.342(9)$  & $4.697(115)$ & $4.690(115)$ & $2.613(63)$ \\
1.50 & $-5.379(26)$ & $-1.365(14)$ & $0.299(10)$ & $3.889(73)$  & $3.878(72)$  & $2.279(42)$ \\
1.75 & $-4.723(22)$ & $-1.211(12)$ & $0.255(8)$  & $3.304(48)$  & $3.289(47)$  & $2.030(29)$ \\
2.00 & $-4.112(20)$ & $-1.061(12)$ & $0.217(7)$  & $2.644(36)$  & $2.679(36)$  & $1.756(24)$ \\
2.25 & $-3.555(19)$ & $-0.932(12)$ & $0.178(8)$  & $2.215(28)$  & $2.213(28)$  & $1.506(19)$ \\
2.50 & $-3.045(18)$ & $-0.815(12)$ & $0.142(8)$  & $1.821(21)$  & $1.818(21)$  & $1.276(15)$ \\
2.75 & $-2.605(17)$ & $-0.701(12)$ & $0.119(7)$  & $1.492(17)$  & $1.487(17)$  & $1.074(12)$ \\
3.00 & $-2.229(18)$ & $-0.601(11)$ & $0.101(7)$  & $1.200(13)$  & $1.203(13)$  & $0.897(10)$ \\
3.25 & $-1.897(19)$ & $-0.513(11)$ & $0.085(7)$  & $0.969(9)$   & $0.968(9)$   & $0.737(7)$ \\
3.50 & $-1.596(21)$ & $-0.441(12)$ & $0.066(8)$  & $0.778(7)$   & $0.777(7)$   & $0.602(5)$ \\
3.75 & $-1.347(23)$ & $-0.377(13)$ & $0.052(9)$  & $0.620(6)$   & $0.618(6)$   & $0.486(5)$ \\
4.00 & $-1.130(23)$ & $-0.327(12)$ & $0.037(8)$  & $0.483(5)$   & $0.483(5)$   & $0.387(4)$ \\
\hline
\end{tabular}
\caption{Results for $X_{AB}^{\lat}$ which are defined in
Eq.~(\ref{eq:SD_projection}). These results are given in units of $10^{-2}$.}
\label{tab:result_npr}
\end{table}

%% file: PT_results.tex
The final elements which are required for our computation of the decay amplitude are the Wilson coefficients and the subtraction constants $Y_{AB}(\mu,\mu_0)$ which first appeared in Eq.\,(\ref{eq:bilocal_MS_RI})\,. The determination of the $Y_{AB}$ is necessarily perturbative since it requires a calculation in the $\MS$ scheme. We outline their determination in Sec.\,\ref{sec:PT_YAB} below with further details presented in Appendix\,\ref{sec:appYAB}. The determination of the Wilson Coefficients is discussed in Sec.\,\ref{subsec:wilsoncoeffs}.

An important aim of this paper is to calculate the decay rate for the process $K^+\to\pi^+\nu\bar{\nu}$ 
without using perturbation theory at the scale of $m_c$ and, as already 
discussed extensively, this requires us to evaluate the matrix elements of bilocal operators. The results are presented in 
Sec.\,\ref{sec:results} below. However, in order to compare these results with those which would be 
obtained in the traditional way for the unphysical quark masses used in our simulations, in this section we integrate out the charm quark reducing the bilocal operators to a local one and use perturbation theory to obtain an estimate of the amplitude. We present the result of this calculation in Sec.\,\ref{subsec:purelyperturbative}, while in Secs.\,\ref{subsec:alphas} and \ref{sec:mcrunning} we discuss the running of $\alpha_s(\mu)$ and $m_c(\mu)$ which are two important elements of the perturbative calculations. The perturbative results obtained by integrating out the charm quark suggest that the contributions from the bilocal and local operators are comparable.

\subsection{Evolution of the strong coupling constant}\label{subsec:alphas}

The evolution of the strong coupling constant $\alpha_s$ from the scale of $M_Z$ to lower scales such as $\mu_c\approx m_c$ has been studied in detail in~Ref.\,\cite{Buras:2006gb}. The
resulting uncertainty in $\alpha_s(\mu_c)$ makes only a negligible contribution to the total uncertainty in $\mathrm{Br}[K^+\to\pi^+\nu\bar{\nu}]$. In our
calculation, we evolve $\alpha_s$ from
$\alpha_s(M_Z)$ to $\alpha_s(\mu_c)$ by solving the renormalization group (RG) equation for $\alpha_s$
numerically.

As the QCD perturbation
theory calculation of the charm quark contribution has been performed at
NNLO~\cite{Buras:2005gr,Buras:2006gb}, we keep to this order and use the 3-loop RG formula for the evolution of the running
coupling constant
\ba
\label{eq:RG_equation}
\mu^2\frac{\partial}{\partial\mu^2}a_s=-\beta_0a_s^2-\beta_1a_s^3-\beta_2a_s^4,
\ea
where $a_s=\alpha_s/(4\pi)$ and the coefficients $\beta_i$ can be found, for example, in
Ref.~\cite{Larin:1993tp} (see  \cite{Olive:2016xmw} for a complete discussion of the running of $\alpha_s$). Solving the RG equation~(\ref{eq:RG_equation}) directly, we have
\ba
\label{eq:RG_equ_solution}
&&g(a_s(\mu_2))-g(a_s(\mu_1))=\log\frac{\mu_2^2}{\mu_1^2},\quad\textrm{where}
\nn\\
&&g(a_s)\equiv\frac{1}{\beta_0a_s}
-\frac{(\frac{\beta_1^2}{\beta_0^2}-2\frac{\beta_2}{\beta_0})\arctan(\frac{\beta_1+2\beta_2a_s}{\sqrt{4\beta_0\beta_2-\beta_1^2}})}
{\sqrt{4\beta_0\beta_2-\beta_1^2}}
-\frac{\beta_1}{2\beta_0^2}\log(\beta_0a_s^{-2}+\beta_1a_s^{-1}+\beta_2).
\ea
Using Eq.\,(\ref{eq:RG_equ_solution}) we can evolve $\alpha_s$ from high to low energy scales following the path $\mu=M_Z\to\mu_b\to\mu_c$.

When a flavor threshold $\mu=\mu_f$ is crossed, the matching
conditions relating $\alpha_s$ with $f$ and $f-1$ active quark flavors are
non-trivial\,\cite{Buras:2006gb}. Using the NNLO matching conditions given in
Ref.~\cite{Buras:2006gb} and choosing the $5\to4$ flavor threshold to be at $\mu_b=5$\,GeV, we 
obtain
\ba
\hspace{-0.5cm}\alpha_s(\muMS)=0.462(11),~0.304(4),~0.255(3),~0.230(2),
\ea
for
$\muMS=$1,\,2,\,3,\,4\,GeV respectively. These results were obtained using the PDG input parameters~\cite{Olive:2016xmw}:
\ba
\alpha_s(M_Z)=0.1185(6),\quad M_Z=91.1876(21)\mbox{ GeV},\quad m_b(m_b)=4.18(3)\mbox{ GeV}.
\ea
In Ref.~\cite{Buras:2006gb}, the threshold scale $\mu_b$ was varied from 2.5\,GeV to 10\,GeV. It was found that this variation affects the charm quark contribution at a level of only $\pm 0.2\%$ compared to the result obtained at $\mu_b=5$\,GeV.

\subsection{Running of the charm quark mass}\label{sec:mcrunning}
Due to the quadratic GIM mechanism, the charm quark contribution to the 
$K^+\to\pi^+\nu\bar{\nu}$ decay amplitude is proportional to the square of the mass of the charm quark.
Thus the running of the charm quark mass plays an important role in the
cancellation of the $\muMS$ scale dependence in the combination of the local and
bilocal contributions.

At the scale $\mu_c\approx m_c$, 
the NNLO expression for the charm quark mass $m_c(\mu_c)$ is given by 
\ba
\label{eq:running_mass}
m_c^2(\mu_c)=\kappa_c\left(1+\frac{\alpha_s(\mu_c)}{4\pi}\xi_c^{(1)}+\left(\frac{\alpha_s(\mu_c)}{4\pi}\right)^2\xi_c^{(2)}\right)m_c^2(m_c)\,,
\ea
where $\kappa_c=(\alpha_s(\mu_c)/\alpha_s(m_c))^{\frac{24}{25}}$ and
$\xi_c^{(1,2)}$ are known coefficients (see Eq.(88) in
Ref.~\cite{Buras:2006gb}).
Here and below we use $m_c(\mu)$ to represent the charm quark
mass computed in the $\MS$ scheme at the scale $\mu$. 

Because of the relatively fast running of $\alpha_s$ at scales of $O(m_c)$, the 
coefficient $\kappa_c$ makes a significant 
impact on the evaluation of local and bilocal Green's functions. 
For example the value of $\kappa_c$ at $\mu_c=3$ GeV is about $40$\% smaller than the value at $\mu_c=1$ GeV. (Even if $\mu_c$ is varied in the range of $2$\,-\,$4$ GeV, $\kappa_c$
still changes by 24\%.) Therefore we include the running of the charm
quark mass and the coefficient $\kappa_c$ in our calculation.
Recall that this calculation is performed with an unphysically light charm-quark mass.
Using the input parameter $m_c(\mbox{2 GeV})=863$ MeV, we obtain $m_c(m_c)=1.080$\,GeV to be compared to the physical value $m_c(m_c)=1.28\pm0.025$\,GeV\,
\cite{Olive:2016xmw}. The charm-quark contribution in our simulation will therefore be suppressed due to the use of an unphysical charm-quark mass.

\subsection{Determination of the Wilson coefficients}\label{subsec:wilsoncoeffs}

In the determination of the Wilson coefficients in the $\MS$ scheme
we follow the procedure given in Ref.\,\cite{Buras:2006gb}.
For the $Z$-exchange diagrams 
$C_1^{\MS}(\mu)$ and $C_2^{\MS}(\mu)$ together with the coefficient
$C_{0,Z}^{\MS}(\mu)$, which is associated with the local operator $Q_0$,
is written as a vector $\vec{C}_Z=(C_+,C_-,C_{0,Z})$. Here $C_\pm = C_2\pm C_1$.
The evolution for $\vec{C}_Z$
can be determined using the equation
\ba
\label{eq:RG_Z}
\vec{C}_Z(\mu)=U_4(\mu,\mu_b)M(\mu_b)U_5(\mu_b,\mu_W)\vec{C}_Z(\mu_W)
\ea
where $\vec{C}_Z(\mu_W)$ indicate the Wilson coefficients at the scale of
$\mu_W=O(M_W)$. (In practice, we take $\mu_W=80.0$ GeV.) 
The values of the coefficients
$\vec{C}_Z(\mu_W)$ are determined by matching the Green's functions in the full and
the effective theory at $\mu_W$ using NNLO QCD perturbation theory. The
evolution matrices $U_5(\mu_b,\mu_W)$, $U_4(\mu,\mu_b)$ and the
$b$-quark threshold matching matrix $M(\mu_b)$ are also
known~\cite{Buras:2006gb}. Thus the values for $\vec{C}_Z(\mu)$ at
$\mu=\mu_c=O(m_c)$ can be determined. At $\mu=2.15$ GeV, we have
$C_1^{\MS}(\mu)=-0.2911$ and $C_2^{\MS}(\mu)=1.1353$. These values have been used in
Table~\ref{tab:npr_local} and Eq.~(\ref{eq:lat_to_RI_matching}) to determine the
Wilson coefficients $C_1^{\lat}(a)$ and $C_2^{\lat}(a)$ for the bare lattice
operators.

For the $W$-$W$ diagram, the vector of Wilson coefficients is constructed
as $\vec{C}_{WW}=(1,C_{0,WW})$. The Wilson coefficient for each
two-quark-two-lepton operator does not run because the anomalous dimension is zero.
Thus it is simply given by $1$. The coefficient $C_{0,WW}$ accounts for
the SD contribution when the two local weak operators approach each other and
is non-trivial. It can be determined using a renormalization group
evolution equation, which takes a similar form to Eq.~(\ref{eq:RG_Z}).

\subsection{Perturbative estimate of the decay amplitude}\label{subsec:purelyperturbative}

In this subsection we digress from the main calculation and estimate the amplitude using the standard procedure of 
integrating out the charm quark and using perturbation theory. This will allow us to determine the difference between 
our non-perturbative computation of long-distance effects and the standard calculation.

Having determined $\vec{C}_Z(\mu)$ and $\vec{C}_{WW}(\mu)$, the next step is
to evaluate the amputated Green's function for the bilocal operators to determine the coefficient $r_{AB}^{\MS}(\mu)$ defined by
\ba
\langle C_A^{\MS}Q_A^{\MS}\, C_B^{\MS}Q_B^{\MS}
\rangle^{\MS}=C_A^{\MS}\,C_B^{\MS}\,r_{AB}^{\MS}\,\langle
Q_0^{\MS}\rangle\,.
\ea
By integrating out the charm quark field, the parameter $r_{AB}^{\MS}(\mu)$ can
be used to describe the bilocal contribution in perturbation theory. At 
$O(\alpha_s^0)$ one has the following contributions to $r_{AB}^{\MS}(\mu)$
\ba
\label{eq:matrix_element_PT}
\begin{array}{ll}
\frac{m_c(\mu)^2}{4\pi^2}\left(1-\ln\frac{\mu^2}{m_c^2(\mu)}\right)\times N_c +O(\alpha_s), &\quad \mbox{from $Z$-exchange diagrams with $Q_1$}\\
\frac{m_c(\mu)^2}{4\pi^2}\left(1-\ln\frac{\mu^2}{m_c^2(\mu)}\right) + O(\alpha_s),&\quad \mbox{from $Z$-exchange diagrams with $Q_2$}\\
\frac{m_c(\mu)^2}{\pi^2}\left(\frac{5}{4}+\ln\frac{\mu^2}{m_c^2(\mu)}+\frac{x_\ell\ln
x_\ell}{1-x_\ell}\right) +O(\alpha_s),&\quad \mbox{from $W$-$W$ diagrams}
\end{array}
\ea
where we have exhibited the $\mu$ dependence of the charm quark mass $m_c(\mu)$. 
$N_c=3$ is
the number of QCD colors.
In the $W$-$W$
diagram, the parameter $x_\ell=m_\ell^2/m_c^2(\mu)$ indicates the non-zero lepton mass correction to the
loop diagram. For the electron and muon this correction can be
neglected given the current precision of the computations. Although the $O(\alpha_s)$
corrections to $r_{AB}^{\MS}(\mu)$ are not shown explicitly in
Eq.~(\ref{eq:matrix_element_PT}), 
they have been calculated and detailed formulas can be found in
Ref.~\cite{Buras:2006gb}. These $O(\alpha_s)$ corrections have been included in our calculation.

\begin{figure}
   \centering
   \includegraphics[width=.8\textwidth]{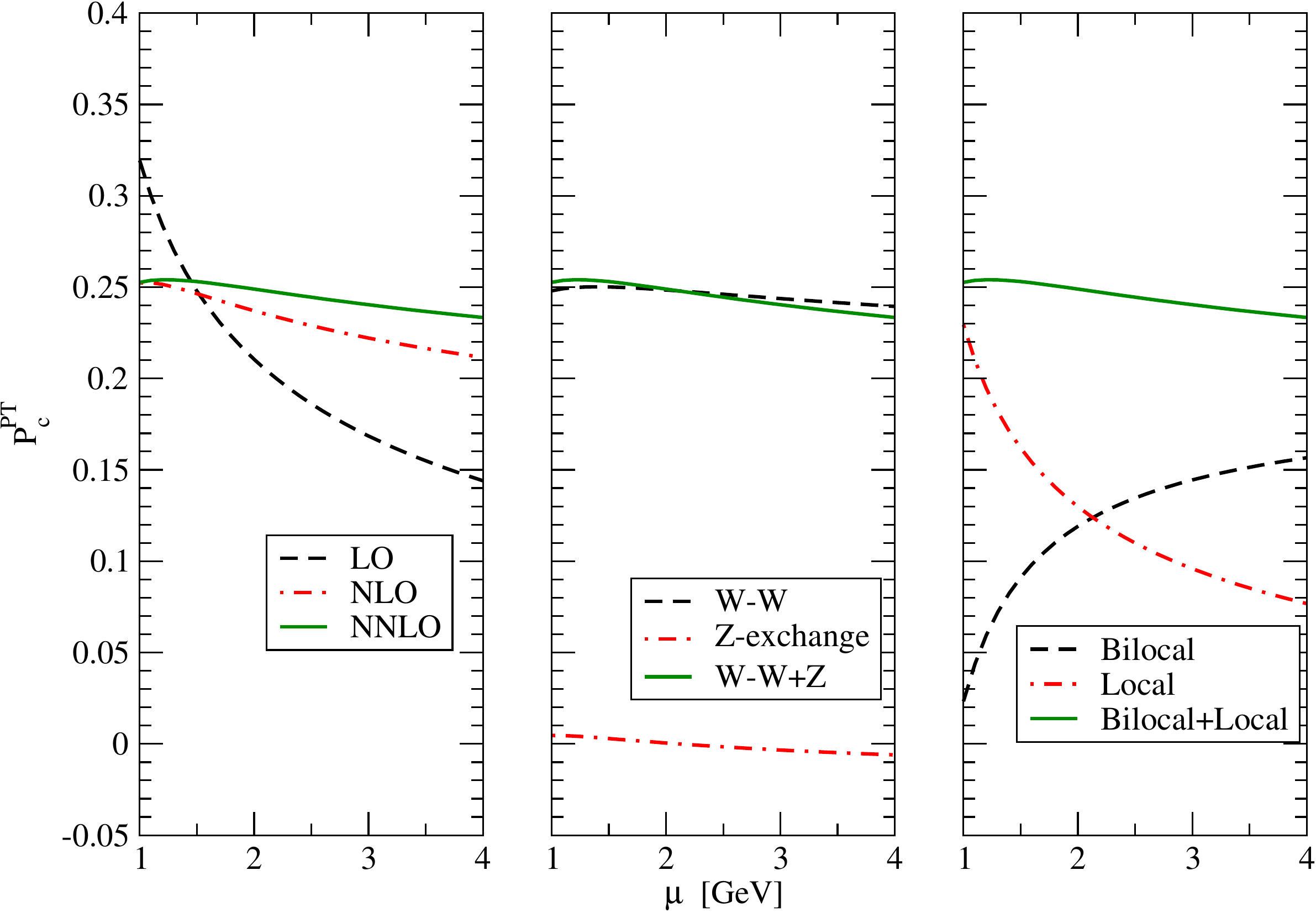}
   \caption{Evaluation of the charm quark contribution $P_c^{\mathrm{PT}}$ at the unphysical
   charm quark mass $m_c(\mbox{2 GeV})=863$ MeV following the procedure given
   in Ref.~\cite{Buras:2006gb}. In the left-hand panel, we show the scale ($\mu$) dependence 
   of total contribution $P_c^{\mathrm{PT}}(\mu)$ at the LO (black dashed curve), NLO (red dash-dotted curve) and 
   NNLO (green solid curve). In the middle panel, we show the NNLO result for 
   the $P_c^{\mathrm{PT}}(\mu)$ by splitting it into the $W$-$W$ ($P_c^{WW}$) and Z-exchange ($P_c^{Z}$) contributions.  
   In the right-hand panel, we compare 
   the bilocal and local contributions as a function of the scale $\mu$;
   both include the NNLO corrections.}
   \label{fig:Pc}
   \end{figure}

Note that in renormalization group improved perturbation theory, the Wilson
coefficients $C_{0,Z}$ and $C_{0,WW}$ contain large logarithms of the form
$\log\frac{\mu^2}{M_W^2}$. These contribute as a LO effect of order 
$O(\alpha_s^{-1})$. The
bilocal contribution $r_{AB}^{\MS}(\mu)$ given in
Eq.~(\ref{eq:matrix_element_PT}) contributes as a
NLO contribution of order $O(\alpha_s^0)$. Both sets of Wilson coefficients
$\vec{C}_Z$ and $\vec{C}_{WW}$ as well as the
parameter $r_{AB}^{\MS}$, have been calculated to NNLO including the
$O(\alpha_s^1)$ corrections. The total charm quark contribution can be written in the form
\begin{equation}\label{eq:Pc_PT}
P_c^\textrm{PT}(\mu)=\frac{1}{\lambda^4}\frac{\pi^2}{M_W^2}\left(C_A^{\MS}(\mu)C_B^{\MS}(\mu)r_{AB}^{\MS}(\mu)+C_{0}^{\MS}(\mu)\right)
\end{equation}
and receives contributions from both the $WW$ and $Z$-exchange diagrams. We write $P_c^{\mathrm{PT}}(\mu)=P_c^{Z}(\mu)+P_c^{WW}(\mu)$
where the superscripts $WW$ and $Z$ denote the contributions from the $WW$ and $Z$-exchange diagrams respectively. We recall that for the $WW$ diagrams one has to average the contributions from the three intermediate leptons.
In Eq.\,(\ref{eq:Pc_PT}) $\lambda\equiv|V_{us}|/\sqrt{|V_{ud}|^2+|V_{us}|^2}$ and $M_W$ is the 
mass of the $W$-boson. We use the values $\lambda=0.22537(61)$ and 
$M_W=80.385(15)$\,GeV taken from the PDG\,~\cite{Olive:2016xmw}. 
At the unphysical charm quark mass $m_c(\mbox{2 GeV})=863$ MeV, $P_c^{Z}(\mu)$
and $P_c^{WW}(\mu)$ at $\MS$ scales $\mu=1$\,-\,4\,GeV are shown in
Fig.~\ref{fig:Pc}.
\bi
\item In the left-hand panel we show the scale ($\mu$) dependence of the total contribution
    $P_c^{\mathrm{PT}}(\mu)$ at LO (indicated by the black dashed curve), 
    NLO (red dash-dotted curve) and NNLO (green solid curve). 
    We see that by including higher-order QCD corrections 
    the scale dependence becomes milder.
\item In the middle panel, we split the total
NNLO result $P_c^{\mathrm{PT}}(\mu)$ into the $W$-$W$ contribution $P_c^{WW}$ 
(indicated by the black dashed curve) and the $Z$-exchange contribution $P_c^{Z}$ 
(red dash-dotted curve). The $W$-$W$ diagrams
dominate $P_c^{\mathrm{PT}}$ with the $Z$-exchange diagrams only making a small contribution.
\item In the right-hand panel, we compare the total bilocal
contribution to $P_c^{\mathrm{PT}}$ (indicated by the black dashed curve) and the
local contribution (red dash-dotted curve) at various scales $\mu$. Both contributions include NNLO corrections. 
At a scale 
$\mu\approx 2$ GeV, the bilocal contribution is of similar size to the local one. 
\ei

We could also compile a figure similar to that shown in Fig.~\ref{fig:Pc} corresponding to the physical charm quark mass, 
$m_{c,\mathrm{phys}}$.
The main difference is that $P_c^{\textrm{PT}}$ would be enhanced by a factor of
$\left(\frac{m_{c,\mathrm{phys}}}{m_{c,\mathrm{unphys}}}\right)^2$, where $m_{c,\mathrm{unphys}}$ is the unphysical mass used in this simulation.

In Fig.~\ref{fig:Pc}, the bilocal contribution is estimated using the perturbation
theory by integrating out the charm quark field. We question 
whether perturbation theory works well at the scale of $\mu=O(m_c)$. We therefore
replace the $r_{AB}^{\MS}$ term by the nonperturbative evaluation of the bilocal
matrix element together with a perturbative matching from RI/SMOM scheme to $\MS$ scheme.
The results are presented in Sec.\,\ref{sec:results}.

\subsection{Determination of the \boldmath$Y_{AB}(\mu,\mu_0)$}
\label{sec:PT_YAB}

The relation between the $\MS$ and RI/SMOM bilocal operators takes the form given in Eq.\,(\ref{eq:bilocal_MS_RI}) which we rewrite here for the reader's convenience:
\ba
\label{eq:O1O2MS}
\{Q_A^{\MS}Q_B^{\MS}\}^{\MS}_\mu=Z_{A}^{\RI\to\MS}(\mu/\mu_0)Z_{B}^{\RI\to\MS}(\mu/\mu_0)\{Q_AQ_B\}^{\RI}_{\mu_0}+Y_{AB}(\mu,\muRI)Q_0^{\RI}(\mu_0),
\ea
where $\mu$ and $\mu_0$ are the $\MS$ and RI/SMOM renormalization scales respectively. For compactness of notation, we have written 
Eq.\,(\ref{eq:O1O2MS}) as if there is no mixing of the operators $Q_A$ and $Q_B$ with other operators. When, as in the case of the $Z$-exchange 
diagrams, there is a mixing then the renormalization constants become matrices, e.g. $Z_A Q_A\to Z_{AC} Q_C$.

In order to determine $Y_{AB}(\mu,\muRI)$ we calculate the amputated Green's functions
for both sides of Eq.~(\ref{eq:O1O2MS}) at $p_i^2=\muRI^2$ and impose the
renormalization condition Eq.~(\ref{eq:renorm_cond}) so that:
   \begin{eqnarray}\label{eq:Ydetermination1}
  \bigl\langle \{Q_A^{\MS} Q_B^{\MS}\}^{\MS}_\mu\bigr\rangle_{p_i^2=\muRI^2}
  =\frac{Z_q^{\RI}(\muRI)}{Z_q^{\MS}(\mu)}\,Y_{AB}(\mu,\muRI)\, \bigl\langle
  Q_0^{\RI}\bigr\rangle_{p_i^2=\muRI^2}.
   \end{eqnarray}
  Here $Z_q^{\RI}$ and $Z_q^{\MS}$ are the quark's wave function renormalization
  constant. In the Landau gauge and setting the renormalization
  scales of both $\MS$ and RI/SMOM schemes to be equal $\mu=\mu_0$, we have
  $Z_q^{\RI}/Z_q^{\MS}=1+O(\alpha_s^2)$~\cite{Sturm:2009kb}. On the right-hand side of Eq.\,(\ref{eq:Ydetermination1}),
  the definition of the RI/SMOM renormalization scheme implies that $\bigl\langle
  Q_0^{\RI}\bigr\rangle_{p_i^2=\muRI^2}=\bigl\langle
  Q_0\bigr\rangle_{p_i^2=\muRI^2}^{(0)}$, where the superscript $(0)$ denotes the {\em
  tree-level} amputated Green's function.

  We write $Y_{A,B}(\mu,\mu_0)\equiv Y_{A,B}(\mu,0)+\Delta
  Y_{AB}(\mu,\mu_0)$, where $Y_{A,B}(\mu,0)$ is exactly given by the
  $r_{AB}^{\MS}(\mu)$ discussed in Sec.\,\ref{subsec:purelyperturbative}.  We present the determination 
  of the $\Delta Y_{AB}$ at $O(\alpha_s^0)$ in Appendix\,\ref{appendix:YAB} and the results are
  given in Eqs.~(\ref{eq:Delta_YAB_WW}), (\ref{eq:Delta_YAB_ZA}) and
  (\ref{eq:Delta_YAB_ZV}). 
  
  In our analysis, we take the expression for $r_{AB}^{\MS}(\mu)$ from
  Ref.~\cite{Buras:2006gb}, where it has been calculated at
  $O(\alpha_s^1)$. We estimate $\Delta Y_{AB}(\mu,\mu_0)$ at $O(\alpha_s^0)$. 
  As we will show later, $\Delta Y_{AB}(\mu,\mu_0)$
  is of comparable size to $r_{AB}^{\MS}(\mu)$. Thus the inclusion of the scale
  dependence of the
  charm quark mass is important for the determination of both $\Delta Y_{AB}(\mu,\mu_0)$ and $r_{AB}^{\MS}(\mu)$. 
  In $r_{AB}^{\MS}(\mu)$, the running charm quark mass only depends on the
  $\MS$ scale $\mu$, while
  in $\Delta Y_{AB}(\mu,\mu_0)$, the charm quark mass also depends on the
  RI/SMOM scale $\mu_0$.
  For simplicity, we choose $\mu=\mu_0$. Note that the mass renormalization
  conversion factors from the RI/SMOM scheme to the $\MS$ scheme have been calculated to
  two-loop order. At $\mu=\mu_0\ge\mbox{2
  GeV}$ these conversion factors only deviate from 1 by a few
  percent~\cite{Gorbahn:2010bf}. 
  We thus neglect the RI/SMOM scale dependence and simply use the $\MS$ charm quark
  mass from Eq.~(\ref{eq:running_mass}) for $\Delta Y_{AB}(\mu,\mu)$.

\begin{figure}
   \centering
   \includegraphics[width=.8\textwidth]{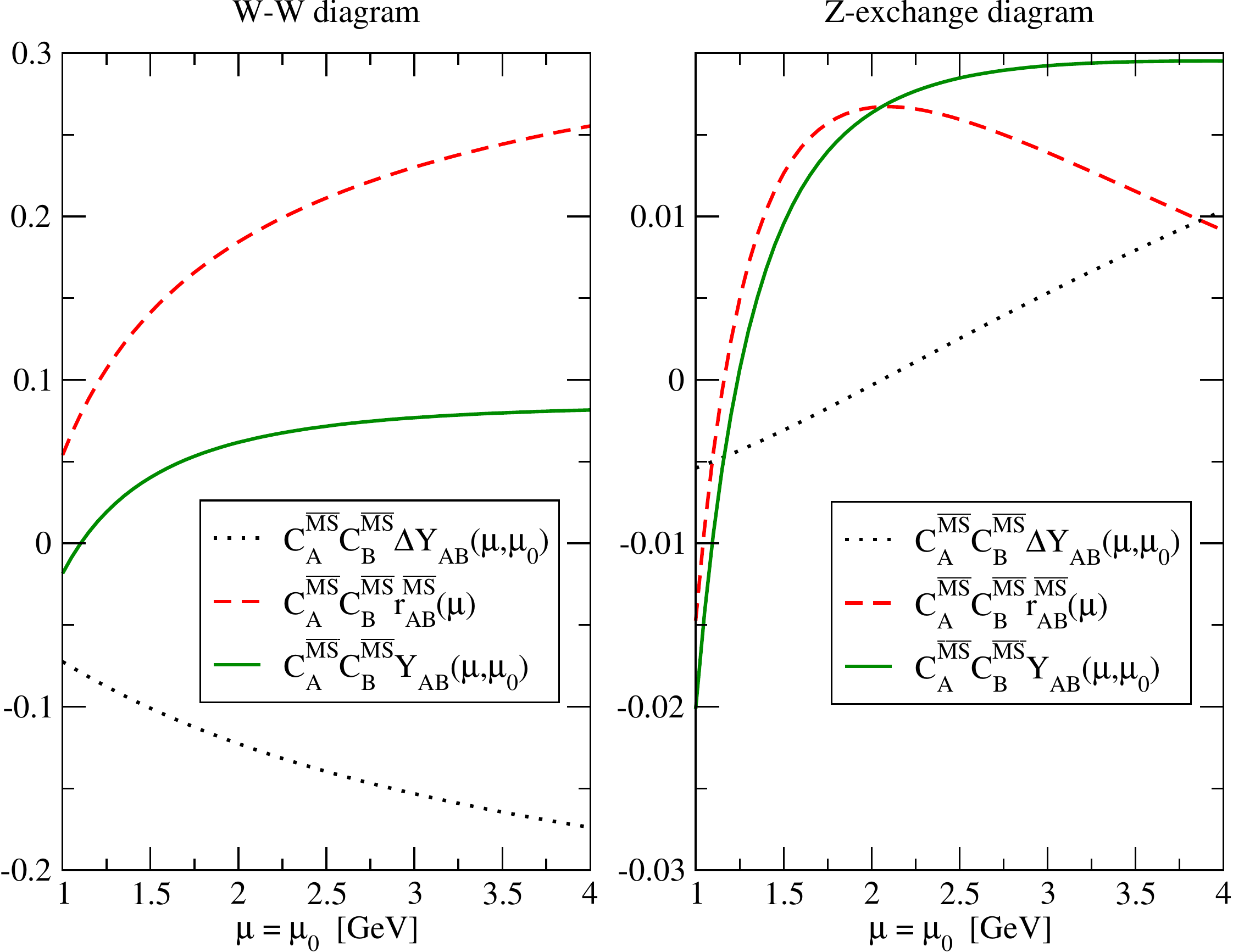}
   \caption{Contributions to $\Delta Y_{AB}(\mu,\mu_0)$, $r_{AB}^{\MS}(\mu)$ and $Y_{AB}(\mu,\mu_0)$, multiplied by the 
   corresponding Wilson Coefficients, from the  $W$-$W$
  diagrams (left panel) and $Z$-exchange diagrams (right panel). They are shown as a 
  function of $\mu=\mu_0$.}
   \label{fig:YAB}
   \end{figure}

  In Fig.~\ref{fig:YAB}, we show the contributions to $C_A^{\MS}C_B^{\MS}\Delta Y_{AB}(\mu,\mu_0)$,
  $C_A^{\MS}C_B^{\MS}r_{AB}^{\MS}(\mu)$ and $C_A^{\MS}C_B^{\MS}Y_{AB}(\mu,\mu_0)$
  as a function of $\mu=\mu_0$ from the $W$-$W$
  diagrams (left panel) and the $Z$-exchange diagrams (right panel).
  Since the magnitude of $\Delta Y_{AB}(\mu,\mu)$ is comparable to
  $r_{AB}^{\MS}(\mu)$, it will be important in future calculations to include the
  $O(\alpha_s)$ correction and the RI/SMOM scale dependence of the charm quark
  mass running in $\Delta Y_{AB}(\mu,\mu_0)$. Another observation from
  Fig.~\ref{fig:YAB} is that the $\ln\frac{\mu^2}{m_c^2}$ dependence, present in each of the terms
  $\Delta Y_{AB}(\mu,\mu)$ and $r_{AB}^{\MS}(\mu)$, cancels at $O(\alpha_s^0)$ in the combination 
  $r_{AB}^{\MS}(\mu)+\Delta Y_{AB}(\mu,\mu)$ (see also Eq.~(\ref{eq:YAB}))\,.
   
%% file: Summary_results.tex
  In the previous sections we have discussed and computed all the ingredients necessary to determine the 
  decay amplitude for the process $K^+\to\pi^+\nu\bar\nu$. Before presenting our final result for the amplitude,
  we briefly summarise how these ingredients are combined to obtain this result. 
  We started in Sec.~\ref{sec:matrix_element} with a calculation of the matrix elements of the local and bilocal
  lattice operators relevant for the rare kaon decays. These computations are naturally non-perturbative. In the
  determination of the matrix elements of bilocal operators new ultraviolet divergences appear when the two local 
  operators $Q_A$ and $Q_B$ which comprise the bilocal operator approach each other. 
  We discuss the subtraction of these additional divergences in Sec.~\ref{sec:SD_NPR},
  introducing and determining the subtraction constants $X_{AB}^{\lat}(\mu_0,a)$ 
  (see Eqs.\,(\ref{eq:renorm_cond}) and (\ref{eq:bilocalRI})). By subtracting these divergences, we define and 
  determine non-perturbatively the matrix element of the bilocal operators renormalised in the RI/SMOM scheme.
  Since Wilson Coefficients are generally calculated in the $\MS$ scheme, we need to convert the RI/SMOM operators 
  into those in the $\MS$ scheme and this is necessarily a perturbative calculation, which we describe 
  in~Sec.~\ref{sec:SD_PT}. The $\RI\to\MS$ conversion of the bilocal operators is characterized by the constants 
  $Y_{AB}(\mu,\mu_0)$ (see Eq.\,(\ref{eq:O1O2MS})). In this way we obtain the matrix elements of the 
  bilocal operators in the $\MS$ scheme without  ``integrating out'' the charm quark. This matrix elements can 
  be written generically in terms of the individual ingredients as follows:
  \ba
  {\mathcal A}_{\mathrm{Bilocal}}^{\MS}&\equiv&\langle\pi^+\nu\bar{\nu}|\{C_A^{\MS}Q_A^{\MS}C_B^{\MS}Q_B^{\MS}\}^{\MS}_\mu|K^+\rangle
  \nn\\
  &=&C_A^{\lat} C_B^{\lat}\langle\pi^+\nu\bar{\nu}|\{Q_A^{\lat}Q_B^{\lat}\}^{\lat}_a|K^+\rangle\nn\\ 
 &&\hspace{0.15in}-C_A^{\lat}C_B^{\lat}X_{AB}^{\lat}(\mu_0,a)\langle\pi^+\nu\bar{\nu}|Q_0^{\lat}|K^+\rangle
 +C_A^{\MS}C_B^{\MS}Y_{AB}(\mu,\mu_0)\langle\pi^+\nu\bar{\nu}|Q_0^{\RI}|K^+\rangle
  \nn\\
    &=&i\,\left[F_{\mathrm{4pt}}(\Delta,s)-2Z_V^{-1}C_A^{\lat}C_B^{\lat}X_{AB}^{\lat}(\mu_0,a)
  f_+(s)+2C_A^{\MS}C_B^{\MS}Y_{AB}(\mu,\mu_0)f_+(s)\right]
  \nn\\
  &&\hspace{0.15in}\times\left[\bar{u}(p_\nu){\slashed
  p}_K(1-\gamma_5)]v(p_{\bar{\nu}})\right]\,.\label{eq:abilocalmsbar}
  \ea
  Depending on the choice of the operators $Q_{\{A,B\}}$, Eq.\,(\ref{eq:abilocalmsbar}) represents contributions to the  
  $W$-$W$ or $Z$-exchange diagrams. 
  The scalar amplitude $F_{\mathrm{4pt}}(\Delta,s)$ is given by $F_{WW}(\Delta,s)$ for $W$-$W$
  diagram and $2F_+^{Z,i}(s)$ ($i=V,A$) for the $Z$-exchange diagram. The variables
  $\Delta$ and $s$ are defined in Eq.\,(\ref{eq:Lorentz_invariant}). The 
  $K_{\ell3}$ form factor $f_+(s)$ is defined in Eq.\,(\ref{eq:local_form_factor}). The results for $f_+(s)$, 
  $F_{WW}(\Delta,s)$ and $F_+^{Z,i}(s)$
  have been given in Tables~\ref{tab:lattice_local_matrix}, \ref{table:WW} and
  \ref{tab:lattice_results_Z_exchange} respectively. The results for
  $X_{AB}^{\lat}(\mu_0,a)$ in the range $\mbox{1\,
  GeV}\le\mu_0\le{4}$\,GeV are listed in
  Table\,\ref{tab:result_npr} and $Z_V=Z_A=0.7163(14)$. For the $Z$-exchange diagrams in 
  Table\,\ref{tab:result_npr} we also give the 
  results with the corresponding Wilson Coefficients (labelled $C_1^\lat Q_1+ C_2^\lat Q_2$).
  The results for $Y_{AB}(\mu,\mu_0)$ for $\mbox{1
  GeV}\le\mu=\mu_0\le\mbox{4 GeV}$ are shown in Fig.~\ref{fig:YAB}.

  It is convenient to define the ratio $R(\Delta,s)$:
  \ba
  \label{eq:result_ratio}
  R(\Delta,s)\equiv \frac{F_{\mathrm{4pt}}(\Delta,s)}{2f_+(s)}.
  \ea
  Since in this calculation we use use a single choice of momenta (see
  Eq.\,(\ref{eq:mom_setup})), we are not able to determine the $\Delta$ and $s$ dependence of
  $R(\Delta,s)$. Here we simply neglect this momentum dependence. 
  
  The bilocal matrix element can be written as
  \ba
  {\mathcal A}_{\mathrm{Bilocal}}^{\MS}&\simeq&
  i\,\left[R(\Delta,s)-Z_V^{-1}C_A^{\lat}C_B^{\lat}X_{AB}^{\lat}(\mu_0,a)+C_A^{\MS}C_B^{\MS}Y_{AB}\right]
  \nn\\
  &&\hspace{0.15in}\times \left[2f_+(s)\bar{u}(p_\nu){\slashed p}_K(1-\gamma_5)v(p_{\bar{\nu}})\right]\,,
  \ea
  where the $\simeq$ symbol on the first line reminds us that the momentum dependence of 
  $R(\Delta,s)$ has been neglected.
  We denote  the sum of the ${\mathcal A}_{\mathrm{Bilocal}}^{\MS}$ from the $WW$ and $Z$-exchange 
  diagrams by ${\mathcal A}_{\mathrm{Bilocal}}^{\MS,\textrm{TOT}}$ and combine it with the contribution from 
  the matrix element of the local operator
  \ba
  {\mathcal
  A}_{\mathrm{Local}}^{\MS}=i\,\left[C_{0,Z}^{\MS}(\mu)+\frac{1}{3}\sum_{\ell,\mu,\tau}C_{0,WW}^{\MS}(\mu)\right]\,
  \left[2f_+(s)\bar{u}(p_\nu){\slashed p}_K(1-\gamma_5)v(p_{\bar{\nu}})\right]
  \ea
  to obtain the total charm quark contributions to the decay amplitude.
  It is conventional to relate the ${\mathcal A}_{\mathrm{Bilocal}}^{\MS,\textrm{TOT}}$ and
  ${\mathcal A}_{\mathrm{Local}}^{\MS}$ to $P_c$ through
  \ba
  \label{eq:Pc_average}
  {\mathcal A}_{\mathrm{Bilocal}}^{\MS,\textrm{TOT}}+ {\mathcal A}_{\mathrm{Local}}^{\MS}
  =\frac{\lambda^4}{\pi^2}M_W^2P_c\left[2f_+(s)\bar{u}(p_\nu){\slashed p}_K(1-\gamma_5)v(p_{\bar{\nu}})\right]\,.
  \ea
  We now separate $P_c$ into two parts: the standard charm-quark estimate
  $P_c^{\mathrm{PT}}$ calculated using perturbation theory (see
  Eq.~(\ref{eq:Pc_PT})) and a
  difference between the full non-perturbative lattice result and the perturbative estimate,
  $P_c-P_c^{\mathrm{PT}}$
  \ba
  P_c-P_c^{\mathrm{PT}}=\frac{1}{\lambda^4}\frac{\pi^2}{M_W^2}\left[R(\Delta,s)-Z_V^{-1}C_A^{\lat}C_B^{\lat}X_{AB}^{\lat}(\mu_0,a)+C_A^{\MS}C_B^{\MS}\Delta
  Y_{AB}(\mu,\mu_0)\right].
  \ea

\begin{figure}
   \centering
   \includegraphics[width=.8\textwidth]{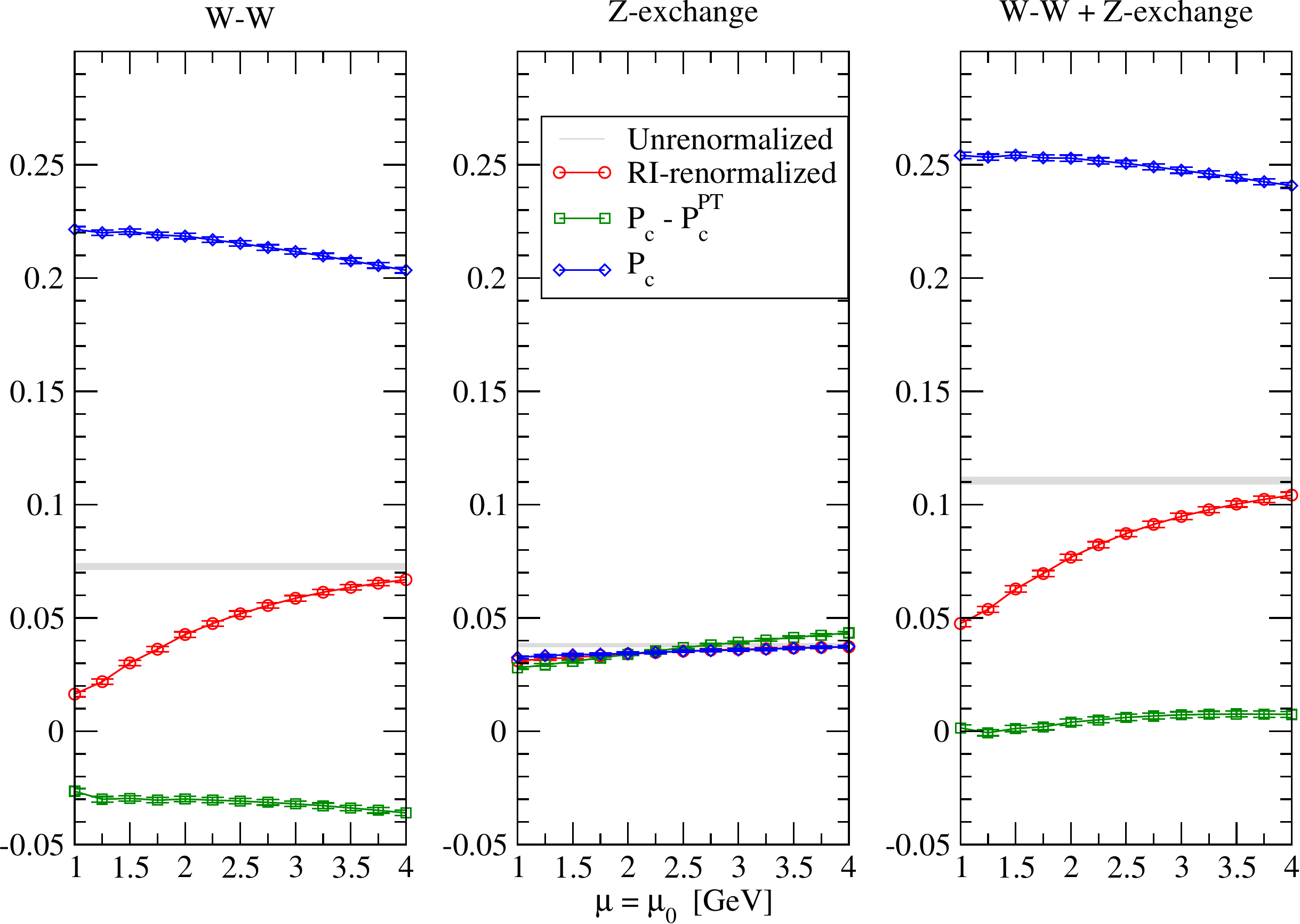}
   \caption{
   The unrenormalized lattice matrix elements
   $\frac{1}{\lambda^4}\frac{\pi^2}{M_W^2} R(\Delta,s)$ (indicated by the gray band), the RI-renormalized 
   matrix elements
   $\frac{1}{\lambda^4}\frac{\pi^2}{M_W^2}[R(\Delta,s)-Z_V^{-1}C_A^{\lat}C_B^{\lat}X_{AB}^{\lat}(\mu_0,a)]$
   (red circles), the total charm-quark contribution $P_c$ (blue diamonds) and the difference
   $P_c-P_c^{\mathrm{PT}}$ (green squares) are shown as a function of
   $\mu=\mu_0$. From left to right, the three panels show the contribution of the $W$-$W$ diagrams,
   the $Z$-exchange diagram and the total, i.e. the sum of the two.}
   \label{fig:Pcu}
   \end{figure}

  In Fig.~\ref{fig:Pcu} we show the unrenormalized quantity
  $\frac{1}{\lambda^4}\frac{\pi^2}{M_W^2} R(\Delta,s)$ (gray band), the RI-renormalized quantity
  $\frac{1}{\lambda^4}\frac{\pi^2}{M_W^2}[R(\Delta,s)-Z_V^{-1}C_A^{\lat}C_B^{\lat}X_{AB}^{\lat}(\mu_0,a)]$
  (red circle),
    the total charm
  contribution $P_c$ (blue diamond) and
  the difference $P_c-P_c^{\mathrm{PT}}$ (green square) as a function of $\mu=\mu_0$.
  From the left to right, three panels show the results for the $W$-$W$ diagrams,
  the $Z$-exchange diagrams and their sum.

  At scales $\mu=\mu_0=1,2,3,4$ GeV, we obtain respectively
  \ba
  \label{eq:Pcu}
  P_c&=&0.2541(13),0.2529(13),0.2476(13),0.2408(13)\,,
  \nn\\
  P_c-P_c^{\mathrm{PT}}&=&0.0015(13),0.0040(13),0.0072(13),0.0074(13)\,.
  \ea
  As shown in Tables~\ref{table:WW} and \ref{tab:lattice_results_Z_exchange} the statistical errors
  in the unrenormalized bilocal matrix are about 1-2\%.
  When these uncertainties propagate to $P_c$, they only appear as
  sub-percent effects, since in $P_c$ the largest contribution
  comes from the perturbation theory.

There is a curious cancellation evident in Fig.\,\ref{fig:Pcu}. The figure shows that the contributions from each of the $WW$ and $Z$-exchange diagrams to $P_c-P_c^{\textrm{PT}}$ clearly deviate from 0 due to non-perturbative effects. However, they have the opposite sign and as a result there is a significant cancellation. For illustration, at $\mu=\mu_0$=2\,GeV, the contribution to $P_c-P_c^{\textrm{PT}}$ from the $WW$ diagram is $-2.99(12)\times 10^{-2}$ and from the $Z$-exchange diagram is $3.39(6)\times10^{-2}$. The sum of the two contributions is about 10 times smaller than each contribution separately. It will be very interesting to check whether such a cancellation persists as the masses of the quarks are changed to their physical values.

  \subsection{Systematic effects}
  Although the statistical errors are well under control, in order to obtain a
  precise calculation of the long-distance contribution to the
  $K^+\to\pi^+\nu\bar{\nu}$ decay amplitude, it is important also to have a good understanding of the
  systematic uncertainties. In this subsection we discuss some of the principle sources of these uncertainties.

  \subsubsection{The {\rm RI/SMOM} and $\MS$ scale dependence}
  As can be seen from Eq.\,(\ref{eq:Pcu}), the systematic uncertainty arising from the
  scale dependence is much larger than the statistical error. There are two main 
  sources of this scale dependence. At small scales
  $\mu=\mu_0\approx 1$ GeV, we expect that higher-order QCD corrections, which
  are not included in our calculation of $\Delta Y_{AB}$, will cause a sizeable effect. At larger
  scales, $\mu=\mu_0\approx 4$\,GeV say, we expect that lattice
  artifacts might be significant.
  We quote the results for $P_c$ and $P_c^{\mathrm{PT}}$ as
  \ba
  P_c=0.2529(13)(32),\quad P_c-P_c^{\mathrm{PT}}=0.0040(13)(32)\,
  \ea
  where the central values correspond to the scale $\mu=\mu_0=2$\,GeV. The first
  error is statistical and the second an estimate of the error implied by the
  residual scale dependence of $P_c$, in the range 1\,GeV$<\mu=\mu_0<3$\,GeV.

  \subsubsection{Contributions from disconnected diagrams}
  The calculation of disconnected diagrams usually suffers from large noise. This
  is also the case for the calculation of the rare kaon decay form factors, where the
  uncertainty of the disconnected diagrams is about 10-30\% while for the connected diagrams,
  the uncertainty is at the level of few percent. This can be seen from
  Table~\ref{tab:lattice_results_Z_exchange}. Fortunately, the size of the form factor
  from the disconnected diagrams,
  $F_0^{Z,A,disc}(s_{\mathrm{max}})=6.0(1.2)\cdot10^{-4}$, is only a few percent of
  that from the connected diagram. It only contributes to $P_{c}$ at the level of
  0.4\%. Here we should point out that since we do not use twisted boundary
  conditions for disconnected diagrams, we only calculate them with the mesons at rest,
  ${\bf p}_K={\bf p}_\pi={\bf 0}$. We thus determine
  $F_0^{Z,A,disc}(s_{\textrm{max}})$ instead of $F_+^{Z,A,disc}(s)$. 
  If we assume that $F_0^{Z,A}(s_{\textrm{max}})$ is a good approximation to
  $F_+^{Z,A}(s)$, then the disconnected diagrams only contribute to $P_c$ at a
  negligible level. Recall that at $s=s_\textrm{max}$ the vector current does not contribute to the amplitude.

  \subsubsection{Finite volume effects}\label{subsubsec:FV}
   
   As explained in Ref.\,\cite{Christ:2016eae}, the main finite volume (FV) effects in
   the lattice calculation of the $K^+\to\pi^+\nu\bar{\nu}$ decay amplitude
   arise from the $K^+\to\pi^+\pi^0\to\pi^+\nu\bar{\nu}$ process for the $Z$-exchange
   diagrams and $K^+\to\pi^0\ell^+\nu\to\pi^+\nu\bar{\nu}$ for the $W$-$W$
   diagrams. The transitions $K^+\to3\pi\to\pi^+\nu\bar{\nu}$ and
   $K^+\to2\pi\ell^+\nu\to\pi^+\nu\bar{\nu}$
   can be neglected due to significant phase space suppression. We therefore
   exclude them from our discussion.
   
   For the transition $K^+\to\pi^+\pi^0$, since the pion mass used in this
   calculation is $420$ MeV (so that $m_K<2m_\pi$), 
   no significant finite-volume effects are expected.
   Nevertheless, we have calculated two-pion scattering energy in the isospin $I=2$
   channel as well as the $K^+\to\pi^+\pi^0$ and
   $\pi^+\pi^0\to\pi^+$ transition amplitude. There are no expected difficulties
   to evaluating the potentially large finite-volume effects by using Lellouch-L\"uscher
   formula when we repeat the calculation at physical quark masses (and therefore with $m_K>2m_\pi$) in the future.

   Here we focus on the transition
   $K^+\to\pi^0\ell^+\nu\to\pi^+\nu\bar{\nu}$ and denote the potentially large, i.e.
   non-exponential, FV correction by $A_{\textrm{FV}}^{\pi^0\ell^+}=A_{WW}(L)-A_{WW}(\infty)$, where $A_{WW}(L)$
   and $A_{WW}(\infty)$ are the contributions to the amplitude from the $WW$ diagrams in finite and infinite 
   volumes respectively. The label 
   {\footnotesize $\pi^0\ell^+$} indicates that the correction comes from the $\pi^0\ell^+$ intermediate state; 
   see Fig.\,\ref{fig:contraction}. The neutrino plays no r\^ole here beyond determining 
   the energy-momentum of the $\pi^0\ell^+$ pair.
   $A_{\textrm{FV}}^{\pi^0\ell^+}$ can be expressed as~\cite{Christ:2016eae}
   \begin{eqnarray}
   \label{eq:WW_FV}
   A_{\textrm{FV}}^{\pi^0\ell^+}&=&\left(\frac{1}{L^3}\sum_{\vec{k}}\int \frac{dk_0}{2\pi}
   -{\mathcal P}\int\frac{d^4k}{(2\pi)^4}\right)
   \left\{A_\alpha^{K^+\to\pi^0}(p_K,k)\frac{1}{k^2+m_\pi^2}
    A_\beta^{\pi^0\to\pi^+}(k,p_\pi)\right\}
   \nn\\
   &&\hspace{1.5cm}\times
   \left\{\bar{u}(p_\nu)\gamma^\alpha(1-\gamma_5)
   \frac{i({\slashed P}-{\slashed k})+m_{\bar{\ell}}}{(P-k)^2+m_{\bar{\ell}}^2}
   \gamma^\beta(1-\gamma_5)v(p_{\bar{\nu}})\right\},
   \end{eqnarray}
   where $k$ is the momentum carried by the intermediate $\pi^0$ and $P=p_K-p_\nu$ is the total
   momentum flowing into the $\pi^0$-$\ell^+$ loop. $A_\alpha^{K^+\to\pi^0}$ and $A_\beta^{\pi^0\to\pi^+}$ represent
   the transition matrix elements indicated by the superscript and $\alpha,\,\beta$ are the Lorentz indices of the 
   weak currents.

   The detailed steps needed to evaluate $A_{\textrm{FV}}^{\pi^0\ell^+}$ are given in
   Appendix~\ref{appendix:FV}.
   Here we only discuss the results. For our current ensemble, with
   $m_\pi=420$ MeV, only the $\ell^+=e^+$ state can satisfy the on-shell
   condition and thus suffers from the non-exponential FV corrections. Our estimate of the 
   FV correction to the scalar amplitude for the electron mode is
   $F_{WW}^e(L)-F_{WW}^{e}(\infty)=1.528\cdot10^{-2}$, which is about 14\% of the
   $F_{WW}^e(L)$ as given in Table~\ref{table:WW}. When this FV correction
   propagates to
   $P_{c}$, it amounts approximately to approximately a 2\% contribution. After including this FV
   correction, we write the results for $P_{c}$ and
   $P_c-P_c^{\mathrm{PT}}$ as
   \ba
   P_c=0.2529(\pm13)(\pm32)(-45),\quad
   P_c-P_c^{\mathrm{PT}}=0.0040(\pm13)(\pm32)(-45).\label{eq:Pcfinal}
   \ea
   Since the calculations of the FV corrections require the determination of $A_\alpha^{K^+\to\pi^0}$ and
   $A_\beta^{\pi^0\to\pi^+}$, which we can only estimate at present, we choose not to decrease the central values in 
   Eq.\,(\ref{eq:Pcfinal}) but to include the estimate of the FV corrections in the uncertainty. 
   In general, the FV corrections depend on the lattice size $L$ and how the momenta
   for the intermediate pion and lepton are assigned and one needs to examine them for each case. In the future, when simulations are 
   performed with physical quark masses, it will be possible to use the calculated or measured values of the $K_{\ell 3}$ and pion 
   form factors at the corresponding momenta to determine the FV corrections reliably.

  \subsubsection{The momentum dependence}
  Using the effective Hamiltonian ${\mathcal H}_{\mathrm{eff},0}$ in
  Eq.~(\ref{eq:eff_Hamiltonian}) and the definition of $P_c$
   in Eq.~(\ref{eq:Pc_average}), one can write the
   $K^+\to\pi^+\nu\bar{\nu}$ decay amplitude as follows
   \ba
   \label{eq:decay_amplitude_local}
   \mathcal{A}(K^+\to\pi^+\nu\bar{\nu})&=&\frac{G_F}{\sqrt{2}}\frac{\alpha
   \lambda^5}{2\pi\sin^2\theta_W}\sum_{\ell=e,\mu,\tau}\left(\frac{\lambda_t}{\lambda^5}
   X_t(x_t)+ \frac{\lambda_c}{\lambda}P_c\right)
\nn\\
&&\hspace{0.15in}\times
   2f_+(s)\bar{u}(p_\nu){\slashed
   p}_K(1-\gamma_5)v(p_{\bar{\nu}}),
   \ea 
   where $X_t(x_t)$ and $P_c$ are the top and charm quark contributions,
   respectively and 
   $f_+(s)$ is the $K_{\ell3}$ form factor. 
   Note that the charm quark contribution $P_c$ generically depends on two variables $\Delta$ and
   $s$. In Eq.~(\ref{eq:result_ratio}) we have taken the ratio between the bilocal matrix
   element and local matrix element and assume this ratio does not have a
   significant $\Delta$ and $s$ dependence. As a consequence, $P_c$ in
   Eq.~(\ref{eq:decay_amplitude_local}) is approximated by a constant. We now examine under what
   circumstance this is a good approximation.

   With this aim in mind, we write out the explicit $\Delta$ and $s$
   dependence for $\mathcal{A}(K^+\to\pi^+\nu\bar{\nu})$ and $P_c$.
   Using the phase space factor for three-body decays\,\cite{Olive:2016xmw}, the decay width for
   $K^+\to\pi^+\nu\bar{\nu}$ can be written as
   \ba
   \Gamma[K^+\to\pi^+\nu\bar{\nu}]=\frac{1}{2^{10}\pi^4m_K^3}
   \int_0^{s_{\mathrm{max}}}ds \,\Delta_{\mathrm{max}}\int
   d\Omega\,
   |\mathcal{A}(\Delta,s)|^2 
   \ea
   where $d\Omega=\sin\theta\,d\phi\,d\theta$ is the element of solid angle of the neutrino's
   momentum in the center-of-mass frame of the $\nu\bar{\nu}$ pair and $\theta$
   indicates the angle between
   the momenta of the pion and neutrino in the same frame. We then have 
   \ba
   s_{\mathrm{max}}=(m_K-m_\pi)^2,\quad
   \Delta_{\mathrm{max}}=\sqrt{(m_K^2+m_\pi^2-s)^2-4m_K^2m_\pi^2},\quad
   \Delta=\Delta_{\mathrm{max}}\cos\theta. 
    \ea
    The square of the amplitude $|\mathcal{A}(\Delta,s)|^2$ is given by
   \ba
   |\mathcal{A}(\Delta,s)|^2&\propto&
   \left[\left(\frac{\mathrm{Im}\,\lambda_t}{\lambda_5}X_t(x_t)\right)^2+\left(\frac{\mathrm{Re}\,\lambda_c}{\lambda}P_c(\Delta,s)+\frac{\mathrm{Re}\,\lambda_t}{\lambda^5}X_t(x_t)\right)^2\right]
\nn\\
   &&\hspace{0.15in}\times 4 f_+(s)^2\left[\Delta_{\mathrm{max}}^2-\Delta^2\right]\,,
   \ea
   where the factor
   $\Delta_{\mathrm{max}}^2-\Delta^2$ arises from the relation
   $|\bar{u}(p_\nu){\slashed
   p}_K(1-\gamma_5)v(p_{\bar{\nu}})|^2=\Delta_{\mathrm{max}}^2-\Delta^2$.

   Assuming that the $\Delta$ and $s$ dependence in $P_c(\Delta,s)$ is mild 
   we perform a Taylor expansion writing
   \ba
   P_c(\Delta,s)=P_c(0,0)+b_\Delta
   \frac{\Delta}{m_K^2}+b_s\frac{s}{m_K^2}+\cdots.
   \ea
   Using this simple expansion as an input, the branching ratio of
   $K^+\to\pi^+\nu\bar{\nu}$ is proportional to
   \ba
   \mathrm{Br}[K^+\to\pi^+\nu\bar{\nu}]\propto 1+0.071\,
   b_\Delta^2+0.202\, b_s\,.
   \ea
   Here, we have used $X_t(x_t)=1.481$,
   $P_c(0,0)=0.404$,
   $\mathrm{Im}\,\lambda_t=1.51\times10^{-4}$,
   $\mathrm{Re}\,\lambda_t=-3.20\times10^{-4}$~\cite{Buras:2015qea}, $\lambda=0.22537$ and the PDG
   values for $m_K$ and
   $m_\pi$\,\cite{Olive:2016xmw}. We also make the approximation that 
   $\mathrm{Re}\,\lambda_c\simeq -\lambda$ and $f_+(s)\simeq 1$.
   If $|b_\Delta|<0.37$ and $|b_s|<0.05$, then the momentum dependence only
   amounts for a sub-percent effect. 
   
   Of course, since the present simulation was performed at a single choice of $(s,\Delta)$ we are unable to 
   estimate the size of the parameters $b_s$ and $b_\Delta$. Nevertheless, the above discussion will be useful in our 
   future studies (see Sec.\,\ref{sec:conclusion}) in which we will determine these parameters and use them to inform our choice 
   of kinematics for simulations at physical quark masses.

%% file: Conclusion.tex
   In this paper we have presented an exploratory lattice QCD calculation of the 
   long-distance contribution to the $K^+\to\pi^+\nu\bar{\nu}$ decay amplitude with a pion mass of $m_\pi=420$\,MeV
   and with a charm quark of mass $m_c^{\MS}(\mbox{2 GeV})=863$\,MeV. The main results have previously been reported in
   Ref.~\cite{Bai:2017fkh}. In this longer version
   we give the details explaining how 
    the bilocal hadronic matrix elements are evaluated and how the three main technical difficulties can be overcome. These are: 
    \begin{enumerate}
    \item[i)] the treatment of the additional ultraviolet divergences which arise in second order perturbation theory 
    when two local operators approach each other;
    \item[ii)] the subtraction of the unphysical terms which appear in Euclidean space and which 
    grow exponentially with the temporal extent of the  
    region of integration over the separation between the two local operators;
    \item[iii)] the correction for potentially large, i.e. non-exponential, finite-volume effects.
    \end{enumerate}
    By using 800 gauge configurations, the statistical uncertainty of the
    lattice result for $P_c$ is
    reduced to sub-percent level. We also make an analysis of the systematic errors,
    which gives us some guidance on how to control these uncertainties in
    future calculations. A curious feature of our results is that there is a very significant cancellation between the contributions
    from the $WW$ and $Z$-exchange diagrams to $P_c-P_c^{\textrm{PT}}$, see Fig.\,\ref{fig:Pcu} and the related discussion. 
    It will be very important and interesting to see if such a cancellation persists as the masses of the quarks are changed towards 
    their physical values in the future simulations discussed below.

    Because of the unphysical quark masses used in this simulation, it is premature to compare our
    current lattice result with perturbative calculations~\cite{Buras:2015qea} and the estimate of LD
    effects from Ref.~\cite{Isidori:2005xm}.
    The technique presented in this work can readily be
    generalized to a future realistic calculation. Such a simulation requires both a small lattice spacing to accommodate a 
    physically heavy charm quark, and a large volume to accommodate physically light pions. We foresee that within four years
    adequate resources will become available to make such a calculation possible with controlled systematic errors.  
    
    We end the discussion with our more immediate plans. We are currently performing a calculation with a lighter pion mass, 
    $m_\pi=170$\,MeV, on a $32^3\times64$ ensemble. This will 
    help us to control the uncertainty from the unphysical pion mass of 420\,MeV which we are currently using and provide 
    information about the $(\Delta,s)$ momentum dependence since the allowed momentum region will be larger.
    
    To include the physical charm quark mass, a fine lattice spacing is
    required. We are planning to use a $64^3\times128$ ensemble with an inverse lattice
  spacing of $1/a=2.38$\,GeV and with physical values for the light, strange and charm
  quark masses. As mentioned above, accurate results with a complete systematic error budget
  should be available within three to four years, which matches well with the experimental
  schedule to measure precisely the $K^+\to\pi^+\nu\bar{\nu}$ branching ratio.

\begin{acknowledgments}
We thank our colleagues in the RBC and UKQCD collaborations for many helpful
discussions.  Z.B., N.C. and X.F. were supported in part by U.S. DOE grant
\#De-SC0011941. C.T.S. (Leverhulme Emeritus Fellow) gratefully acknowledges
support from the Leverhulme Trust and was also supported in part by UK STFC
Grants ST/L000296/1 and ST/P000711/1. X.F. is supported in part by the National
Natural Science Foundation of China (NSFC) under
Grant No. 11775002 and by the Thousand Talents Plan for Young Professionals. A.P. is supported in part by UK STFC grants ST/L000458/1 and ST/P000630/1 
and by the European Research Council (ERC) under the European Union's Horizon 2020 research and innovation programme by grant agreement No 757646.
A.L. was supported by an EPSRC Doctoral Training Centre Grant No. EP/G03690X/1.
\end{acknowledgments}

%% file: Appendix_lepton_prop.tex
\label{appendix:lepton_prop}

The internal lepton is treated as an overlap fermion in this calculation.
We employ the overlap quark action from Refs.~\cite{Capitani:2002mp,Aoki:2008tq}, with the Dirac operator
defined as
\ba
D(0)&=&\rho\,(1+\gamma_5\,\textmd{sgn}\,[H_W(-\rho)]),\quad H_W=\gamma_5 D_W
\nn\\
D(m)&=&\left(1-\frac{m}{2\rho}\right)D(0)+m\,,
\ea
where $D_W$ is the Wilson Dirac operator. Here we set the Wilson parameter
$r=1$. The parameter $\rho$ introduced into the overlap fermion action is
equivalent to the five-dimensional domain wall height $M_5$ in the domain wall
fermion action. $m$ is the lepton mass.

It is useful to write the propagator of the free overlap fermion in momentum space
\ba
S(p)=\frac{1}{2}\frac{(\rho-\frac{m}{2})X^\dagger(p)+(\rho+\frac{m}{2})\omega(p)}
{\left(\rho^2+\frac{m^2}{4}\right)\omega(p)+\left(\rho^2-\frac{m^2}{4}\right)b(p)}\,,
\ea
where
\ba
X(p)&=&i\sum_\mu\gamma_\mu\sin p_\mu+r\sum_\mu(1-\cos p_\mu)-\rho
\nn\\
b(p)&=&r\sum_\mu(1-\cos p_\mu)-\rho
\nn\\
\omega(p)&=&\sqrt{\sum_\mu\sin^2 p_\mu+\left(r\sum_\mu(1-\cos p_\mu)-\rho\right)^{\!\!2}}\,.
\ea

When $0<\rho<2r$ there is no pole at $p_4=\pi+iE$, since 
\ba
b(p)=\left(r(1+\cosh E)+r\sum_i(1-\cos p_i)-\rho\right)>0
\ea
and the constraint $\omega(p)+b(p)=0$ (corresponding to the massless case) cannot be satisfied. 
Thus the fermion doubling problem is solved and the correct spectrum of massless fermions is obtained in
the range $0<\rho<2r$. We therefore only need to consider the pole at $p=({\bf p}, iE_a)$,
which satisfies the relation
\ba
\label{eq:overlap_pole}
\sum_\mu\sin^2 p_\mu = -\frac{\rho^2 m^2}{(\rho^2+ m^2/4)^2}b^2(p)\equiv-\bar{m}^2b^2(p)\,.
\ea

\begin{figure}
   \centering
   \includegraphics[width=.8\textwidth]{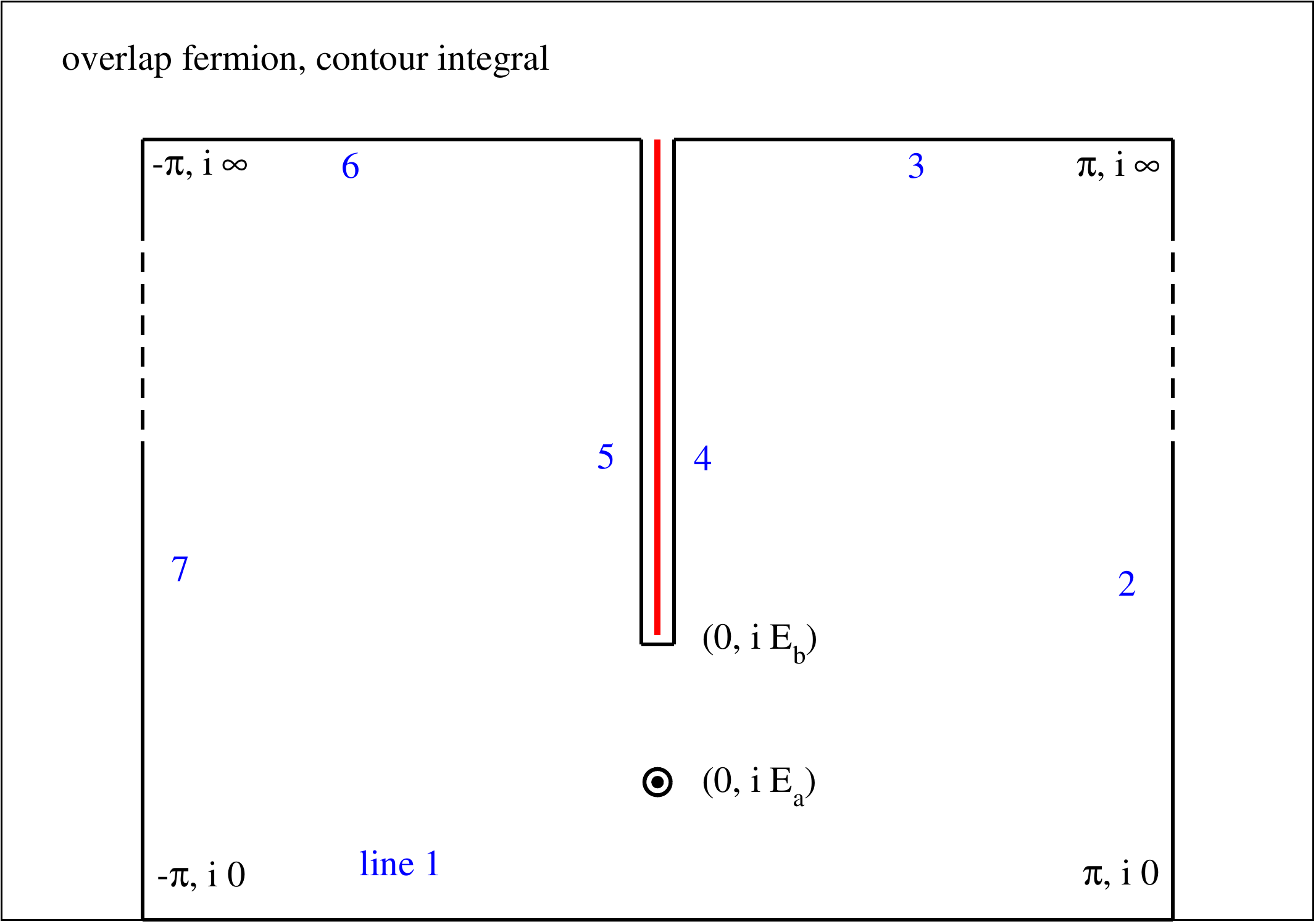}
   \caption{Line 1 is the contour of the integral on the left-hand side of  
   Eq.\,(\ref{eq:Cauchy}) for $\mathbf{p}=\mathbf{0}$. For $t>0$, 
   we close the contour in the upper-half plane picking up the residue of the pole at
   $p_4=iE_a$ and the contributions from the cut starting at $(0,iE_B)$ leading to the three contributions on the 
   right-hand side of Eq.\,{(\ref{eq:Cauchy})}. }
   \label{fig:overlap}
   \end{figure}

We next perform the Fourier transform in the time direction and convert the propagator 
to the momentum-time representation
\ba
S({\bf p},t)=
\int_{-\pi}^{\pi} \frac{dp_4}{2\pi}\,S({\bf p},p_4)e^{ip_4 t}.
\ea
Here the integral $\int_{-\pi}^{\pi}\frac{dp_4}{2\pi}$ is
used to obtain the propagator with infinite time extension.
$S({\bf p},t)$ can be determined using Cauchy integration. Note that
the square root in
$\omega(p)$ brings in two branch cuts, one from a starting point $+iE_b$
to $+ i\infty$ and the other from $-iE_b$ to $-i\infty$, where $p=({\bf p},iE_b)$ is the zero of $\omega(p)$. 
So the contour of the
Cauchy integral should exclude these branch cuts as shown in
Fig.\,\ref{fig:overlap}. For $t>0$ we have
\ba\label{eq:Cauchy}
\int_{-\pi}^{\pi}\frac{dp_4}{2\pi}\, f(p_4)=i\,\mathrm{res}\{
f\}_{p_4=iE_a}+\int_{+iE_b+\epsilon}^{+i\infty+\epsilon}\frac{dp_4}{2\pi}\,f(p_4)
-\int_{+iE_b-\epsilon}^{+i\infty-\epsilon}\frac{dp_4}{2\pi}\,f(p_4),\quad
\ea
with $f(p_4)= S({\bf p},p_4)e^{ip_4t}$. In the first term on the right-hand side
$\mathrm{res}\{
f\}_{p_4=iE_a}$ is the residue of $f(p_4)$ at the pole $p_4=iE_a$. 
For $t<0$, we can choose the contour
along the branch cut $-iE_b$ to $-i\infty$ to determine $S({\bf p},t)$. 

The propagator $S({\bf p},t)$ can be written in two parts: the residue of the
pole at $p_4=\pm iE_a$, $S_a({\bf p},t)$, and the contribution from the branch
cuts, $S_b({\bf p},t)$. 
We first focus on the contribution from the pole and find
\ba
S_a({\bf p},t)=\frac{{\mathcal C}\left(\mathrm{sgn}(t) \sinh E_a\gamma_t-i\sum_i \sin
p_i \gamma_i\right)+{\mathcal M}}{2{\mathcal E}}e^{-E_a|t|}\,,
\ea
with
\ba
\cosh E_a&=&\frac{\bar{m}^2r\left(\rho-r-r\sum_i(1-\cos p_i)\right)+{\mathcal
U}}{1-\bar{m}^2r^2}
\nn\\
{\mathcal E}&=&\frac{\sinh
E_a}{\rho}\left(\rho^2+\frac{m^2}{4}\right)\frac{{\mathcal U}}{\omega({\bf
p},iE_a)}
\nn\\
{\mathcal U}^2&=&\bar{m}^2\left(\rho-r-r\sum_i(1-\cos
p_i)\right)^{\!\!2}+(1-\bar{m}^2r^2)\left(1+\sum_i\sin^2 p_i\right)
\nn\\
{\mathcal C}&=&1-\frac{m}{2\rho}
\nn\\
{\mathcal M}&=&\frac{m}{\rho+m/2}\omega({\bf p},iE_a).
\ea
In order to achieve $O(a)$ improvement the propagator is modified as follows:
\ba
\hat{S}({\bf p},t)=\left(1-\frac{D(0)}{2\rho}\right)S({\bf p},t)=\frac{1}{{\mathcal
C}}S({\bf p},t)-\frac{1}{2\rho-m}\,.
\ea
This modification cancels the coefficient ${\mathcal C}$ in $S({\bf p},t)$.
The mass term ${\cal M}$ and $1/(2\rho-m)$  do not contribute to this calculation because of the $V-A$ structure of the
two weak operators. We therefore write the modified propagator $\hat{S}_a({\bf p},t)$ as
\ba
\hat{S}_a({\bf p},t)&\sim&\frac{\mathrm{sgn}(t)\sinh E_a\gamma_t-i\sum_i\sin
p_i\gamma_i}{2{\mathcal E}}e^{-E_a|t|}
\nn\\
&=&\frac{\sinh E_a}{{\mathcal E}}\frac{\mathrm{sgn}(t)\sinh
E_a\gamma_t-i\sum_i\sin p_i\gamma_i}{2\sinh E_a}e^{-E_a|t|}\,.
\ea
We define the wave function normalization factor by $Z_\ell=\frac{\sinh E_a}{\mathcal
E}\bigl|_{\bf p=0}$. As the lepton mass approaches $0$, 
$Z_\ell\to1$. For large lepton masses, e.g. when $\ell=\tau$, we multiply $\hat{S}_a({\bf p},t)$
by $Z_\ell^{-1}$ in order to make the propagator have a closer
form to the continuum one. Another subtelty is that at ${\bf p}=0$, the energy
$E_a$ deviates from the input mass parameter $m$. For $\rho=r=1$, we
have
\ba
m=2\operatorname{tanh}\frac{E_a}{2}\bigg|_{\bf p=0}.
\ea
We tune the parameter $m$ for each lepton, $e$, $\mu$ and $\tau$, to ensure that the pole mass
$E_a$ at ${\bf p=0}$ takes the physical value of the mass of the lepton.

The branch-cut contribution is suppressed at large $t$. Its integral representation is 
\ba
S_b({\bf
p},t)=\int_{E_b}^\infty\frac{dE}{2\pi}\frac{(\rho-m/2)(\rho^2+m^2/4)\omega\sinh E}
{(\rho^2+m^2/4)^2\omega^2+(\rho^2-m^2/4)^2b^2}\,e^{-E|t|}\,.
\ea

%% file: Appendix_YAB.tex
\label{appendix:YAB}

In this appendix we evaluate the amputated Green's function in
Eq.~(\ref{eq:amputated_Green_F})
using naive dimensional regularization (NDR) with a fully anticommuting $\gamma^5$.
The external momentum $p_i$ are
given by Eq.~(\ref{eq:extern_mom}). Since the external legs are amputated, at $O(\alpha_s^0)$
only the momentum $p=p_{\mathrm{loop}}$ enters as a parameter in the 4-momentum
integral. 

For the $W$-$W$ diagram, we have
\ba
\Gamma^{WW}(p)&=&\Gamma_u^{WW}(p)-\Gamma_c^{WW}(p)\quad\textrm{where}
\label{eq:WWGIM}\\
\Gamma^{WW}_q(p)&\equiv&\mu^{\varepsilon}\int
\frac{d^Dk}{(2\pi)^D}\gamma_\mu^LS_q(-k)\gamma_\nu^L\otimes
\gamma_\mu^LS_\ell(k+p)\gamma_\nu^L,\label{eq:WWintegral}
\ea
$D$ is the number of space-time dimensions, $\varepsilon=4-D$ and the factor $\mu^\varepsilon$ ensures that $\Gamma_q^{WW}$ has the correct dimensions. In the integrand on the right-hand side of Eq.\,(\ref{eq:WWintegral}) $S_q(k)=\frac{1}{i{\slashed k}+m_q}$ is the quark propagator with $q=u,c$
and $S_\ell(k)=\frac{1}{i{\slashed k}+m_\ell}$ is the lepton propagator. The
gamma matrix $\gamma_{\mu}^L$ ($\gamma_\mu^R$) is defined as $\gamma_\mu^L\equiv
\gamma_\mu(1-\gamma_5)$ ($\gamma_\mu^R\equiv \gamma_\mu(1+\gamma_5)$). 

As a result of the GIM cancellation in Eq.\,(\ref{eq:WWGIM}), $\Gamma^{WW}(p)$ is logarithmically divergent 
($\Gamma^{WW}_u$ and $\Gamma^{WW}_c$ are separately quadratically divergent).
The standard way to evaluate $\Gamma^{WW}_q$ requires the use of the $\gamma$
matrix algebra in the $D$ dimension. 
However, if we perform the subtraction
$\Delta \Gamma^{WW}\equiv \Gamma^{WW}(p) - \Gamma^{WW}(0)$
then $\Delta \Gamma^{WW}$ is a finite quantity. We can then let $D\to 4$ and
calculate $\Delta \Gamma^{WW}$ directly using the 4-dimensional $\gamma$-matrix
algebra. 

The conversion term $Y_{AB}^{WW}(\mu,\mu_0)$ is given by
\ba
Y_{AB}^{WW}(\mu,\mu_0)=\mu^{\varepsilon}\, \int \frac{d^Dk}{(2\pi)^D}
\frac{\mathrm{Tr}[\gamma_\mu^LS_u(-k)\gamma_\nu^L\gamma_\rho^R]
    \mathrm{Tr}[\gamma_\mu^LS_\ell(k+p)\gamma_\nu^L\gamma_\rho^R]}
    {\mathrm{Tr}[\gamma_\mu^L\gamma_\rho^R]\mathrm{Tr}[\gamma_\mu^L\gamma_\rho^R]}-\{u\to c\}\,,
\ea
where $\mu_0^2=p_i^2$. Here we retain our general notation with the operators denoted by $A$ and $B$ on the left hand side, but with the specific operators for the $W$-$W$ diagrams included on the right-hand side (see Eq.\,(\ref{eq:OWW})).

At zero external momentum
$Y_{AB}^{WW}(\mu,0)=r_{AB}^{WW}(\mu)$, where $r_{AB}^{WW}(\mu)$ is given by
Eq.~(\ref{eq:matrix_element_PT}). The difference between
$Y_{AB}^{WW}(\mu,\mu_0)$ and $Y_{AB}^{WW}(\mu,0)$ is finite and
we evaluate it at order $O(\alpha_s^0)$:
\ba
\label{eq:Delta_YAB_WW}
\Delta Y_{AB}^{WW}(\mu,\mu_0)&\equiv& Y_{AB}^{WW}(\mu,\mu_0)-Y_{AB}^{WW}(\mu,0)
\nn\\
&=&16
\left[\left(I_1(0,m_\ell,p)-I_1(m_c,m_\ell,p)\right)-\left(I_1(0,m_\ell,0)-I_1(m_c,m_\ell,0)\right)\right],\label{eq:I1terms}
\ea
where $p^2=\mu_0^2$ and 
\ba
I_1(m_1,m_2,p)\equiv\frac{1}{16\pi^2}\int_0^1dx\,(\Delta-A)\ln\frac{\Delta}{\mu^2}\,,
\ea
$x$ is a Feynman parameter, 
$\Delta=x(1-x)p^2+xm_1^2+(1-x)m_2^2$ and $A=2x(x-1)p^2-xm_1^2-(1-x)m_2^2$.

For the $Z$-exchange diagram with the insertion of the axial vector current, we evaluate the amputated Green's function writing
\ba
\label{eq:Y_Z_A}
\Gamma^{Z,A}&=&\Gamma^{Z,A}_u-\Gamma^{Z,A}_c
\nn\\
\Gamma_q^{Z,A}&=&
\left\{
\begin{array}{ll}
q_A\,\mu^{\varepsilon}\int \frac{d^Dk}{(2\pi)^D}\,
\gamma_\mu^LS_q(k+p)\gamma_\nu\gamma_5S_q(k)\gamma_\mu^L\otimes \gamma_\nu^L, & ~\mbox{for the $Q_2$ operator}
\\
-q_A\,\mu^{\varepsilon}\int \frac{d^Dk}{(2\pi)^D}\,
\mathrm{Tr}[\gamma_\mu^LS_q(k+p)\gamma_\nu\gamma_5S_q(k)]\gamma_\mu^L\otimes
\gamma_\nu^L, & ~\mbox{for the $Q_1$ operator}
\end{array}
\right.
\ea
where $q_A=-T_3^u=-\frac{1}{2}$ and $T_3^u$ is the weak isospin for the up-type
quarks.

Performing the projection and evaluating $\Delta Y_{AB}^{Z,A}\equiv
Y_{AB}^{Z,A}(\mu,\mu_0)-Y_{AB}^{Z,A}(\mu,0)$, we find
\ba
\label{eq:Delta_YAB_ZA}
\Delta
Y_{AB}^{Z,A}=
\left\{
\begin{array}{ll}
2q_A\left[\left(I_2(0,p)-I_2(m_c,p)\right)+I_2(m_c,0)\right], & ~ \mbox{for the
$Q_2$ operator} \\
2q_A N_c\left[\left(I_2(0,p)-I_2(m_c,p)\right)+I_2(m_c,0)\right], & ~ \mbox{for the
$Q_1$ operator} \label{eq:I2terms}
\end{array}
\right.
\ea
where
\ba
I_2(m,p)=\frac{1}{16\pi^2}\int_0^1dx\,\left[3x(1-x)p^2+4m^2\right]\ln\frac{x(1-x)p^2+m^2}{\mu^2}.
\ea

For the insertion of the vector current, we can simply replace $\gamma_\nu\gamma_5\to\gamma_\nu$ and $q_A\to
q_V=T_3^u-2Q_{\mathrm{em},u}\sin^2\theta_W$ in Eq.~(\ref{eq:Y_Z_A}), where
$Q_{\mathrm{em},u}$ is the electric charge for up-type quarks and $\theta_W$ is the Weinberg angle. We have
\ba
\label{eq:Delta_YAB_ZV}
\Delta Y_{AB}^{Z,V}(\mu,\mu_0)=
\left\{
\begin{array}{ll}
-2q_V\left[\left(I_3(0,p)-I_3(m_c,p)\right)+I_3(m_c,0)\right], & ~ \mbox{for the 
$Q_2$ operator} \\
-2q_V N_c\left[\left(I_3(0,p)-I_3(m_c,p)\right)+I_3(m_c,0)\right], & ~
\mbox{for the $Q_1$ operator}\label{eq:I3terms}
\end{array}
\right.
\ea
where
\ba
I_3(m,p)=\frac{1}{16\pi^2}\int_0^1dx\,3x(1-x)p^2\ln\frac{x(1-x)p^2+m^2}{\mu^2}.
\ea
Note that the contributions to $\Delta Y_{AB}$ are finite and $\log \mu^2$ cancels in each of the expressions in Eqs.\,(\ref{eq:I1terms}),
(\ref{eq:I2terms}) and (\ref{eq:I3terms}) at lowest order.

At large RI/SMOM scales, we have
\ba
\Delta Y_{AB}^{WW}(\mu,\mu_0)&\xrightarrow[]{\mu_0^2\gg m_c^2}&
\frac{m_c^2}{\pi^2}\left(-\frac{x_\ell \ln x_\ell}{1-x_\ell} -\ln\frac{\mu_0^2}{m_c^2}\right)
\nn\\
\Delta Y_{AB}^{Z,A}(\mu,\mu_0)&\xrightarrow[]{\mu_0^2\gg m_c^2}&
\frac{m_c^2}{4\pi^2}\left(-\frac{5}{4}+\ln\frac{\mu_0^2}{m_c^2}\right),\quad
\mbox{for the $Q_2$ operator}
\nn\\
\Delta Y_{AB}^{Z,V}(\mu,\mu_0)&\xrightarrow[]{\mu_0^2\gg
m_c^2}&q_V\frac{3}{8}\frac{m_c^2}{\pi^2},\quad \mbox{for the $Q_2$ operator}.\label{eq:YABresults}
\ea
For the $Z$-exchange diagram, the results for the $Q_1$ operator are obtained by simply multiplying those for  
the $Q_2$ operator in Eq.\,(\ref{eq:YABresults}) by a factor of $N_c$.

Combining $\Delta Y_{AB}(\mu,\mu_0)$ with $r_{AB}^{\MS}(\mu)$ and taking $\mu_0=\mu$ we have
\ba
\label{eq:YAB}
Y_{AB}^{WW}(\mu,\mu_0)\bigl|_{\mu=\mu_0}&\xrightarrow[]{\mu_0^2\gg
m_c^2}&\frac{5}{4}\frac{m_c^2}{\pi^2}
\nn\\
Y_{AB}^{Z,A}(\mu,\mu_0)\bigl|_{\mu=\mu_0}&\xrightarrow[]{\mu_0^2\gg
m_c^2}&-\frac{1}{16}\frac{m_c^2}{\pi^2},~ \mbox{for the $Q_2$ operator}
\nn\\
Y_{AB}^{Z,V}(\mu,\mu_0)\bigl|_{\mu=\mu_0}&\xrightarrow[]{\mu_0^2\gg
m_c^2}&q_V\frac{3}{8}\frac{m_c^2}{\pi^2},~ \mbox{for the $Q_2$ operator}.
\ea

%% file: Appendix_FV.tex
    \label{appendix:FV}

    We rewrite the expression in Eq.\,(\ref{eq:WW_FV}) in a more general form:
   \begin{eqnarray}
   I_{FV}=I(L)-I(\infty)=\left(\frac{1}{L^3}\sum_{\vec{k}}\int \frac{dk_4}{2\pi}
   -{\mathcal P}\int\frac{d^4k}{(2\pi)^4}\right)\frac{f(k_0,{\bf
   k})}{(k^2+m_1^2)((P-k)^2+m_2^2)},
   \end{eqnarray}
   where $P=p_K-p_{\nu}$, $m_1=m_\pi$ and $m_2=m_{\bar{\ell}}$ in our
   calculation.
   For the moving frame (${\bf P}\neq{\bf 0}$) and non-identical
   particles ($m_1\neq m_2$), the finite volume correction can be written as
   \ba
   \label{eq:I_FV}
   I_{FV}=\frac{1}{2E^*}\sum_{lm}f_{lm}^*(p^*)c_{lm}^P(p^*)
   \ea
   where the superscript $*$ indicates the center-of-mass frame. The energy $E^*$ is
   the total energy in the center-of-mass frame, satisfying $E^{*2}=P_0^2-{\bf
   P}^2$, where the Minkowski and Euclidean energies, $P_0$ and $P_4$ respectively, are related by $P_0=-iP_4$. The momentum $p^*$ satisfies the on-shell condition
   $E^*=\sqrt{m_1^2+p^{*2}}+\sqrt{m_2^2+p^{*2}}$.
   The Lorentz boost factor $\gamma$ is given by $\gamma\equiv P_0/E^*$.
   Under the Lorentz transformation from the moving frame to the center-of-mass frame, the
   function $f({\bf k})$ changes as $f({\bf k})\to f^*({\bf k^*})$. 
   The potentially large finite volume effects appear when the two particles in the
   intermediate state are both on-shell. In this case, the function $f({\bf k})$
   corresponds to the on-shell physical transition and thus is Lorentz invariant:
   $f({\bf k})=f^*({\bf k^*})$. In Eq.\,(\ref{eq:I_FV}) $f^*_{lm}(k^*)$ is coefficient in the partial wave expansion of 
   the function $f^*({\bf k^*})$:
   \ba
   f^*({\bf k}^*)=\sum_{lm}f^*_{lm}(k^*)k^{*l}Y_{lm}(\Omega_{\bf k^*}),\quad
   k^*\equiv|{\bf k^*}|=p^*.
   \ea
   The function $c_{lm}^P(p^*)$ is given by~\cite{Kim:2005gf,Briceno:2012yi}
   \ba
   \label{eq:c_lm}
   c_{lm}^P(p^*)=\frac{1}{\gamma}\left(\frac{1}{L^3}\sum_{\bf
       k\in\frac{2\pi}{L}{\bf n}}-{\mathcal
   P}\int\frac{d^3{\bf k}}{(2\pi)^3}\right)\frac{|\hat{\bf
       k}|^{l}Y_{lm}(\Omega_{\hat{\bf k}})}{|\hat{\bf k}|^{2}-p^{*2}}\,,
   \ea
   where the momentum $\hat{\bf k}$ is defined as 
   \ba
   \hat{\bf k}=\gamma^{-1}\left[{\bf k}_\parallel-\frac{\bf
   P}{2}\left(1+\frac{m_1^2-m_2^2}{E^{*2}}\right)\right]+{\bf k}_\perp,\quad {\bf k}_\parallel=\frac{{\bf k}\cdot{\bf
   P}}{|{\bf P}|^2}{\bf P},\quad {\bf k}_\perp={\bf k}-{\bf k}_\parallel.
   \ea
   The subscripts $\parallel$ and $\perp$ refer to parallel to and perpendicular to $\mathbf{P}$ respectively.
   Each of the two terms in Eq.\,(\ref{eq:c_lm}) is separately ultraviolet divergent but the difference is convergent.
   The divergence can be regulated by  
   introducing an exponential factor $e^{\alpha(p^{*2}-|\hat{\bf k}|^{2})}$ with
   $\alpha>0$ to the summand/integrand in Eq.\,(\ref{eq:c_lm}).  By using the heat kernel method
   proposed by L\"uscher~\cite{Luscher:1990ux}, one can evaluate $c_{lm}^P(p^*)$ in the limit of
   $\alpha\to 0^+$. 

   Once $c_{lm}^P(p^*)$ is determined, 
   the remaining task is to evaluate $f_{lm}^*(k^*)$. The scalar amplitude $f(\mathbf k)$ is defined from the 
   transition amplitude
   \ba
   {\mathcal
   A}^{K^+\to\pi^0\ell^+\nu}&=&i\, A_\alpha^{K^+\to\pi^0}(p_K,k)A_\beta^{\pi^0\to\pi^+}(k,p_\pi)\bar{u}(p_\nu)\gamma^\alpha(1-\gamma_5)
   ({\slashed P}-{\slashed k})\gamma^\beta(1-\gamma_5)v(p_{\bar{\nu}})
   \nn\\
   &=& i\, f({\bf k})\,\bar{u}(p_\nu){\slashed p}_K (1-\gamma_5)v(p_{\bar{\nu}})\,.
   \ea
   The 
   scalar amplitude $f({\bf k})$ is then obtained from ${\mathcal A}_{K^+\to\pi^0\ell^+\nu}$ through the projection
   \ba
   f({\bf k})=\frac{\mathrm{Tr}[{\slashed
   A}^{K^+\to\pi^0}(1-\gamma_5)
   ({\slashed P}-{\slashed k}){\slashed A}^{\pi^0\to\pi^+}
   (1-\gamma_5){\slashed
   p}_{\bar{\nu}}{\slashed p}_K(1-\gamma_5){\slashed
   p}_\nu]}{\mathrm{Tr}[{\slashed p}_K(1-\gamma_5){\slashed
   p}_{\bar{\nu}}{\slashed p}_K(1-\gamma_5){\slashed p}_\nu]}.
   \ea
   We now make the following approximations:
   $A_\alpha^{K^+\to\pi^0}(p_K,k)\simeq i(p_K+k)_\alpha$ and 
   $A_\beta^{\pi^0\to\pi^+}(k,p_\pi)\simeq i(p_\pi+k)_\beta$, which correspond to setting the $K_{\ell 3}$ and pion form-factors to 1.
   (In future simulations with physical quark masses these approximations can be relaxed by using the measured or computed form
   factors at the corresponding momentum transfers.)
   After performing the Lorentz
   boost, we have
   \ba
   f^*({\bf k}^*)=\frac{-2\mathrm{Tr}\left[({\slashed p}_K^*+{\slashed
           k}^*)({\slashed P}^*-{\slashed
   k}^*)({\slashed p}_\pi^*+{\slashed k}^*)(1-\gamma_5){\slashed
   p}_{\bar{\nu}}^*{\slashed p}_K^*(1-\gamma_5){\slashed p}_\nu^*\right]}
   {\mathrm{Tr}[{\slashed p}_K^*(1-\gamma_5){\slashed p}_{\bar{\nu}}^*{\slashed
   p}_K^*(1-\gamma_5){\slashed p}_\nu^*]}.
   \ea
   Finally, after performing the partial wave expansion for $f^*({\bf k}^*)$, the finite volume
   corrections are given by Eq.~(\ref{eq:I_FV}).